\documentclass[rmp,aps,twocolumn,floatfix]{revtex4}
\usepackage{amssymb}
\usepackage{amsmath}
\usepackage{graphicx}
\usepackage{color}

\usepackage{mathrsfs,bbm}
\usepackage{comment}
\newcommand{\bm}[1]{\boldsymbol #1}
\newcommand{\lh}{\mbox{$\neg$}}
\newcommand{\rh}{\reflectbox{$\neg$}}
\usepackage{txfonts}
\DeclareSymbolFont{symbols}{OMS}{cmsy}{m}{n}
\DeclareSymbolFont{largesymbols}{OMX}{cmex}{m}{n}

\newcommand{\TR}{\text{Tr}}
\newcommand{\BK}{{\bm k}}
\newcommand{\Vk}{{\bm k}}

\newcommand{\tb}{{\bar t}}
\newcommand{\myast}{{{}\hspace*{-0.1em}\ast\hspace*{-0.1em}{}}}
\newcommand{\mydagger}{{\dagger}}
\newcommand{\phdagger}{{\phantom{\dagger}\!}}
\newcommand{\bra}[1]{\langle{#1}|}
\newcommand{\ket}[1]{|{#1}\rangle}

\newcommand{\expval}[1]{\langle{#1}\rangle}
\newcommand{\CS}{\mathcal{S}}

\newcommand{\delC}{\delta_\CC}
\newcommand{\intC}{\int_\CC}

\newcommand{\tmax}{t_{\text{max}}}
\newcommand{\CC}{\mathcal{C}}
\newcommand{\TC}{\mathcal{T}_{\CC}}
\newcommand{\gtrc}{\succ}

\newcommand{\convz}{\ast}

\newcommand{\Gwnull}{Q}\newcommand{\Gweins}{R}
\newcommand{\gweins}{r}
\renewcommand{\dh}{dh}
\newcommand{\jd}{\Gamma_\text{\dh}}
\newcommand{\fth}{F_\text{th}}

\renewcommand{\bar}{\overline}

\newcommand{\Tr}{\text{Tr}}

\begin{document}
\author{Hideo Aoki}
\affiliation{Department of Physics, University of Tokyo, Hongo, Tokyo 113-0033, Japan}
\author{Martin Eckstein}
\affiliation{Max Planck Research Department for Structural Dynamics, University of Hamburg-CFEL, Hamburg, Germany}
\author{Marcus Kollar}
\affiliation{Theoretical Physics III, Center for Electronic Correlations and Magnetism, Institute of Physics, University of Augsburg, 86135 Augsburg, Germany}
\author{Takashi Oka}
\affiliation{Department of Applied Physics, University of Tokyo, Hongo, Tokyo 113-8656, Japan}
\author{Naoto Tsuji}
\affiliation{Department of Physics, University of Tokyo, Hongo, Tokyo 113-0033, Japan}
\author{Philipp Werner}
\affiliation{Department of Physics, University of Fribourg, 1700 Fribourg, Switzerland}

\title{Nonequilibrium dynamical mean-field theory and its applications}

\begin{abstract}
The study of nonequilibrium phenomena in correlated lattice systems has developed into one of the most active and exciting branches of condensed matter physics.  This 
research field provides rich new insights that could not be obtained from the study of equilibrium situations, and the theoretical understanding of the physics often requires the development of new concepts and methods. On the experimental side, ultrafast pump-probe spectroscopies enable studies of excitation and relaxation phenomena in correlated 
electron systems, while ultracold atoms in optical lattices provide a new way to control and measure the time evolution of interacting lattice systems with 
a vastly different characteristic time scale compared to electron systems. A theoretical description of these phenomena is challenging because, first, the quantum-mechanical time evolution of many-body systems out of equilibrium must be computed and second, strong-correlation effects which can be of 
nonperturbative nature must be addressed. This review discusses the nonequilibrium extension of the dynamical mean field theory (DMFT), which treats quantum fluctuations in the time domain and works directly in the thermodynamic limit. The method reduces the complexity of the calculation via a mapping to a self-consistent impurity problem, which becomes exact in infinite dimensions. Particular emphasis is placed on a detailed derivation of the formalism, and on a discussion of numerical techniques, which enable solutions of the effective nonequilibrium DMFT impurity problem. Insights gained into the properties of the infinite-dimensional Hubbard model under strong nonequilibrium conditions are summarized. These examples illustrate the current ability of the theoretical framework to reproduce and understand fundamental nonequilibrium phenomena, such as the dielectric 
breakdown of Mott insulators, photodoping, and collapse-and-revival oscillations in quenched systems. Furthermore, 
remarkable novel phenomena have been predicted by the nonequilibrium DMFT simulations of correlated lattice systems, 
including dynamical phase transitions and field-induced repulsion-to-attraction conversions.  
\end{abstract}

\maketitle

\tableofcontents

%\input{introduction}
%%%%%%%%%%%%%%%%%%%%%%%%%%%%%%%%%%%%%%%%%%%%%%%%%%%%%%%%%%%%%%%%%%%%%%%%%%%%%%%%
\section{Introduction}

\paragraph{Strongly correlated systems out of equilibrium}
There is a growing realization that nonequilibrium physics is a major avenue in condensed matter physics.  Of particular interest are nonequilibrium phenomena in 
strongly correlated electron systems. This class of materials has been intensively studied since the discovery of high-$T_c$ superconductivity in the cuprates.   
Already in equilibrium, strong electronic correlations bring about  a tantalizing variety of novel phenomena, such as
metal-to-Mott insulator transitions and transitions to magnetic and superconducting states. If such a system 
is driven out of equilibrium, we can expect even richer physics, of which only a tiny fraction has been 
discovered so far, and of which even less can be considered as being ``understood''.
The present article reviews a recently developed theoretical approach to study those strongly correlated many-body systems out of equilibrium, 
namely the nonequilibrium dynamical mean-field theory (DMFT), and illustrates its strength and versatility with numerous applications 
that have led to new physical insights in several cases.

Over the last two decades, 
DMFT has greatly contributed to our present understanding of strongly correlated systems in equilibrium, in particular, to Mott physics \cite{Georges96, Kotliar2006}.  
It provides the exact solution of lattice models in the infinite-dimensional limit \cite{Metzner1989}. 
The method treats spatial correlations in a mean-field manner, which allows a self-consistent formulation in terms of an effective single-site impurity problem \cite{Georges1992a}, 
but accurately treats the temporal quantum fluctuations that are essential for 
describing strong-correlation phenomena such as the Mott transition. 
Another virtue of DMFT was realized when \citet{Schmidt2002} proposed  
a nonequilibrium generalization of DMFT, by introducing the Keldysh formalism (see Sec.~\ref{steady-states}) 
to describe nonequilibrium steady states of correlated electrons driven by time-periodic external fields.
While the set-up considered in this pioneering paper (a spatially uniform scalar potential) did not correctly capture the effect of an applied electric field, nor the dissipation mechanism
which is essential for the description of nonequilibrium steady states (see Sec.~\ref{steady-states}), it laid the groundwork 
for the formalism we now call nonequilibrium DMFT.   
A general formulation of the nonequilibrium DMFT and its application to 
an electric field driven lattice system 
was then given by \citet{Freericks2006}, 
who employed the Kadanoff-Baym formalism (see Sec.~\ref{kadanoff-baym}) to describe general transient real-time evolutions from a thermal initial state. 

\begin{figure}[t]
\begin{center}
\includegraphics[width=7cm]{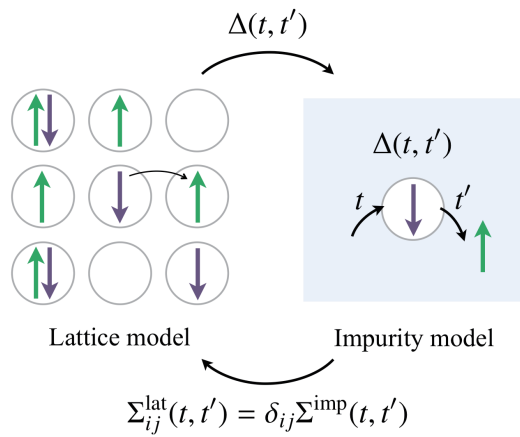}
\caption{A schematic picture of the nonequilibrium DMFT formalism. $\Delta(t,t')$ is the hybridization function, while $\Sigma(t,t')$ is the self-energy.}
\label{intro:dmft}
\end{center}
\end{figure}

The essential approximation underlying DMFT, both in and out of equilibrium, is the local nature (or momentum-independence) of the self-energy $\Sigma$. 
This approximation allows to map the lattice problem onto an impurity problem with a single correlated site embedded in an uncorrelated bath. 
The hybridization between the impurity site and the bath (the dynamical mean field) is represented by
a two-time function $\Delta(t,t')$, which is subject to a self-consistency condition (see Fig.~\ref{intro:dmft}).
Once a self-consistent solution has been obtained, the impurity self-energy $\Sigma^{\rm imp}$, which is local but time-dependent, yields the approximate lattice self-energy, so that
the DMFT approximation reads
\begin{equation}
\Sigma_{ij}^\text{lat}(t,t')\approx \delta_{ij}\Sigma^\text{imp}(t,t').
\end{equation}
In a time-dependent problem, the self-energy $\Sigma$ becomes a function of two time arguments $(t,t')$, not just the time difference, and the theory thus incorporates an overall temporal evolution of correlated systems, as well as 
quantum fluctuations.

The DMFT formalism represents the lattice system as a collection of local entities (atoms or sites) rather than in terms of 
extended Bloch states, and is thus well suited to treat strong local interactions such as the on-site Hubbard interaction $U$
in a non-perturbative manner. 
While the reduction from a correlated lattice system to 
an impurity model is a drastic simplification, the quantum impurity model is still a highly non-trivial 
many-body system, which must be solved with suitable numerical methods.  Over the past 
several years, the DMFT formalism and various numerical techniques for solving the effective impurity model 
have been extended to time evolutions (see Sec.~\ref{methods}), and subsequently applied to a broad range of problems, including 
electric-field- and quench-induced phenomena (see Sec.~\ref{applications}).

What are the advantages of the nonequilibrium DMFT over other methods 
for studying nonequilibrium phenomena in correlated systems?
A naive approach would be to solve the time-dependent Schr\"odinger equation for a many-body wavefunction numerically,
which is, however, quite restricted in terms of the system size, due to exponential growth of the Hilbert space dimension.
For one-dimensional systems, the time-dependent density 
matrix renormalization group (DMRG) method \cite{White2004a,Daley04,Schollwoeck2005} and its variants have been widely adopted to accurately simulate 
the temporal evolution of relatively large (or infinite-size) systems.
The restrictions here are the one-dimensionality and the accessible time range, which is severely limited, since
entanglement grows rapidly in highly excited systems.  
The nonequilibrium DMFT, by contrast, is formulated directly in the thermodynamic limit 
and can, in principle and in practice, access longer times. The main limitation of the nonequilibrium DMFT lies in the local approximation for the self-energy,
which may not be appropriate in low dimensional systems where spatially non-local correlations can become relevant.
These nonlocal correlations may however be incorporated into the DMFT formalism through cluster extensions or diagrammatic extensions (Sec.~\ref{subsec:extensions}). 
Within DMFT the nonequilibrium problem is thus approached by starting from a solution
which captures the local dynamics in high dimensions correctly, and
and then trying to build in nonlocal correlations.

An alternative approach that has conventionally been used employs
quantum kinetic or quantum Boltzmann equations~\cite{RammerBook},
based on the nonequilibrium Green's function formalism. It is usually
derived from a weak-coupling perturbation expansion, in combination
with a semiclassical approximation or gradient expansion. On long
time scales quantum Boltzmann equations describe the relaxation towards
a thermal state, while the fast dynamics on short time scales is not
captured.
While nonequilibrium DMFT can be compared or combined with these
methods, it has the advantage that it is nonperturbative and can
capture both the short-time and long-time evolutions for any strength
of the interaction.

\paragraph
{Physical background}

Before we start the detailed discussion of nonequilibrim DMFT, we briefly overview the evolution of nonequibrium 
physics in a broader context.   
The previous decade has witnessed a remarkable development in the field of ultrafast time-resolved spectroscopies in solids, 
in which an intense pump laser pulse is used to drive the system into highly excited states, while the temporal 
evolution of the system is tracked with subsequent probe pulses. 
The ``pump-probe" technique has enabled the study of excitation and relaxation processes in 
correlated electron systems on their intrinsic microscopic time scale, 
defined by the electron hopping between the crystal lattice sites \cite{Wall11}. 
In strongly correlated materials, quantum fluctuations, inherent in 
correlated electronic states, 
are highly intertwined, which makes it difficult to resolve the origin of 
given physical properties.  
Real-time spectroscopy introduces a ``new dimension'' 
on top of energy and  
momentum, and can provide an additional perspective on the correlated 
system by disentangling 
complicated electronic and lattice processes in, e.g., the cuprates \cite{DalConte2012}.
Often, the relaxation pathways in complex materials are not at all intuitive, and their study may 
lead to new concepts for the description and understanding of quantum many-body systems with no simple relation to 
the familiar equilibrium physics. 
Pioneering experiments in the field include 
photo-induced insulator-to-metal transitions in correlated Mott and charge-transfer insulators 
\cite{Ogasawara00,Perfetti2006,Perfetti2008,Okamoto07,Iwai03,Kuebler2007a}, the pump-induced 
melting and 
recovery of charge density waves \cite{Schmitt2008,Hellmann2010,Petersen2011} 
with studies combining structural and electronic dynamics \cite{Eichberger10}, and ultrafast dynamics induced in 
ferromagnets \cite{Beaurepaire1996} or antiferromagnets \cite{Ehrke2011}, to name only a few. 

Remarkably, ultrafast pump-probe spectroscopies have not only unveiled the response to strong external fields, but also provide means to {\em manipulate} phases of correlated electron systems. 
One manifestation of strong correlations, in equilibrium, is the Mott insulator, where 
the large cost in energy of putting two electrons on the same site leads to a charge excitation gap and inhibits conduction.  
Using an intense 
laser pulse, one can excite electrons across the charge gap, which drives the system into a nonequilibrium but relatively long-lived 
conducting state \cite{Ogasawara00,Perfetti2006,Okamoto07,Iwai03}.  
Such a process, sometimes called photo-doping \cite{NasuPPT}, is a typical example of 
a pathway to new phases, where mobile carriers are introduced {\it in situ}, 
as distinct from techniques employed in equilibrium, where the 
carrier concentration is typically controlled by chemical doping \cite{Imada1998}. 

\begin{figure*}[tb]
\centering 
\includegraphics[height=4.7cm]{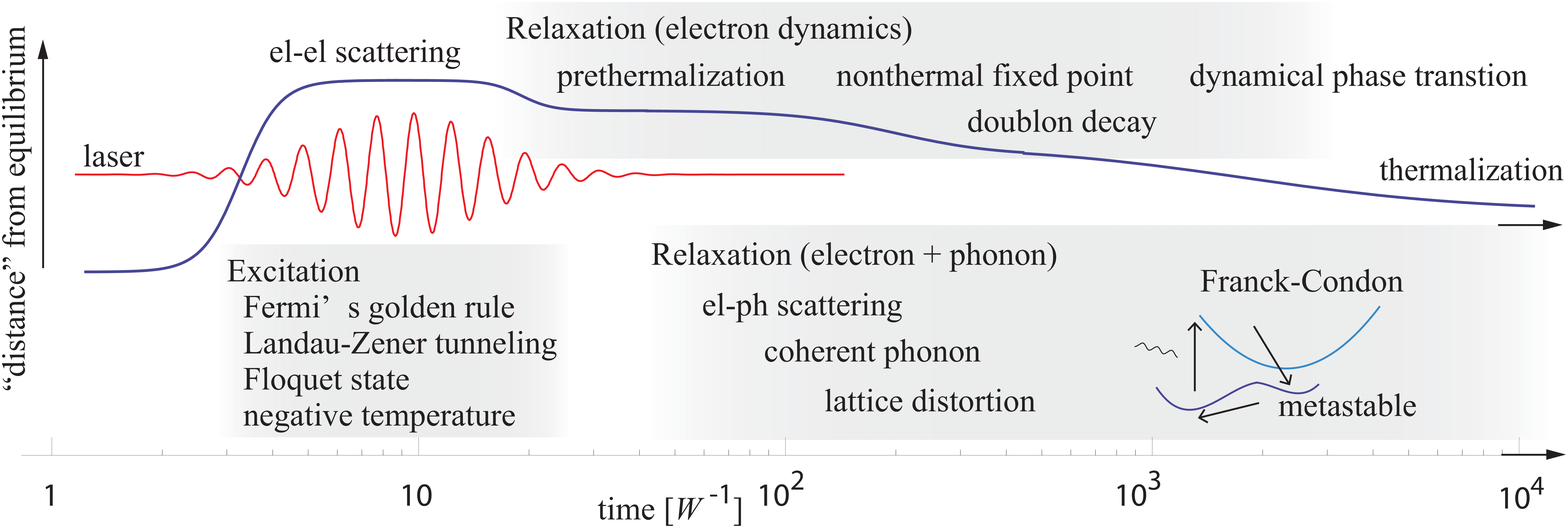}\includegraphics[height=4.7cm]{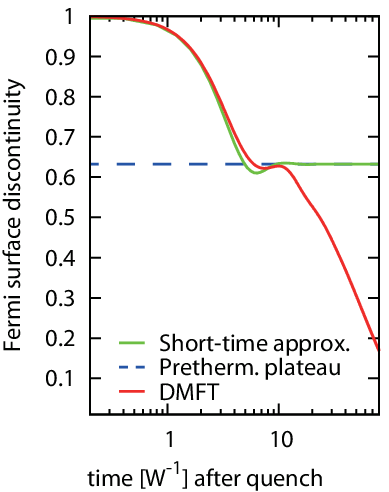}
\caption{Left panel: Schematic time evolution of the system in a
  pump-probe experiment with various physical processes (see text).  Right panel:
  Comparison of a short-time approximation
  \cite{Moeckel2008a,Eckstein2010}, which approaches a
  prethermalization plateau, and nonequilibrium DMFT \cite{Tsuji2013},
  which also describes the crossover towards a thermal state, for a
  sudden switching-on of the Hubbard interaction to $U$ $=$ $0.375W$.}
\label{fig:timescale}
\end{figure*}

A major difficulty in describing strongly correlated systems 
is the huge dimension of the Hilbert space, which 
is a problem already in equilibrium and becomes an even more serious challenge in 
nonequilibrium. One factor that makes the treatment of time-evolving 
quantum many-body systems challenging is the wide range of relevant time scales. One 
might first expect that strong interactions would help to quickly restore an equilibrium state after a perturbation, due 
to fast inter-particle scattering. However, contrary to the naive expectation, the dynamics of correlated 
systems generally exhibits a variety of time scales, which can be orders of magnitude different from the intrinsic microscopic 
time scale of the system, as sketched in the left panel of Fig.~\ref{fig:timescale}. The initial dynamics of a system excited by pumping 
is governed by the electronic degrees of freedom. The excitation during photo-irradiation takes place via Fermi's 
golden rule (linear-response theory) or the Schwinger mechanism (Landau-Zener tunneling in strong fields), 
depending on whether or not the photon energy is larger than the energy gap. 
During the laser application, the 
system may also reach a nonequilibrium time-periodic steady state (a so-called Floquet state, see 
Sec.~\ref{floquet intro}) for which the effective (temporal-Fourier transformed) Hamiltonian can drastically 
differ from the original one. 

After the pulse irradiation, electronic relaxation processes set in (Fig.~\ref{fig:timescale}). In Mott insulators, 
e.g., doublons (doubly-occupied sites) and holes, which are created in the first stage, start to annihilate in pairs. The 
relaxation time of doublons in a gapped system scales as \cite{Strohmaier10}
$\tau\propto 
W^{-1}
\exp
[\alpha (U/W)\ln(U/W)]
$,
where $U$ is the on-site Coulomb repulsion, $W$ is the electronic 
bandwidth, and $\alpha \sim O(1)$ a dimensionless constant.  
Thus one can see that there emerges a new time scale, which can be orders of magnitude longer than the intrinsic time scales ($W^{-1}$ and $U^{-1}$). 
Even in the course of thermalization of correlated metals, different time scales may emerge due to  
prethermalization \cite{Berges2004a}, 
the passage by nonthermal fixed points \cite{Berges2008}, 
and dynamical phase transitions (Sec.~\ref{subsec:parameterchanges}). At a certain point, the relaxation process enters a second phase (Relaxation (electron+phonon) in 
Fig.~\ref{fig:timescale}), where classical degrees of freedom such as lattice distortions start to play a role. This 
regime can be understood within the Frank-Condon picture \cite{NasuPPT} (inset of Fig.~\ref{fig:timescale}). New 
time scales can also appear through criticality in the dynamics of long-range order, such as spin-density 
waves or  superconductivity, which may behave classically on a long time 
scale (time-dependent Ginzburg-Landau 
picture), but are predicted to traverse through metastable supercritical phases \cite{Mathey2010,Tsuji2012} on intermediate time scales.

Another unique feature of correlated systems is that an external perturbation may cause cooperative 
changes through many-body interactions, and even drive the system into hidden states which are 
not accessible via adiabatic or thermal pathways \cite{Ichikawa2011}.
While photo-doping often puts the system in a highly excited state in which the effect of correlations can be smeared, more recently much lower photon energies 
(in the terahertz range) are being used to control material properties 
by selectively driving certain optical phonon modes. With this technique it is possible to control 
metal-insulator transitions \cite{Rini2007a,Caviglia2012}, or to induce superconductivity in a 
stripe-ordered cuprate \cite{Fausti11} on ultrafast time scales. An intriguing further step in this direction
would be to stabilize otherwise unstable many-body states by a continuous driving, and thus 
design material properties by external modulations.  
An example of this type of nonequilibrium control, namely the sign conversion of the interaction strength by AC electric fields, will be discussed in Sec.~\ref{subsubsec:acfields}.

The nonequilibrium phenomena and underlying concepts considered here are quite universal. One manifestation of this universality is the fact that 
phenomena known from condensed matter physics are now being realized with cold atomic gases in optical lattices
\cite{Bloch2008a}. 
Although these dilute gases are a totally different class of systems, they provide an almost ideal realization 
of the many-body lattice models that have long been studied as 
low-energy effective theories of real materials.  
Cold gases are unique in terms of their controllability, which is 
currently unthinkable in electron systems. 
For example, one can tune the inter-particle interaction almost arbitrarily using a Feshbach resonance, 
or by changing the lattice potential depth, and thus realize the Mott metal-insulator transition for both  
bosonic \cite{Greiner2002b} and fermionic \cite{Joerdens2008a,Schneider2008a} 
atomic systems. 
The basic time scale for the temporal evolution is orders of magnitudes longer ($\sim$ 1 ms) than that 
for correlated electron systems ($\sim$ 1 fs), making it much easier to keep track of the time evolution. 
Furthermore, cold atom systems may usually be regarded as isolated from the environment on the 
time scale of typical experiments. These unique properties make cold atom systems a valuable testing ground for the study of nonequilibrium physics. 

\begin{figure}[t]
\centering 
\includegraphics[width=6cm]{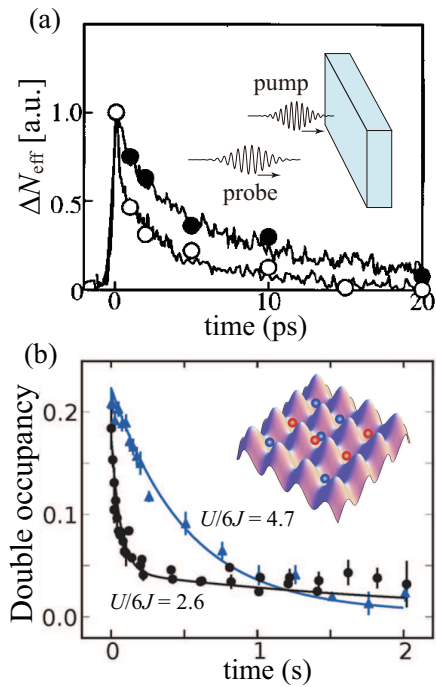}
\caption{
(a) Temporal evolution of the density of photo-induced carriers 
in a correlated electron system (a Ni complex) for two values 
of the excitation density (from \onlinecite{Iwai03}). The inset schematically shows a pump-probe 
experiment. 
(b) Temporal evolution of the doublon occupation in a 
cold-atom system on an optical lattice  (as schematically depicted in the inset)
(from \onlinecite{Strohmaier10}).
}
\label{fig:experiment}
\end{figure}

Figure~\ref{fig:experiment} shows an obvious parallelism between condensed matter and cold-atom systems. 
The top panel plots the number of excited carriers in a 
Mott insulator as a function of time after photoexcitation \cite{Iwai03}, while the bottom panel 
plots the relaxation of the double occupancy, i.e., the probability that a single site is occupied by two 
fermions with opposite (hyperfine)spins, generated by a periodic modulation of 
the optical lattice \cite{Strohmaier10}. 
Doubly occupied sites play the role of carriers in a Mott insulating 
background, so that the two panels plot essentially the same quantity.  
While it has been 
difficult to measure the doublon density directly in electronic systems, 
one can see for the cold-atom system that the double occupancy 
decays exponentially, and that the relaxation time changes significantly 
as the interaction strength is varied. 
In both systems the bottleneck for the decay of the excited carriers is the transformation of a high energy excitation 
into many low-energy excitations via many-body processes \cite{Sensarma2010a,Lenarcic2012preprint}. 
Although the absolute time scales in the two systems are vastly different, a physical understanding of basic 
nonequilibrium phenomena can thus be developed along similar lines.
Furthermore, with cold gases in optical lattices, it has been demonstrated that non-perturbatively 
strong external fields of oscillating  \cite{Struck2011} or DC nature \cite{Simon2011} can be used 
not only to change the state of the system, but to modify its microscopic Hamiltonian
in a controlled fashion. While this is yet to be realized for condensed-matter systems, 
interdisciplinary interactions between the fields of condensed-matter and cold-atom physics may help to 
achieve this goal in the near future.  

The study of nonequilibrium many-body physics extends to broad  
areas involving high-energy physics, as stressed in the concluding section 
of this review.
A long-standing issue, in both condensed-matter and 
high-energy physics, concerns the thermalization in isolated quantum 
systems \cite{Deutsch1991a,Srednicki1994a,Rigol2008a,Polkovnikov2011RMP}. It is a highly nontrivial and 
deep question how and when thermalization takes place as a result of the unitary time evolution of 
a quantum system. Integrable systems usually do not relax to the Gibbs 
ensemble, but rather to a generalized Gibbs ensemble~\cite{Rigol2007}
which also fixes the large number of constants of motion on average.
%although this scenario can only be applied to integrable systems.  
Motivated by cold-atom experiments, 
these questions have been addressed particularly in the context of quantum 
``quench" problems  
\cite{Kollath2007,Manmana2007,Calabrese2006a,Cazalilla2006,Barmettler2008a,Moeckel2008a,
Eckstein2008a,Eckstein09,cassidy_clark_11,Polkovnikov2011RMP,Dziarmaga2010a}, where 
a parameter in the Hamiltonian is suddenly changed 
to generate a nonequilibrium dynamics. After a quench, correlated systems often 
exhibit ``prethermalization" \cite{Berges2004a,Moeckel2008a}, i.e., relax to a state in which certain local observables look nearly thermalized, even though  
the whole momentum distribution still deviates from the thermal one. 
The right panel of Fig.~\ref{fig:timescale} shows the Fermi surface
discontinuity in the momentum distribution after a moderately large
interaction quench in the Hubbard model in infinite dimensions. The
weak-coupling expansion \cite{Moeckel2008a,Eckstein2010} describes the
transient behavior up to the prethermalization time scale. While the 
nonequilibrium DMFT result \cite{Tsuji2013} agrees with the perturbative treatment for
short times, it also describes the crossover towards the thermal state.
This thermalization process may be approximately reproduced by a quantum kinetic approach \cite{Stark2013}.
We will discuss these topics in more  
detail in Sec.~\ref{subsec:parameterchanges}.

Another arena of nonequilibrium physics is correlated systems in strong DC fields.  
There, one of the simplest 
questions, in the regime beyond the linear-response, is to ask what will happen when we apply a strong electric field to an insulator.
While a weak field  causes only a polarization of the system, stronger fields 
will induce a dielectric 
breakdown, and lead to a nonzero current. In a band insulator \cite{okaLMP}, 
valence and 
conduction bands may be modeled, around the band gap, by a two-band Hamiltonian 
for the valence and conduction bands. When we 
apply a constant electric field ${\bm E}$, the wave vector evolves, 
in the Bloch picture, according to ${\bm k}=
{\bm k}(0)-e{\bm E}t/\hbar$ (in a temporal gauge with a vector potential 
taking care of the field). Non-adiabatic 
transitions from the lower to the upper band can thus occur, in accord with the nonadiabatic Landau-Zener quantum tunneling \cite{Landau1932a,Zener1932a}, when the field exceeds a scale set by the gap. 
The situation is totally different for the breakdown in correlated electron 
systems, where the relevant gap is a many-body 
(Mott) gap. Here, the theoretical description becomes a formidable problem, 
since there are two non-perturbative effects involved: 
the Landau-Zener tunneling which is already non-perturbative (with regards 
to the electric field $E$), and the Mott transition which is 
also non-perturbative (with regards 
to the interaction $U$).  The dielectric breakdown in 
Mott insulators can then be understood as a field-induced quantum tunneling of 
many-body states 
across the Mott gap, which results in a finite doublon-hole creation rate in a strong field
\cite{Oka03,Okadmrg05,Oka05,okaLMP,Okabethe}. Hence there is 
a continuous crossover from the AC laser excitation ($\sim$ 
photon energy $\Omega$) across the gap, to the physics in strong DC fields with field strength $E$: 
quantum tunneling dominates the nonlinear DC regime, while (generally multi-) photon absorption dominates the AC regime. 

%When the total energy of a 
%continuously driven isolated system exceeds the upper limit of the equilibrium total energy, one may arrive at a 
%negative absolute temperature state \cite{Ramsey1956,Klein1956,Tsuji11,TsujiOkaAokiWerner2012} as described in Sec.~\ref{applications}, which provides an example for a 
%situation totally inaccessible in thermal equilibrium. 

One of the ultimate goals in the field of strongly correlated nonequilibrium 
physics is to induce some kind of long-range ``order'' that emerges in systems driven 
out of equilibrium.
In this context, an important and still open theoretical issue
is how to characterize a nonequilibrium phase transition and the associated critical behavior \cite{HohenbergHalperin1977}.
It has been argued that there exist certain universality classes for quantum phase transitions in low dimensional
systems driven out of equilibrium \cite{Feldman2005,Mitra2006}. In addition to the
criticality at the phase transition point, one may further pose the question of whether one can realize quasistationary
``nonequilibrium phases'' that are thermally inaccessible through adiabatic pathways. One idea along this line is
the concept of a ``nonthermal fixed point" \cite{Berges2008}, where the system does not immediately relax to a
thermal final state after excitation, but is trapped for a while in a nonthermal quasisteady state.

%%%%%%%%%%%%%%%%%%%%%%%%%%%%%%%%%%%%%%%%%%%%%%%%%%%%%%%%%%%%%%%%%%%%%%%%%%%%%%%%%%%%%%%
\section{Methods}
\label{methods}

\subsection{Nonequilibrium Green's function approach}
\label{noneq green function}

There exists a variety of methods to deal with the problem of a time-evolving
quantum many-body system, ranging from direct wave-function-based techniques
such as exact diagonalization and DMRG, quantum master equations  \cite{BreuerPetruccioneBook}, or quantum
kinetic equations \cite{RammerBook}, to the Keldysh formalism for nonequilibrium Green's
functions \cite{Schwinger1961, KadanoffBaym1962, Keldysh1964}. The nonequilibrium Green's function method
is an extension of the standard equilibrium formulation 
on the imaginary-time axis~\cite{AbrikosovGorkovDzyaloshinskiBook}. Using the
Keldysh formalism, many theoretical techniques which have been developed for
the study of strongly correlated systems, including DMFT, can be straightforwardly
adapted to nonequilibrium on a formal level.

The nonequilibrium Green's function approach 
is applicable to 
arbitrary time evolutions of correlated systems,
and does not involve 
any assumption on the statistical distribution of particles out of equilibrium,
since the time evolution of the distribution function is determined by the initial condition (initial-value problem).
A different formulation 
is needed 
if one focuses on nonequilibrium steady states of open systems, 
where driving by an external force is balanced by dissipation to an external heat bath.
By assuming that the system has arrived at a nonequilibrium steady state, so that the Green's functions do not change any more
as a function of `average time', they are determined by the boundary condition introduced by the heat bath (boundary-value problem).
In this case, the formulation is greatly simplified because one can drop the `average time' dependence of Green's functions
as well as correlations between the time evolving state and the initial state (initial correlations).
We will review the general formulation of nonequilibrium Green's
functions (Kadanoff-Baym formalism) in Sec.~\ref{kadanoff-baym}
and then discuss a more specific formulation (Keldysh formalism) for nonequilibrium steady states in Sec.~\ref{steady-states}.

\subsubsection{Kadanoff-Baym formalism for time evolution from a thermal initial state}
\label{kadanoff-baym}
%\input{kadanoff-baym.tex}
%%%%%%%%%%%%%%%%%%%%%%%%%
\newcommand{\kk}{{\bm k}}

\paragraph{Contour-ordered formulation}
\label{contour-ordered}

Let us consider a general quantum system driven out of equilibrium by an external field, whose time evolution is described by 
a time-dependent Hamiltonian $H(t)$. Initially (at $t=0$) the system is assumed to be in a mixed state described by a density matrix,
\begin{align}
\rho(0)
&=
    \frac{1}{Z}e^{-\beta \mathcal{H}(0)},
\label{rho0}
\end{align}
where $\beta=1/T$ is the inverse temperature (with $k_B=1$), $\mathcal{H}(t)=H(t)-\mu N(t)$ [$\mu$ is the chemical potential, 
$N(t)$ is the number operator for the particles], and $Z={\rm Tr}\, e^{-\beta \mathcal{H}(0)}$ is the equilibrium partition function.
At $t=0$ we switch on a driving field, and the system starts to evolve from its initial state. The time evolution of the density 
matrix is determined by the von Neumann equation,
\begin{align}
i\frac{d}{dt}\rho(t)
  &=
    [\mathcal{H}(t),\rho(t)],
\label{von Neumann}
\end{align}
where the bracket `$[,]$' represents the commutator, and $\hslash=1$.
Formally, one can write down the solution of Eq.~(\ref{von Neumann}) as
\begin{align}
\rho(t)
&= U(t,0)\, \rho(0) \, U(0,t),
\label{formal rho}
\end{align}
where we have defined the unitary evolution operator,
\begin{align}
U(t,t')
  &=
    \begin{cases}
      \vspace{.2cm}
      \displaystyle
      \mathcal{T} \exp\left(-i\int_{t'}^t d\bar{t}\, \mathcal{H}(\bar{t})\right) 
      & t>t'
      \\
      \displaystyle
      \bar{\mathcal{T}} \exp\left(-i\int_{t'}^t d\bar{t}\, \mathcal{H}(\bar{t})\right)
      & t<t'
    \end{cases}.
\label{U def}
\end{align}
Here $\mathcal{T}$ ($\bar{\mathcal{T}}$) denotes the (anti-)time-ordering operator, i.e., it arranges the operators 
so that an operator with time argument $t$ comes left (right) to operators with earlier time arguments $t'$ 
$<t$. Note that the Hamiltonians at different times do in general not commute with each other 
($[\mathcal{H}(t),\mathcal{H}(t')]\neq 0$). With this ordering, the evolution operator satisfies a fusion rule, $U(t,t')U(t',t'')=U(t,t'')$,
and becomes unitary, $U(t,t') [U(t,t')]^\dagger=U(t,t')U(t',t)=1$.

\begin{figure}[t]
\begin{center}
\includegraphics[width=7.5cm]{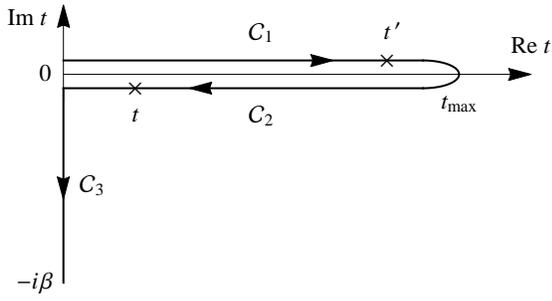}
\caption{The L-shaped contour $\CC=\CC_1\cup \CC_2\cup \CC_3$ 
in the Kadanoff-Baym formalism. The arrows indicate the contour ordering. For example,  
$t$ lies ahead of $t'$ in the ordering ($t\succ t'$).}
\label{L-shaped contour}
\end{center}
\end{figure}

Using the time-dependent density matrix (\ref{formal rho}), the expectation value of an observable $\mathcal{O}$ 
measured at time $t$ is given by
\begin{align}
\langle\mathcal{O}(t)\rangle
  &=
    {\rm Tr}\,[\rho(t)\mathcal{O}].
\label{expectation value}
\end{align}
By substituting $\rho(0)$ in Eq.~(\ref{formal rho}) with Eq.~(\ref{rho0}) and considering $\rho(0)$ as the evolution along 
the imaginary time axis from $0$ to $-i\beta$  (with imaginary-time ordering), we can express Eq.~(\ref{expectation value}) in a more 
convenient form,
\begin{align}
\langle \mathcal{O}(t) \rangle
  &=
    \frac{1}{Z} {\rm Tr}\,[U(t,0) e^{-\beta \mathcal{H}(0)} U(0,t) \mathcal{O}]
\nonumber
\\
  &=
    \frac{1}{Z} {\rm Tr}\,[ U(-i\beta,0) U(0,t) \mathcal{O} U(t,0)].
\label{expectation value2}
\end{align}
In the second line, we have permuted the operators under the trace.
If one reads the operators from right to left, one can see that the operators follow the time ordering of $0\to t\to 0\to -i\beta$.
This motivates us to adopt an L-shaped contour $\CC$ with three branches, $\CC_1$: $0\to \tmax$, $\CC_2$: $\tmax\to 0$, 
and $\CC_3$: $0\to -i\beta$, as shown in Fig.~\ref{L-shaped contour}, where $\tmax$ is the maximal time up to which one wants 
to let the system evolve \cite{KadanoffBaym1962}. Then the expectation value (\ref{expectation value2}) can be written as
\begin{align}
\langle \mathcal{O}(t) \rangle
  &=
    \frac{\displaystyle{\rm Tr}
    \big[
    \mathcal{T}_{\CC}\, e^{-i\int_{\CC} d\bar{t}\, \mathcal{H}(\bar{t})}\mathcal{O}(t)
    \big]
    }
    {\displaystyle{\rm Tr}
    \big[
    \mathcal{T}_{\CC} \,e^{-i\int_{\CC} d\bar{t}\, \mathcal{H}(\bar{t})}
    \big]
    },
\label{rho contour}
\end{align}
where $\mathcal{T}_{\CC}$ is a contour-ordering operator that arranges operators on the contour $\CC$ in the order 
 $0\to \tmax\to 0\to -i\beta$ (as indicated by the arrows in Fig.~\ref{L-shaped contour}), $\mathcal{O}(t)$ indicates 
that the operator $\mathcal{O}$ is inserted at time $t$ on the contour $\CC$ (we are working in the Schr\"odinger 
picture), and we have used the fact that the evolution along the forward ($\CC_1$) and backward ($\CC_2$) 
contours cancels if no other operator is inserted, so that $e^{-\beta\mathcal{H}(0)}=\mathcal{T}_{\CC} \exp\left(-i\int_{\CC} 
d\bar{t}\, \mathcal{H}(\bar{t})\right)$.

The contour-ordered formalism reveals its full power when it is applied to higher-order correlation functions,
\begin{align}
\label{expOO}
  \langle \mathcal{T}_{\CC}\, \mathcal{A}(t)\mathcal{B}(t')\rangle
    &\equiv
      \frac{1}{Z}{\rm Tr}\left[
      \mathcal{T}_{\CC}\,
      e^{-i\int_{\CC} d\bar{t}\, \mathcal{H}(\bar{t})}
      \mathcal{A}(t)\mathcal{B}(t')
      \right].
\end{align}
Here $\mathcal A$ and $\mathcal B$ are combinations of particle creation and annihilation operators.
We call them ``fermionic'' if they contain odd number of fermion creation or annihilation operators, and ``bosonic'' otherwise.
In this expression, $t$ and $t'$ can lie anywhere on $\CC$, and the contour-ordered product of two 
operators $\mathcal{A}$ and $\mathcal{B}$ is defined as
\begin{align}
  \mathcal{T}_{\CC}\, \mathcal{A}(t)\mathcal{B}(t')
    =
      \theta_{\CC}(t,t')\mathcal{A}(t)\mathcal{B}(t') \pm\theta_{\CC}(t',t)\mathcal{B}(t')\mathcal{A}(t), 
  \label{contour ordering}
\end{align}
where $\theta_{\CC}(t,t')=1$ when $t'$ comes earlier than $t$ in the contour ordering (denoted by $t\succ t'$, see 
Fig.~\ref{L-shaped contour}) and $0$ otherwise ($t\prec t'$). The sign $\pm$ is taken to be minus when the 
operators $\mathcal{A}$ and $\mathcal{B}$ are both fermionic and plus otherwise. Whenever an operator appears in a contour-ordered product, one has to specify which branch its time argument lies on. For $t=t'$ (on the same branch of $\mathcal C$), we adopt a
normal ordering convention, which puts all creation operators to the left of all annihilation operators
\cite{AbrikosovGorkovDzyaloshinskiBook}, unless the ordering is irrelevant (when $t$ or $t'$ is integrated over)
or explicitly indicated. 

Contour-ordered correlation functions provide a concise way to keep track of both spectral information and occupation 
functions in a many-particle system out of equilibrium. Before explaining this in further detail (Sec.~\ref{nonequilibrium green}), 
let us comment on the relation between the formulation presented here and the one used in field theories at zero temperature, 
where one usually works with a single-branch time axis ranging from $-\infty$ to $\infty$. The latter is possible due to Gell-Mann 
and Low's theorem \cite{Gell-MannLow1951, FetterWaleckaBook}, which states that the ground states $|\Psi(0)\rangle$ and 
$|0\rangle$ of the interacting and noninteracting system are related by $|\Psi(0)\rangle=U(0,-\infty)|0\rangle$ where the 
interaction is adiabatically turned on from $t=-\infty$ to $t=0$ (the ground states are assumed to be nondegenerate).
Then the expectation value 
(\ref{expectation value}) is given by
\begin{align}
  \langle \mathcal{O}(t) \rangle
    &=
      \langle 0|U(-\infty,\tmax) U(\tmax,t) \mathcal{O}(t) U(t,-\infty)|0\rangle.
  \label{expectation value 0}
\end{align}
We can similarly assume that the interaction is adiabatically switched off in the far future and let  $\tmax\to\infty$. Then, the 
wavefunction goes back to the noninteracting ground state $|0\rangle$ up to a phase factor $e^{iL}$ ($L$ is a real number) 
\cite{Gell-MannLow1951, FetterWaleckaBook}: $U(\infty,-\infty)|0\rangle=e^{iL}|0\rangle$. Taking its Hermite conjugate and 
inserting it to Eq.~(\ref{expectation value 0}) gives an expression for the expectation value,
\begin{align}
  \langle \mathcal{O}(t) \rangle
    &=
      \frac{\langle 0| 
      \mathcal{T} 
      e^{-i\int_{-\infty}^{+\infty} d\bar t \,\mathcal{H}(\bar t)}
      \mathcal{O}(t)|0\rangle}
      {\langle 0|
      \mathcal{T} 
      e^{-i\int_{-\infty}^{+\infty} d\bar t \,\mathcal{H}(\bar t)}
      |0\rangle},
  \label{single branch}
\end{align}
in which the time argument $t\in (-\infty,\infty)$ moves on a single branch of the real time axis. However, for general nonequilibrium 
systems one cannot use Eq.~(\ref{single branch}), since the initial state $|0\rangle$ would be driven into excited states and never 
return after the whole time evolution, i.e., $U(\infty,-\infty)|0\rangle\neq e^{iL}|0\rangle$. This forces one to use the analogy of 
Eq.~(\ref{rho contour}) instead of Eq.~(\ref{single branch}), with a round trip ($-\infty\to\infty\to -\infty$), and $\langle 0| \cdots |0\rangle$ 
instead of the $\text{Tr}$. The idea of this multi-branch formalism was originally introduced by \citet{Schwinger1961} and 
\citet{Keldysh1964}. They assumed that many-body interactions are adiabatically switched on from a noninteracting initial 
state, so that there should be no correlation between the initial state and the time-evolving state. Under this condition it is sufficient to 
consider two branches $\CC_1$ and $\CC_2$ for the time axes \cite{KamenevBook}. After that, their approach has been extended to 
arbitrary initial states with initial correlations taken into account \cite{Danielewicz1984a, Danielewicz1984b,Wagner1991} by 
employing the triple-branch contour depicted in Fig.~\ref{L-shaped contour}.

\paragraph{Contour-ordered Green's functions}
\label{nonequilibrium green}

Single-particle Green's functions are the fundamental objects of many-body theories. They describe single-particle excitations 
as well as statistical distributions of particles, and play a central role in the formulation of nonequilibrium DMFT, which will be 
reviewed in Sec.~\ref{Sec_DMFT}. We define the nonequilibrium Green's function as the contour-ordered expectation value,
\begin{align}
G(t,t')
&\equiv
-i\langle\mathcal{T}_{\CC}\, c(t)\, c^\dagger(t')\rangle,
\label{green def}
\end{align}
where $c^\dagger$($c$) is a creation (annihilation) operator of particles, and $t,t'\in \CC$. For simplicity, spin and orbital indices  
associated with the operators are not shown. Due to the three branches, on which the time arguments $t$ and $t'$ can lie, the 
Green's function has $3\times 3=9$ components: $G(t,t') \equiv G_{ij}(t,t')$ ($t\in \CC_i$, $t'\in \CC_j$, $i,j=1,2,3$). Conventionally 
we express them in a $3\times 3$ matrix form,
\begin{align}
  \hat{G}
    &=
      \begin{pmatrix}
        G_{11} & G_{12} & G_{13} \\
        G_{21} & G_{22} & G_{23} \\
        G_{31} & G_{32} & G_{33}
      \end{pmatrix}.
  \label{G 3x3}
\end{align}
In general, one can shift the operator with the largest real-time argument (e.g., $t'$ in Fig.~\ref{L-shaped contour}) from $\CC_1$ to 
$\CC_2$ (and vice versa), because the time-evolution along $\CC_1$ and $\CC_2$ to the right of that operator cancels. This kind of 
redundancy implies the following relations among the components of the matrix (\ref{G 3x3}),
\begin{subequations}
\begin{align}
  \label{redundancy-1}
  G_{11}(t,t')
    &=
      G_{12}(t,t') \quad ({\rm for}\; t\le t'),
  \\
  G_{11}(t,t')
    &=
      G_{21}(t,t') \quad ({\rm for}\; t>t'),
  \\
  G_{22}(t,t')
    &=
      G_{21}(t,t') \quad ({\rm for}\; t<t'),
  \\
  \label{redundancy-4}
  G_{22}(t,t')
    &=
      G_{12}(t,t') \quad ({\rm for}\; t\ge t'),
  \\
  G_{13}(t,\tau ')
    &=
      G_{23}(t,\tau '),
  \\
  G_{31}(\tau,t')
    &=
      G_{32}(\tau,t').
\end{align}
\label{redundancy}
\end{subequations}
Equations (\ref{redundancy-1})-(\ref{redundancy-4}) can be summarized as
\begin{align}
G_{11}+G_{22}
  &=
    G_{12}+G_{21}. 
\label{linear dependence1}
\end{align}
The violation of this relation at $t=t'$ in the normal ordering convention is negligible under the time-integrations used below.
The relations (\ref{redundancy}) thus allow one to eliminate three components out of nine in the Green's function (\ref{G 3x3}). 
To this end, let us introduce six linearly independent ``physical'' Greens functions, called the retarded ($G^R$), advanced 
($G^A$), Keldysh ($G^K$), left-mixing ($G^{\lh}$), right-mixing ($G^{\rh}$), and Matsubara Green's function ($G^M$). 
They are explicitly given by
\begin{subequations}
\label{physical green def}
\begin{align}
G^R(t,t')
  &=
     \tfrac12(G_{11}-G_{12}+G_{21}-G_{22})
\nonumber
\\
  &=
    -i\theta(t-t') \langle [c(t),c^\dagger(t')]_\mp \rangle,
\\
G^A(t,t')
  &=
     \tfrac12(G_{11}+G_{12}-G_{21}-G_{22})
\nonumber
\\
  &=
    i\theta(t'-t) \langle [c(t),c^\dagger(t')]_\mp \rangle,
\label{advanced green def}
\\
G^K(t,t')
  &=
     \tfrac12(G_{11}+G_{12}+G_{21}+G_{22})
\nonumber
\\
  &=
    -i \langle [c(t),c^\dagger(t')]_\pm \rangle,
\label{keldysh green function}
\\
G^{\lh}(t,\tau ')
  &=
     \tfrac12(G_{13}+G_{23})
  =
    \mp i \langle c^\dagger(\tau ') c(t) \rangle,
\\
G^{\rh}(\tau, t')
  &=
     \tfrac12(G_{31}+G_{32})
  =
    -i\langle c(\tau) c^\dagger(t') \rangle,
\\
G^M(\tau,\tau ')
  &=
    -iG_{33}
  =
    -\langle \mathcal{T}_\tau\, c(\tau) c^\dagger(\tau ') \rangle.
\end{align}
In the above formulas, we choose the upper (lower) sign if the operators $c$ and $c^\dagger$ are bosonic (fermionic), $[,]_{-(+)}$
denotes an (anti-)commutator, $t,t'\in \CC_1\cup\CC_2$, $\tau,\tau '\in\CC_3$, $\theta(t)$ is a step function, and $\mathcal{T}_\tau$
is the time-ordering operator on the imaginary time axis. Note that the anti-commutator is used for bosonic operators while the 
commutator is used for fermionic operators in $G^K$ (\ref{keldysh green function}).
For convenience, we also define the lesser and greater Green's functions:
\begin{align}
\label{gles def}
  G^<(t,t')
    &=
      G_{12}
    =
      \mp i \langle c^\dagger(t') c(t) \rangle,
  \\
\label{ggtr def}
  G^>(t,t')
    &=
      G_{21}
    =
      -i \langle c(t) c^\dagger(t') \rangle,
\end{align}
\end{subequations}
which are related to the retarded, advanced, and Keldysh Green's functions via
\begin{subequations}
\begin{align}
  G^<
    &=
      \tfrac12
      (G^K-G^R+G^A),
  \\
  G^>
    &=
      \tfrac12
      (G^K+G^R-G^A).
\end{align}
\end{subequations}
In addition to the redundancy (\ref{redundancy}), the components of (\ref{G 3x3}) are related via their hermitian 
conjugates. For the physical Green's function components, conjugation yields
\begin{subequations}
\begin{align}
  G^{<,>,K}(t,t')^\ast
    &=
      -G^{<,>,K}(t',t),
  \\
  G^R(t,t')^\ast
    &=
      G^A(t',t) ,
  \\
  G^{\lh}(t,\tau' )^\ast
    &=
      \mp G^{\rh}(\beta-\tau' ,t),
  \label{h.c. G13}
\end{align}
\label{hermite conjugate}
\end{subequations}
where we take the upper (lower) sign in Eq.~(\ref{h.c. G13}) for bosons (fermions). Finally,  if a fermionic system has 
particle-hole symmetry, the Green's function is antisymmetric,
\begin{align}
G(t,t')
&=
-G(t',t).
\label{particle-hole symmetry}
\end{align}
In addition to these symmetries, it follows from the cyclic invariance of the trace and the definition of 
$\mathcal{T}_\CC$ that $G(t,t')$ satisfies a boundary condition on $\CC$ in both arguments,
\begin{subequations}
\label{kk-boundary}
\begin{align}
G(0^+,t) &= \pm  G(-i\beta,t), 
\\
G(t,0^+) &= \pm  G(t,-i\beta), 
\end{align}
\end{subequations}
where $0^+ \in \CC_1$ and $-i\beta \in \CC_3$ denote the two end-points of $\CC$, and the upper (lower) sign corresponds 
to the case of bosons (fermions). 

The Matsubara component $G^M$ plays a somewhat special role, since it is always translationally invariant $G^M(\tau,\tau') 
\equiv G^M(\tau-\tau')$ ($\mathcal{H}$ does not depend on imaginary time). Furthermore, it is real (hermitian), $G^M(\tau)^*
=G^M(\tau)$, and as a consequence of Eq.~(\ref{kk-boundary}) it is periodic (antiperiodic) for bosons (fermions), 
$G^M(\tau)=\pm G^M(\tau+\beta)$. One can thus use its Fourier decomposition in terms of Matsubara frequencies,
\begin{subequations}
\label{mat FT}
\begin{align}
G^M(\tau,\tau') 
&
=
T\sum_n e^{-i \omega_n (\tau-\tau')} G^M(i\omega_n),
\\
G^M(i\omega_n) 
&=
\int_0^\beta d\tau  \,e^{i \omega_n \tau} G^M(\tau).
\end{align}
\end{subequations}

Using the physical Green's function components instead of the full matrix (\ref{G 3x3}) can be quite beneficial in numerical simulations, 
since with this one almost automatically exploits the symmetries and redundancies and thus reduces the amount of data to be handled 
(see Sec.~\ref{KB-numerics}). Moreover, the components (\ref{physical green def}) are often used to interpret the results of calculations 
since they have an intuitive interpretation, which originates from their physical meaning in equilibrium: When $\mathcal{H}$ does not 
depend on time, real-time components of $G$ depend on the time difference only and can be represented via their Fourier transform. The 
imaginary part of the retarded (or advanced) Green's function gives the single-particle spectral function \cite{AbrikosovGorkovDzyaloshinskiBook},
\begin{align}
  A(\omega)
    &=
      -\frac{1}{\pi}{\rm Im}\, G^R(\omega) 
      =
      \frac{1}{\pi}{\rm Im}\, G^A(\omega),
\end{align}
which represents the density of single-particle excitations at energy
$\omega$ of the many-body state, as can be seen from the Lehmann representation \cite{MahanBook},
\begin{equation}
\label{lehmann representation}
A(\omega) = \frac{1}{Z}\sum_{mn} \big(\mp e^{-\beta E_n}+e^{-\beta E_m}\big) |\langle n | c^\dagger | m \rangle|^2 \delta (\omega - E_n + E_m).
\end{equation}
Out of equilibrium, one can still define the spectral function using
the partial Fourier transformation,
\begin{equation}
A(\omega,t_{\rm av})=-\frac{1}{\pi}{\rm Im}\int dt_{\rm rel} e^{i\omega t_{\rm rel}}G^R(t,t'),
\end{equation}
with $t_{\rm av}=(t+t')/2$ and  $t_{\rm rel}=t-t'$, which satisfies the sum rule
\begin{equation}
\int d\omega A(\omega,t_{\rm av})=1. 
\end{equation}
Higher moment sum rules have also been derived \cite{Turkowski2006,Turkowski2008}).
These relations hold exactly in and out of equilibrium, so that they are quite useful in benchmarking calculations.

In equilibrium, all components of $G$ can be related to the spectral function 
\begin{align}
G(t,t') = -i \int d\omega \,e^{-i\omega(t-t')}\,A(\omega)\, [\theta_\CC(t,t') \pm f(\omega)],
\label{G in equilibrium}
\end{align}
where $f(\omega)= 1/(e^{\beta \omega}\mp 1)$ is the Bose (Fermi) occupation function. Equation (\ref{G in equilibrium}) follows from the 
analytic properties of the Green's function components as a function of the  {\em time-difference}, together with the Kubo-Martin-Schwinger 
boundary condition (\ref{kk-boundary}) \cite{Kubo1957, MartinSchwinger1959}, and can also be read off a Lehmann representation. 
In particular, the imaginary part of the lesser (greater) Green's function thus yields the density of {\it occupied} ({\em unoccupied}) states,
\begin{subequations}
\label{un and occupied}
\begin{align}
      \mp{\rm Im}\, G^<(\omega)
     &= 2\pi\,A(\omega)f(\omega) 
     \equiv  2\pi N(\omega),
  \label{occupied}
\\
      -{\rm Im}\, G^>(\omega)
     &= 2\pi\,A(\omega)[1\pm f(\omega)].
  \label{unoccupied}
\end{align}
\end{subequations}
In essence, Eq.~(\ref{un and occupied}) is the fluctuation-dissipation relation \cite{Kubo1957, MahanBook}  for single-particle excitations,
\begin{subequations}
\begin{align}
  G^K(\omega)
    &=
      F(\omega)[G^R(\omega)-G^A(\omega)],
  \label{fluctuation-dissipation}
\end{align}
where 
\begin{align}
F(\omega) = 1 \pm 2f(\omega)
  &=
    \begin{cases}
      \coth\left(\frac{\beta\omega}{2}\right) & \mbox{for bosons}\\
      \tanh\left(\frac{\beta\omega}{2}\right) & \mbox{for fermions}
    \end{cases}.
\label{coth, tanh}
\end{align}
\end{subequations}
In equilibrium, the density of occupied (unoccupied) states is often taken as a first approximation to understand 
(inverse) photoemission spectroscopy in correlated materials. Similarly, intensities for time-resolved (inverse) 
photoemission spectroscopy can be obtained from the real-time Green's functions $G^<(t,t')$ and $G^>(t,t')$ 
(Sec.~\ref{dmft observables}).

Let us conclude this subsection with the remark that all relations concerning contour-ordered Green's functions remain valid 
if one replaces the time-ordered exponential in Eq.~(\ref{expOO}) by a more general action, 
for example  
\begin{align}
\label{G with action 001}
&\mathcal{S}
= 
-i\intC \!dt \, \mathcal{H}(t) 
 -i  \intC \!dt\, dt'\, c^\dagger(t) \Delta(t,t') c(t'),
\end{align}
where $\Delta(t,t')$ is a function on the contour with the same boundary and symmetry properties as the Green's function. 
In this case, the contour-ordered Green's function is defined as
\begin{align}
G(t,t') 
= 
-i
\langle\mathcal{T}_\CC\, c(t)c^\dagger(t')\rangle_{\mathcal{S}},
\end{align}
where the expectation value of observables with respect to $\mathcal{S}$ is
\begin{align}
\langle \cdots \rangle_{\mathcal{S}}=
\frac{{\rm Tr}[\TC\, \exp(\mathcal{S}) \cdots]}{{\rm Tr}[\TC\, \exp(\mathcal{S})]}.
\label{cntr-expval}
\end{align}
The action (\ref{G with action 001}) arises naturally when parts of a system are traced out in order to derive an effective description of the rest. Expressions like Eq.~(\ref{cntr-expval}) can conveniently be rephrased in terms of path integrals over Grassmann variables 
\cite{Negele1988a,KamenevBook}, but throughout this review we stay with the equivalent formulation in terms of 
time-ordered expectation values.

\paragraph{Noninteracting contour-ordered Green's functions}
\label{noninteracting Green}

\begin{table}
\begin{tabular}{|c|l|}
\hline
1) &
$
\begin{array}{ll}
t^\pm,\,t \in [0,\tmax]: & \text{point on $\CC_{1,2}$}\\
-i\tau,\,\tau \in [0,\beta]: & \text{point on  $\CC_3$}\\
\end{array}
$\\[1mm]
\hline
2) &
\hspace*{0.5mm}
$\displaystyle
\intC dt\,g(t) = 
\int_{0}^{\tmax} \!\!dt\,g(t^+)
-\int_{0}^{\tmax} \!\!dt\,g(t^-)
-i\int_{0}^{\beta} \!\!d\tau\, g(-i\tau)
$
\\[1mm]
\hline
3) &
\hspace*{0.5mm}
$\displaystyle
[a \convz b](t,t') = \intC \!d\tb\, a(t,\tb) b(\tb,t')
$
\\[1mm]
\hline
4) &
\hspace*{0.5mm}
$\displaystyle
\partial_t g(t)
=
\left\{
\begin{array}{ll}
\partial_t g(t^\pm) & t\in \CC_{1,2}\\
i\partial_\tau g(-i\tau) & t=-i\tau\in\CC_3 
\end{array}
\right.
$
\\[1mm]
\hline
5) &
\hspace*{0.5mm}
$\displaystyle
\theta_\CC(t,t')
=
\left\{
\begin{array}{ll}
1 & t\gtrc t' \\
0 & \text{else}
\end{array}
\right.
$
\\[1mm]
\hline
6) &
\hspace*{0.5mm}
$\displaystyle
\delC(t,t')
=
\partial_t \theta_\CC(t,t')$,
$\displaystyle
\intC\! d\tb\, \delC(t,\tb) g(\tb) 
= g(t)$ $\forall g(t)$
\\[1mm]
\hline
\end{tabular}
\caption{Notation for the contour calculus used in this text:
2) and 3): Contour integration and convolution. 
4): Time derivative (not a derivative along the contour).
5) and 6): Contour theta and delta functions.  }
\label{ky-ccalc}
\end{table}

In this paragraph we discuss equations of motion for noninteracting Green's functions. For a tight-binding model $\mathcal{H}_0(t)= 
\sum_{\kk} [\epsilon_{\kk}(t)-\mu] c_{\kk}^\dagger c_{\kk}$ one can directly compute the time-derivatives of $G_{0,\kk}(t,t')=-i\langle 
\mathcal{T}_\CC c_\kk(t) c_\kk^\dagger(t')\rangle$,
\begin{subequations}
\label{EoM}
\begin{align}
&\big[
i\partial_t  + \mu-\epsilon_\kk(t)  
\big]
G_{0,\kk}(t,t') 
=
\delta_\CC(t,t'),
\label{EoM-1}
\\
&G_{0,\kk}(t,t') 
\big[ -i\overleftarrow{\partial_{t'}}  + \mu - \epsilon_\kk(t') \big]
=
\delta_\CC(t,t'),
\label{EoM-2}
\end{align}
\end{subequations}
where we used the notation for contour calculus introduced in Table \ref{ky-ccalc} ($f(t)\overleftarrow{\partial_t}
\equiv \partial_tf(t)$ is acting to the left). The two equations are equivalent, and each determines $G_{0,\kk}$
uniquely if solved with the boundary condition (\ref{kk-boundary}) \cite{Turkowski2005a},
\begin{align}
G_{0,\kk}(t,t') &= 
-i [\theta_{\CC}(t,t') \pm f(\epsilon_{\kk}(0)-\mu)]
e^{-i \int_{t'}^{t} d\bar t [\epsilon_\kk(\bar t)-\mu]}.
\label{noninteracting green}
\end{align}
The two equations of motion (\ref{EoM}) can be rephrased by introducing the inverse of the Green's function
\begin{align}
\label{G0inv_k}
G_{0,\kk}^{-1}(t,t')
=
\big[
i\partial_t  + \mu - \epsilon_{\kk}(t) \big] \delta_\CC(t,t'),
\end{align}
which is a differential operator on the contour. Equations (\ref{EoM}) then simply read $G_{0,\kk}^{-1} \convz G_{0,\kk}=
G_{0,\kk}\convz G_{0,\kk}^{-1}= \delta_\CC$, where the star ($\convz$) denotes a convolution (Table \ref{ky-ccalc}).
Closed equations of motion can also be derived for the general case in which the action is nonlocal in time
[cf.~Eq.~(\ref{G with action 001})], and $\mathcal{H}_0$ is not diagonal in orbitals, 
\begin{equation}
\mathcal{S}
= 
-i\intC \!dt \, \mathcal{H}_0(t) 
 -i\sum_{ij}  \intC \!dt\, dt'\, c_i^\dagger(t) \Delta_{ij}(t,t') c_j(t'),
\end{equation}
with $\mathcal{H}_0(t)= \sum_{ij} [v_{ij}(t)-\mu\delta_{ij}] c_{i}^\dagger c_{j}$.
In this case, both $G_0$ and $G_0^{-1}$ are matrices in orbital indices, and 
\begin{align}
\label{G0inv}
(G_0^{-1})_{ij}(t,t')
\!=\!
\big[
\delta_{ij}(i\partial_t  + \mu) - v_{ij}(t) 
\big]\delta_\CC(t,t') - \Delta_{ij}(t,t').
\end{align}
For $\Delta\neq 0$, the solution of the equation of motion $G_0^{-1}\convz G_0 = \delta_\CC$ in general requires 
a numerical technique, which will be discussed in Sec.~\ref{KB-numerics}.

\paragraph{Dyson equation}
\label{quantum boltzmann}
To describe nonequilibrium correlated systems using Green's functions, one has to take account of self-energy corrections 
$\Sigma$ to the noninteracting Green's function $G_0$. In the language of Feynman diagrams, the self-energy is the 
sum of all one-particle irreducible diagrams of the interacting Green's function $G$, i.e., diagrams that cannot be separated 
into two parts by cutting a single $G_0$ lines. (The diagram rules are the same for imaginary time ordered and contour-ordered
Green's functions when imaginary time integrals over internal vertices are replaced by contour integrals.) The self-energy 
is defined on the contour $\CC$ (Fig.~\ref{L-shaped contour}), so that it satisfies symmetry and boundary conditions analogous 
to that of the Green's functions (Sec.~\ref{nonequilibrium green}). The fully interacting Green's function 
$G=G_0 + G_0\convz \Sigma \convz G_0+ G_0\convz  \Sigma  \convz G_0 \convz\Sigma \convz G_0 + \cdots$ 
is then given by the Dyson equation,
\begin{subequations}
\label{dyson}
\begin{align}
\label{dyson nocc}
G 
&= G_0 +  G_0 \convz \Sigma \convz G
\\
\label{dyson-cc}
&= G_0 +  G \convz \Sigma \convz G_0.
\end{align}
\end{subequations} 
To evaluate the self-energy is truly a nonequilibrium quantum many-body problem, and one generally needs additional techniques, 
which will be explained in the following sections. Once the self-energy is fixed, the full Green's function is determined from one of the 
two equivalent integral equations (\ref{dyson}), which is still a formidable numerical task that will be discussed in the remainder of this 
section. 

We can transform the Dyson equation and its conjugate from its integral form into a differential form by convoluting with the 
operator $G_0^{-1}$ from the left [Eq.~(\ref{dyson nocc})], or right [Eq.~(\ref{dyson-cc})], respectively,
\begin{align}
[G_0^{-1} - \Sigma] \convz G = 
G \convz [G_0^{-1} - \Sigma] =\delta_{\mathcal{C}}.
\label{inv dyson}
\end{align}
The result is conveniently expressed by the definition $G^{-1} = G_0^{-1} - \Sigma$.
In this abstract notation, the Dyson equation is identical to the form used in equilibrium \cite{MahanBook}. However, equation 
(\ref{inv dyson}) has conceptually a very different meaning for Matsubara and contour-ordered Green's functions. 
Using a differential form for $G_0^{-1}$ analogous to Eq.~(\ref{G0inv_k}),
%Using Eq.~(\ref{G0inv}), 
one can see that the two Eqs.~(\ref{inv dyson})  are integral-differential equations of the generic form,
\begin{subequations}
\label{ky-cinv-v01}
\begin{align}
&[i\partial_t - h(t) ] G(t,t') - \int_\CC \!d\bar t\, \Sigma(t,\bar t) G(\bar t,t') 
= \delta_\CC(t,t'),
\\
&G(t,t')[-i\overleftarrow{\partial_{t'}} - h(t') ] - \int_\CC \!d\bar t\,  G(t,\bar t) \Sigma(\bar t,t') 
= \delta_\CC(t,t').
\end{align}
\end{subequations}
The time derivative $\partial_t G$ in these equations is related to the value of $G$ at different times via the convolution $\Sigma\convz G$.
The equations are causal, and thus provide a non-Markovian time-propagation scheme for $G$, in which the self-energy takes the role of 
a memory kernel (Sec.~\ref{KB-numerics}). On the imaginary branch, on the other hand, the same equations provide a boundary value 
problem for the (Matsubara) Green's functions of an equilibrium state (which play the role of an initial value for the time propagation). 
The  solution of the integral-differential equation (\ref{ky-cinv-v01}) has numerous applications in various areas of physics, including 
condensed matter physics, nuclear physics, high-energy physics, and cosmology  \cite{BonitzBook,BonitzSemkatBook,PNGFIII,PNGFIV}. 
The biggest challenge is to deal with the memory effects in a proper way. Traditionally one tries to reduce the memory depth by deriving 
quantum Boltzmann equations \cite{KadanoffBaym1962,RammerBook,Mahan1984a,Haug2008a}, or by using decoupling schemes like 
the generalized Kadanoff-Baym ansatz \cite{Lipavski1986}. While those approaches usually work well for weakly interacting systems 
or in the semiclassical limit, one must account for the full memory when dealing with the ultrafast time-evolution in strongly correlated systems. 

\paragraph{Numerical Solution}
\label{KB-numerics} 
In this section, we discuss the numerical solution of the generic contour equation (\ref{ky-cinv-v01}). 
On a suitable time grid,  with $N$ time slices $\Delta t = \tmax/N$ on $\CC_{1,2}$ and $M$ imaginary time slices $\Delta\tau=\beta/M$
on $\CC_3$,
the operator $(i\partial_t -h(t))\delta_\CC(t,t')-\Sigma(t,t')$ can be written as a $(2N+M+1)$-dimensional matrix (with some care to 
correctly discretize the singular operators $\delta_\CC(t,t')$ and $\partial_t \delta_\CC(t,t')$), such that the solution for $G$ becomes a 
matrix inversion \cite{Freericks2008}. On the other hand, there is a slightly tedious but rather powerful approach which is based on an 
equivalent set of integral-differential equations for the physical components (\ref{physical green def}) of $G$, known as Kadanoff-Baym 
equations \cite{KadanoffBaym1962,BonitzBook} This approach, which has been introduced to nonequilibrium DMFT by \citet{Tran2008}, 
interprets Eq.~(\ref{ky-cinv-v01}) as a non-Markovian time-propagation scheme, which can be 
of great value both conceptually and numerically, and it automatically exploits the symmetries (\ref{redundancy}) 
and (\ref{hermite conjugate}).

\begin{figure}[t]
\begin{center}
\includegraphics[width=6.5cm,bb=0 160 800 600,clip=true]{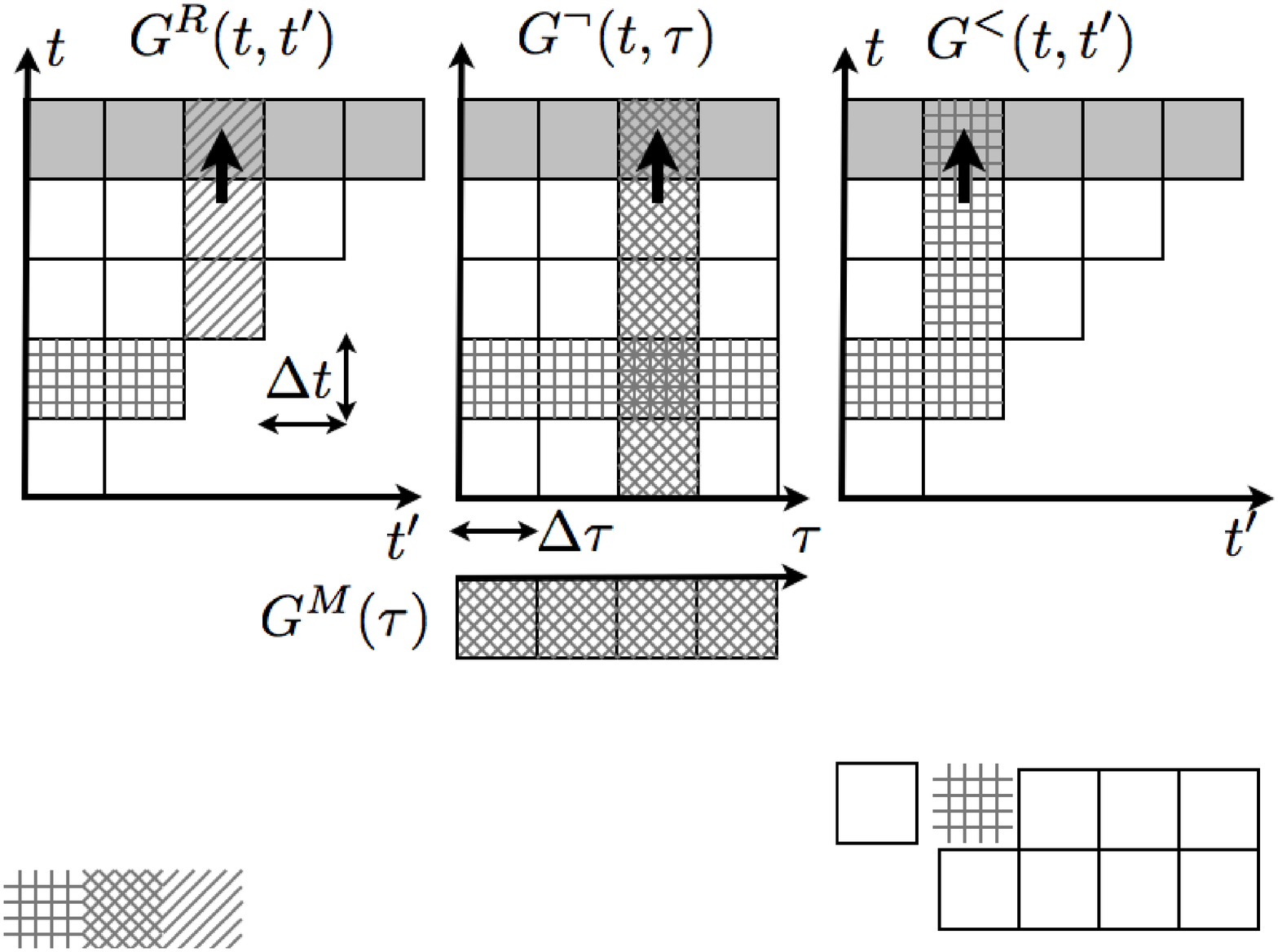}
\caption{Causal structure of the Kadanoff-Baym equation (\ref{kadanoff-baym equation}), when $G$ is computed 
on one time-slice of an equally spaced grid (shaded area): The computation of derivatives for the propagation from 
the previous time-slice (bold arrows) addresses $G$ only at earlier times, marked by a diagonal pattern for $\partial_tG^R$, 
diamond pattern for $\partial_tG^\neg$, and checkerboard pattern for $\partial_tG^<$. Hermitian symmetries (\ref{hermite 
conjugate}) are used to relate $G^A$ with $G^R$, $G^<(t,t')$ with $G^<(t',t)$,  and 
left-mixing with right-mixing components.
}
\label{KB propagation}
\end{center}
\end{figure}

For this procedure we choose a subset of the components (\ref{physical green def}) that completely parametrize $G$, taken into account 
the symmetries (\ref{redundancy}) and (\ref{hermite conjugate}), e.g., $G^M$, $G^R$, $G^{\lh}$, and $G^<$. Using their definition 
(\ref{physical green def}) and the definition of the convolution (Table \ref{ky-ccalc}), one can express the physical components of 
a convolution $\Sigma\convz G$ in terms of the components of $G$ and $\Sigma$ (using the Langreth rules \cite{Langreth1976}), and hence derive four coupled
integral equations,
\begin{subequations}
\begin{align}
&[-\partial_\tau-h(0^-)] G^M(\tau)
-\!\!
\int_0^\beta \!\!d\bar{\tau}\, \Sigma^M(\tau-\bar{\tau}) G^M(\bar{\tau})
  =
    \delta(\tau),
\label{kadanoff-baym M}
\\
&[i\partial_t-h(t)]G^R(t,t')
-\int_{t'}^t \!\!\!d\bar{t}\, \Sigma^R(t,\bar{t})G^R(\bar{t},t') =
    \delta(t-t'),
\label{kadanoff-baym R}
\\
&[i\partial_t\!-h(t)]G^{\!\lh}(t,\tau ')
-
\!\!\int_0^t \!\!d\bar{t}\, \Sigma^R(t,\bar{t})G^{\!\lh}(\bar{t},\tau')
\!=\! Q^{\!\lh}(t,\tau'),
\label{kadanoff-baym Left}
\\
&[i\partial_t-h(t)]G^<(t,t')
-\int_0^t \!\!d\bar{t}\, \Sigma^R(t,\bar{t})G^<(\bar{t},t')
= Q^{<}(t,t'),
\label{kadanoff-baym L}
\end{align}
with
\begin{align}
&Q^{\!\lh}(t,\tau '),
=    \int_0^\beta d\bar{\tau}\, \Sigma^{\lh}(t,\bar{\tau})G^M(\bar{\tau},\tau '),
\label{Q Left}
\\
 &Q^{<}(t,t')=
     \int_0^{t'} \!d\bar{t}\, \Sigma^<(t,\bar{t})G^A(\bar{t},t')
    -i\int_0^\beta \!d\bar{\tau}\, \Sigma^{\lh}(t,\bar{\tau}) G^{\rh}(\bar{\tau},t').
\label{Q les}
\end{align}
\label{kadanoff-baym equation}
\end{subequations}
Here, the integral limits take into account that retarded functions vanish for $t<t'$. Together with the boundary 
condition (\ref{kk-boundary}) these ``Kadanoff-Baym equations'' determine $G$ uniquely. To see how these 
Kadanoff-Baym equations represent the above mentioned time-propagation scheme, one may first notice that 
Eq.~(\ref{kadanoff-baym M}) for $G^M$ is decoupled from the rest: It must be solved with the boundary condition 
$G^M(\tau)=\pm G^M(\tau+\beta)$ for bosons ($+$) or fermions ($-$). Hence one can solve it by Fourier transformation 
(\ref{mat FT}), and its solution is the Green's function of the initial equilibrium state,
\begin{align}
\label{ky-cinv-mat-w}
G^M(i\omega_n) = [i\omega_n - h(0^-) -\Sigma^M(i\omega_n)]^{-1},
\end{align}
independent of the subsequent perturbation of  the system.
The remaining equations (\ref{kadanoff-baym R}) - (\ref{kadanoff-baym L}) have an inherent causal 
structure: If $G_t=\{ G^R(t,t'), G^{\lh}(t,\tau),G^<(t,t') | \,0\le  \tau\le \beta,t'\le t\}$ denotes the values of 
$G$ at ``time slice $t$'', then $\partial_t G_t$ is determined by $G^M$ (the initial state), $\Sigma_t$, 
and $G_{t'}$ for $t'\le t$ (Fig.~\ref{KB propagation}). One can therefore always solve (\ref{kadanoff-baym R}) - 
(\ref{kadanoff-baym L}) by successively increasing $t$. Furthermore, when one keeps the second time 
argument of $G$ fixed in Eqs.~(\ref{kadanoff-baym R}) - (\ref{kadanoff-baym L}) one obtains a set of 
one-dimensional integral-differential equations of the type 
\begin{equation}
\label{volterra diff}
\frac{d}{ds} y(s) = q(s) + p(s)y(s) + \int_0^s \!d\bar s\,  k(s,\bar s) y(\bar s), 
\end{equation}
i.e., Volterra equations of the second kind \cite{BrunnervanderHouwenBook}. The causal structure of this equation is 
evident from the limits of the integral. For example, for Eq.~(\ref{kadanoff-baym R}) we define $y(s) = G^R(t+s,t)$, 
and $k(s,\bar s)=\Sigma^R(t+s,t+s-\bar s)$, and the initial condition is provided by $y(s<0)=0$. In practice,
one thus solves a large number of coupled Volterra equations, for which very stable and accurate high-order 
algorithms can be found in the literature \cite{NumericalRecipesC, LinzBook, BrunnervanderHouwenBook}.
In Appendix \ref{Appendix A},
we show one of them, namely the implicit Runge-Kutta method or the collocation method.

Detailed descriptions of the numerical implementation of the Kadanoff-Baym equations
can be found in \cite{Stan2009,Koehler1999,Eckstein2010,Balzer2013}. In general, the required resources scale like 
$\mathcal{O}(M^2)$ for memory and $\mathcal{O}(M^3)$ for CPU time, where $M$ is the number of time-discretization steps. 
In particular the memory can be a limiting factor, (when Green's functions carry many orbital indices), such that efficient 
shared memory or even (distributed-memory) parallelization schemes can become necessary \cite{Balzer2013}. 
To validate the accuracy of the numerics, checking nonequilibrium sum rules  \cite{Turkowski2006,Turkowski2008}
can be very helpful.

Let us conclude this section by mentioning two other generic contour equations, which appear frequently in nonequilibrium DMFT.
These are 
\begin{align}
\label{ky-vie2}
[1 + F ] \convz  G &= Q,
\\
\label{ky-vie1}
F \convz  G &= Q,
\end{align}
to be solved for $G$, where $Q$ and $F$ are contour Green's functions. Both equations have a causal structure analogous  
to Eq.~(\ref{ky-cinv-v01}). Their numerical solution differs only in that there is no derivative term, and the existence of the source 
term on the right hand. Hence an analogous time-propagation scheme exists \cite{Eckstein2010}, in which 
Eqs.~(\ref{ky-vie2}) and (\ref{ky-vie1}) are reduced to Volterra integral equations of the second and first kind, respectively
\cite{BrunnervanderHouwenBook},
\begin{align}
y(s) + \int_0^s \!d\bar s\,  k(s,\bar s) y(\bar s) 
&= 
q(s) ,
\\
\int_0^s \!d\bar s\,  k(s,\bar s) y(\bar s) 
&= 
q(s).
\end{align}
A this point, it may be interesting to note that in general high-order accurate propagation schemes for Volterra equations of 
the first kind are unstable \cite{NumericalRecipesC, BrunnervanderHouwenBook}. Fortunately, within nonequilibrium DMFT
one can always rewrite the equations in order to take advantage of the ``stabilizing $1$'' in Eq.~(\ref{ky-vie2}).

\subsubsection{Keldysh formalism for nonequilibrium steady states}
\label{steady-states}
\begin{figure}[t]
\begin{center}
  \includegraphics[width=7cm]{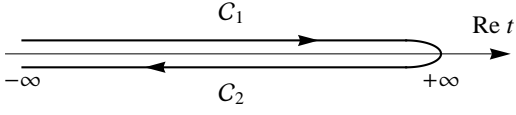}
  \caption{The Keldysh contour $\mathcal{C}_K=\mathcal{C}_1\cup\mathcal{C}_2$ with the two branches ranging from $-\infty$ to $\infty$.}
  \label{keldysh contour}
\end{center}
\end{figure}

%\input{keldysh.tex}
%%%%%%%%%%%%%%%%%%%%%%%%%%%%%%%
\paragraph{Keldysh formalism}
An alternative and simpler approach to nonequilibrium Green's functions is the Keldysh formalism \cite{Keldysh1964}, which 
is particularly well suited to describe nonequilibrium steady states of open systems, in which the energy supplied to the system 
by an external driving field is balanced by the energy flowing out to the environment. In the original Keldysh theory, it is assumed 
that the initial state is noninteracting, and an interaction is adiabatically turned on from $t=-\infty$. Since no interaction vertex 
can be inserted on the imaginary-time axis, correlations between the initial state and the time-evolving state represented by the 
mixed self-energy $\Sigma^{\lh}$ and $\Sigma^{\rh}$ 
(initial correlations) 
vanish. The imaginary-time contour $\CC_3$ is thus decoupled from $\CC_1$ and $\CC_2$,
and one can restrict oneself to the real-time branches $\CC_K=\CC_1\cup\CC_2$ (Keldysh contour, Fig.~\ref{keldysh contour}). The 
contour-ordered Green's function is closed on a $2\times 2$ subspace of (\ref{G 3x3}),
\begin{align}
  \hat{G}
    &=
      \begin{pmatrix}
        G^{11} & G^{12} \\
        G^{21} & G^{22}
      \end{pmatrix}.
\end{align}
One can conveniently carry out the transformation to physical Green's functions 
(\ref{physical green def}) by introducing the matrices
\begin{align}
  L
    &=
      \frac{1}{\sqrt{2}}
      \begin{pmatrix}
        1 & -1  \\
        1 &  1 
      \end{pmatrix}
  ,\quad
  \tau_3
    =
      \begin{pmatrix}
        1 &  0  \\
        0 & -1  \\
      \end{pmatrix}.
  \label{L tau3}
\end{align}
$L$ is a unitary matrix, while $\tau_3$ represents the sign of the measure $dt$ that appears in an integral along 
$\CC_K$. Using these, we perform a linear transformation (Keldysh rotation \cite{Keldysh1964, LarkinOvchinnikov1975, RammerBook}), 
obtaining
\begin{align}
  \underline{G}
  \equiv 
  L \tau_3 \hat{G} L^\dagger
    &=
      \begin{pmatrix}
        G^R & G^K \\
          0 & G^A
      \end{pmatrix}.
\end{align}
Convolutions along $\CC_K$ are then simply written as one-dimensional time integrals, 
$\underline{A\convz B}\,(t,t') = \int_{-\infty}^{\infty} d\bar t \,\underline{A}(t,\bar t)\,\underline{B}(\bar t,t')$.

In general, initial correlations are supposed to be relevant in a realistic situation, since the interaction always exists in the initial state.
However, in a dissipative system coupled to an external heat bath the initial correlation is expected to disappear in the long-time limit, 
since the large number of degrees of freedom in the heat bath would influence the long-time dynamics, and wipe out the information of 
the initial state and initial transient dynamics. In this case, the Keldysh formalism is applicable to the nonequilibrium steady state
without the use of adiabatic switching of interactions. Although the independence on the initial state is assumed to be true in 
general, it is extremely hard to prove this fact rigorously for a given model, as it is ultimately related to the fundamental question 
of thermalization of the system (\cite{Polkovnikov2011RMP}, see also Sec.~\ref{subsec:parameterchanges}).

\paragraph{Free-fermion bath}
\label{buttiker}

To describe a nonequilibrium steady state in a dissipative system, one may consider a system coupled to an environment (open system),
\begin{align}
\label{h with environment}
  H_{\rm tot}
    &=
      H_{\rm s}+H_{\rm mix}+H_{\rm bath}.
\end{align}
Here $H_{\rm s}$ and $H_{\rm bath}$ are Hamiltonians of the system and environment, respectively, and $H_{\rm mix}$ represents
a coupling between them. $H_{\rm bath}$ is assumed to have a much larger number of degrees of freedom
than $H_{\rm s}$, so that $H_{\rm bath}$ acts as a heat bath for the system. When the system is excited
by an external field, the energy injected from the field flows to the bath, resulting in the energy dissipation.

\begin{figure}[t]
\begin{center}
  \includegraphics[width=7cm]{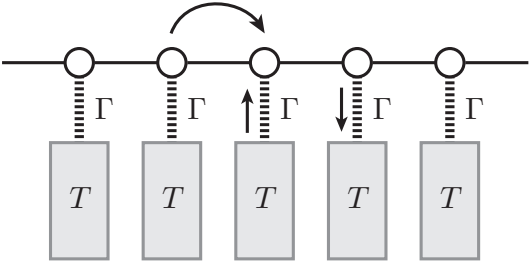}
  \caption{Schematic representation of a free-fermion bath model. Each site of the system is connected to
  an electrode with a coupling $\Gamma$ at temperature $T$. The arrows indicate the motion of particles.}
  \label{buttiker model}
\end{center}
\end{figure}

The simplest model of the heat bath, that can be solved analytically, is a free-fermion bath 
\cite{Tsuji09,Amaricci2012,Aron12,Werner2012Kondo,Han2012}
defined by
\begin{align}
  H_{\rm mix}
    &=
      \sum_{i,p} V_p (c_i^\dagger b_{i,p}+b_{i,p}^\dagger c_i),
  \label{h_mix}
  \\
  H_{\rm bath}
    &=
      \sum_{i,p} (\epsilon_{b,p}-\mu_b)b_{i,p}^\dagger b_{i,p},
  \label{h_b}
\end{align}
where $c_i^\dagger$ ($c_i$) creates (annihilates) the system's fermions, 
$b_{i,p}^\dagger$ ($b_{i,p}$) creates (annihilates) fermionic degrees of freedom of the bath, 
$\epsilon_{b,p}$ is the bath level energy, and $V_p$ is the hybridization between the system and the mode $p$ of the bath.
The thermal bath is assumed to be equilibrated with temperature $T$. 
The chemical potential of the bath ($\mu_b$) is determined such that no current flows between the bath and the system.
This model is equivalent to the one where an electrode is attached to every site of the system's lattice,
such that the fermions can hop between the lattice and the electrodes (Fig.~\ref{buttiker model}).
In other words, the model can be seen as a set of coupled quantum dots connected to independent electrodes
(cf. B\"{u}ttiker model \cite{Buttiker1985, Buttiker1986}).
This is one explicit realization of a grand canonical ensemble, in which the system is coupled to a particle reservoir. 

The bath's degrees of freedom can be analytically integrated out \cite{FeynmanVernon1963, CaldeiraLeggett1981},
since 
$H_\text{mix}+H_\text{bath}$ is 
quadratic in $b_{i,p}^\dagger$ and $b_{i,p}$.
To this end, we introduce the contour-ordered Green's function of the bath,
\begin{align}
  G_{\rm bath}(p;t,t')
    &=
      -i\langle \mathcal{T}_{\mathcal{C}_K}\, b_{i,p}(t)b_{i,p}^\dagger(t') \rangle_\text{bath}.
\end{align}
Since the bath's fermions are noninteracting, $G_{\rm bath}$ can be evaluated exactly. After integrating out $b_{i,p}^\dagger$ and $b_{i,p}$,
one can show that the effect of the bath is an additional self-energy correction to the system Green's function,
\begin{align}
  \Sigma_{\rm bath}(t,t')
    &=
      \sum_p V_p G_{\rm bath}(p;t,t') V_p,
  \label{sigma_diss}
\end{align}
which is local in space (or does not have momentum dependence in $\bm k$-space).
As a result, the Dyson equation for the system Green's function can be symbolically written as 
\begin{align}
\underline{G}
=
\big[
\underline{G_0}^{-1}
-
\underline{\Sigma_\text{bath}}
-\underline{\Sigma}
\big]^{-1}
\label{dyson with dissipation}
\end{align}
in the Keldysh formulation. 
Since the bath Green's function has a time translation invariance,
it is convenient to represent the self-energy $\Sigma_{\rm bath}$ in real frequency space.
The retarded component of $\Sigma_{\rm bath}$ is 
\begin{align}
\Sigma_{\rm bath}^R(\omega)
&=
\sum_p \frac{V_p^2}{\omega+\mu_b-\epsilon_{b,p}+i\eta},
\end{align}
where $\eta$ is a positive infinitesimal. By using the formula $\frac{1}{\omega+i\eta}=\mathcal{P}\frac{1}{\omega}-i\pi\delta(\omega)$
(with $\mathcal{P}$ the principal value), one can divide $\Sigma_{\rm bath}^R(\omega)$ into real and imaginary parts.
The imaginary part defines a spectral function of the heat bath, 
\begin{align}
\Gamma(\omega)
&=
\sum_p \pi V_p^2 \delta(\omega+\mu_b-\epsilon_{b,p}).
\label{gamma}
\end{align}
A simple treatment of dissipation is to 
omit the $\omega$ dependence of $\Sigma_{\rm bath}^R(\omega)$. This corresponds to considering
a flat density of states [$\Gamma(\omega)=\Gamma$] for the reservoir. 
The presence of the imaginary part of $\Sigma_{\rm bath}^R$
is a manifestation of irreversibility, i.e., dissipation of energy (and particles) via the fermionic bath.
The real part of $\Sigma_{\rm bath}^R$,
a potential shift due to the coupling to the bath,
can be absorbed into the chemical potential $\mu$ of the system, so that only the imaginary part
affects the dynamics of the system. Recall that the bath is always in equilibrium with temperature $T=\beta^{-1}$. 
Hence the fluctuation-dissipation relation
(\ref{fluctuation-dissipation}) holds for the bath Green's function, which determines the Keldysh component
of the dissipation term. Thus, one has
\begin{align}
\begin{pmatrix}
\Sigma_{\rm bath}^R(\omega) & \Sigma_{\rm bath}^K(\omega) \\
0 & \Sigma_{\rm bath}^A(\omega)
\end{pmatrix}
&=
\begin{pmatrix}
-i\Gamma & -2i\Gamma F(\omega) \\
0 & i\Gamma
\end{pmatrix}
\label{sigma bath}
\end{align}
with $F(\omega)=\tanh(\beta\omega/2)$ (\ref{coth, tanh}).
The retarded and advanced components are damping terms that cause a relaxation,
while the Keldysh component represents thermal fluctuations from the heat bath.
Although the model defined by Eqs.~(\ref{h_mix}) and (\ref{h_b}) might look somewhat artificial, one can consider it
as a phenomenological treatment of dissipation with two parameters $\Gamma$ and $T$, 
where the damping rate is simply parametrized by a constant $\Gamma$, and the heat-bath temperature by $T$ 
(analogous to a relaxation time approximation in the classical Boltzmann equation \cite{MahanBook}).

The Keldysh component of $\Sigma_{\rm bath}$ in Eq.~(\ref{dyson with dissipation}) works as a 
boundary condition and allows one to determine
a distribution function for the nonequilibrium steady state. 
If it is absent, the Keldysh Green's function reads $G^K=G^R [-(G_0^{-1})^K+\Sigma^K] G^A$,
where $(G_0^{-1})^K\equiv -G_0^R{}^{-1}G_0^K G_0^A{}^{-1}$ contains information
about the initial condition. In general, $(G_0^{-1})^K$ is proportional
to $i\eta$ in frequency space, and vanishes in the limit of $\eta\to 0$ (since the noninteracting systems
is dissipationless). Thus one ends up with $G^K=G^R \Sigma^K G^A$ (Keldysh equation), which is, however,
{\it homogeneous} in the sense that it does not have a source term to determine the Keldysh component 
of the Green's function. Due to the lack of an input for the distribution function, it is impossible to find 
a unique solution for the distribution. On the other hand,
in the Dyson equation (\ref{dyson with dissipation})
we have a nonzero source term $\Sigma_{\rm bath}^K(\omega)=-2i\Gamma F(\omega)$ in the Keldysh component,
which acts as a boundary condition. Then we can totally neglect the noninteracting term $(G_0^{-1})^K$,
i.e., the initial condition is wiped out by dissipation.
Equation (\ref{dyson with dissipation}) becomes
an {\it inhomogeneous} equation, enabling one to determine the nonequilibrium steady state from the boundary condition.
This mechanism allows a description of nonequilibrium steady states in dissipative systems within the Keldysh formalism.

\subsection{Nonequilibrium dynamical mean-field theory}
\label{Sec_DMFT}

%\input{dmft_formalism.tex}
%%%%%%%%%%%%%%%%%%%%%%%%%%%%%%
\newcommand{\vect}[1]{{\bm #1}}
\newcommand{\Veck}{{\vect{k}}}
\newcommand{\Vecr}{{\vect{r}}}
\newcommand{\VecA}{{\vect{A}}}
\newcommand{\VecE}{{\vect{E}}}
\newcommand{\VecR}{{\vect{R}}}
\newcommand{\Vecq}{{\vect{q}}}
\newcommand{\Vecv}{{\vect{v}}}
\newcommand{\Vecx}{{\vect{x}}}
\newcommand{\Vecz}{{\vect{z}}}
\newcommand{\Vecj}{{\vect{j}}}
\newcommand{\CCdt}{\partial_t^\CC}

\newcommand{\qq}{{\bm q}}
\newcommand{\pp}{{|\!|}}

\subsubsection{Overview of equilibrium DMFT}
\label{dmft_formalism::sec-eq}

Many of the ideas underlying nonequilibrium DMFT are direct generalizations of the conventional equilibrium DMFT formalism \cite{Georges96}, so it is worthwhile to start this chapter with a 
brief overview of equilibrium DMFT and its foundations. 
Static mean-field theories, such as  the Weiss mean-field theory for a spin system $H = \sum_{ij} J_{ij}{\bm S}_i\cdot {\bm S}_j 
+ \sum_i {\bm h_i}\cdot {\bm S}_i$, have been known for a long time. In the latter, the spin $\langle {\bm S}\rangle$ at a given 
site $i$ is determined by the (thermal) average $\text{Tr} [e^{-\beta H_\text{eff}} {\bm S} ]/Z$, taken with an effective 
single-site Hamiltonian $H_\text{eff} = ({\bm h}_i + {\bm h}_\text{mf}) \cdot {\bm S}$ which describes one spin in the average 
(``Weiss'') field ${\bm h}_\text{mf}=\sum_{j} J_{ij} \langle {\bm S}_j \rangle$ due to the interaction with its neighbors. 
For interacting electrons on a lattice, the simplest static mean-field theory is the Hartree approach, which 
approximates the Coulomb interaction between the particles by an averaged time-independent potential. 
Electrons can thus avoid each other only by forming a static long-range 
order, which is clearly not the true story:  Interacting electrons correlate their motion {\em in time}, so that they 
almost never occupy the same orbital simultaneously. A theory of Mott insulators and correlated metals (in particular, 
the paramagnetic state) must necessarily keep track of these non-trivial time-dependent correlations. DMFT can 
achieve this goal because it is not an effective theory for the electron density, but for the local frequency-dependent 
Green's function $G(\omega)$, which contains the information about these time-dependent fluctuations. Besides this 
important difference, a formal analogy to the static mean-field theory remains:  $G(\omega)$ is obtained from an 
effective model that involves only one site of the lattice. This site is coupled to a ``fluctuating Weiss field'' 
$\Delta(\omega)$ that resembles the exchange of particles with the rest of the lattice and must be 
determined self-consistently as a functional of $G$.

In the following we first state and then discuss the equilibrium DMFT equations for the case of a single-band 
Hubbard model (generalizations are being discussed below),
\begin{align}
\label{dmft_formalism::h-hubbard}
H &=
\sum_{\langle ij \rangle,\sigma} 
v_{ij}\,
c_{i\sigma}^\dagger
c_{j\sigma}
+
\sum_{i} 
H_\text{loc}^{(i)},
\\
\label{dmft_formalism::hint-hubbard}
H_\text{loc}^{(i)}
&=
U
\big(c_{i \uparrow}^\dagger c_{i \uparrow}-\tfrac12 \big)
\big(c_{i \downarrow}^\dagger c_{i \downarrow}-\tfrac12 \big).
\end{align}
Here, $c_{i\sigma}^\dagger$ ($c_{i\sigma}$) creates (annihilates) an electron with spin $\sigma$ in a Wannier orbital at 
site $i$ of a crystal lattice, $v_{ij}$ is the hopping matrix element (we use the symbol $v$ in order to avoid later confusion 
with time $t$), and electrons interact via a local Coulomb interaction $U$. The key approximation of DMFT is that the 
electronic self-energy is taken to be local in space,
\begin{equation}
\label{dmft_formalism::local-sigma-1}
\Sigma_{ij}(i\omega_n) = \delta_{ij}\, \Sigma_{ii}(i\omega_n),
\end{equation}
(where $\sigma$ is suppressed for simplicity). Furthermore, it is assumed that $\Sigma_{ii}(i\omega_n)$ and 
the local Green's function $G_{ii}(\tau) = - \langle \mathcal{T}_\tau c_{i\sigma}(\tau) c_{i\sigma}^\dagger (0)\rangle$ $=$ $T\sum_n 
e^{-i\omega_n \tau} G(i\omega_n)$ can be computed from an effective impurity model with action \cite{Georges1992a}
\begin{equation}
\label{dmft_formalism::imp-action-1}
\mathcal{S}_i =
- \int_0^\beta
\!\!\! d\tau H_\text{loc}(\tau)
-
\!
\int_0^\beta 
\!\!\!d\tau d\tau'
\sum_{\sigma}
\!
c_\sigma^\dagger(\tau)
\Delta_i(\tau-\tau')
c_\sigma(\tau'),
\end{equation}
where $\Delta$ is the hybridization to a fictitious bath. One has
\begin{align}
\label{dmft_formalism::imp-g-1}
&G_{ii}(\tau) 
= 
-\text{Tr} \big[\mathcal{T}_\tau e^{\mathcal{S}_i} c(\tau) c^\dagger(0)\big]/Z,
\\
\label{dmft_formalism::imp-dyson-1}
&G_{ii}(i\omega_n) ^{-1}
=
i\omega_n + \mu - \Delta_i(i\omega_n) -\Sigma_{ii}(i\omega_n),
\end{align}
where the second equation is the Dyson equation for the impurity model that defines the 
self-energy $\Sigma_{ii}(i\omega_n)$. Because $\Sigma_{ii}$ and $G_{ii}$ are related by the 
Dyson equation of the lattice model,
\begin{equation}
\label{dmft_formalism::latt-dyson-1}
G^{-1}_{ij}(i\omega_n) = \delta_{ij}\big[ i \omega_n + \mu - \Sigma_{ii}(i\omega_n)\big] - v_{ij},
\end{equation}
the auxiliary quantity $\Delta$ can be eliminated to close the equations. The hybridization
function $\Delta(\omega)$ plays the role of a (frequency-dependent) Weiss field, and Eqs.~(\ref{dmft_formalism::imp-dyson-1}) 
and (\ref{dmft_formalism::latt-dyson-1}) provide the (implicit) functional relation between $G(\omega)$ and $\Delta(\omega)$. 
Closed relations $\Delta[G]$ can be obtained, for example, for hopping models on the Bethe lattice, 
in particular for nearest-neighbor hopping $v_*/\!\sqrt{z}$ with
coordination number $z\rightarrow \infty$, which gives \cite{Georges96} 
\begin{equation}
\label{bethe selfconsistency}
\Delta(i\omega_n)=v_*^2G(i\omega_n).
\end{equation} 

For a translationally invariant system, with $\Sigma_{ii}(i\omega_n)\equiv  \Sigma(i\omega_n)$, Eq.~(\ref{dmft_formalism::latt-dyson-1}) 
can be solved for $G_{ii}$ in the form
\begin{align}
G_{ii}(i\omega_n) 
&= 
\frac{1}{L}\sum_{\kk} G_{\kk}(i\omega_n)
=
\frac{1}{L}\sum_{\kk} \frac{1}{i\omega_n + \mu - \Sigma(i\omega_n) - \epsilon_\kk }
\nonumber
\\
&= 
\int d\epsilon \,
\frac{D(\epsilon)}{i\omega_n + \mu - \Sigma(i\omega_n) - \epsilon },
\end{align}
where $L$ is the number of lattice sites,
$G_{\kk}(\tau) = - \langle \mathcal{T}_\tau c_{\kk\sigma}(\tau) c_{\kk\sigma}^\dagger (0)\rangle$ is the momentum-resolved
Green's function, and $D(\epsilon)=\frac{1}{L}\sum_\kk \delta(\epsilon-\epsilon_\kk)$ is the local density of states. 
An important case is the semielliptic density of states 
\begin{align}
D(\epsilon)=\frac{1}{2\pi v_\ast^2} \sqrt{4v_\ast^2-\epsilon^2},
    \label{eq:semielliptic}
\end{align} 
which corresponds to 
nearest-neighbor hopping on 
the Bethe lattice with infinite coordination number and thus implies Eq.~(\ref{bethe selfconsistency}).

The starting point for the derivation of the DMFT equations (\ref{dmft_formalism::local-sigma-1})-(\ref{dmft_formalism::latt-dyson-1}) 
has been the limit of infinite dimensions \cite{Metzner1989}. A meaningful limit $d\to\infty$ is obtained when the hopping 
matrix elements are rescaled such that the average kinetic energy remains finite, and the physically relevant competition 
between kinetic and interaction energy is preserved. For a hypercubic lattice with nearest neighbor hopping, one chooses
\begin{equation}
\label{dmft_formalism::dinf}
v=\frac{v_\ast}{\sqrt{2d}},
\end{equation}
where $v_*$ is kept constant as $d\to\infty$. While the limit $d\to\infty$ leads to many simplifications in fermionic 
lattice models \cite{Vollhardt1991,Vollhardt1992}, its most important consequences are arguably the local nature of perturbation theory  \cite{Metzner1989} and in particular the locality of 
the self-energy [Eq.~(\ref{dmft_formalism::local-sigma-1})]~\cite{Muellerhartmann1989a,Muellerhartmann1989b}. These properties lead 
to the mean-field equations (\ref{dmft_formalism::local-sigma-1})-(\ref{dmft_formalism::latt-dyson-1}), which 
provide the exact solution of the Hubbard model in the limit of infinite dimensions \cite{Georges1992a,Jarrell1992}. 
A proof can be given in various ways, including a linked cluster expansion around the atomic limit 
\cite{Metzner1990,Georges96}, a field-theoretical approach \cite{Janis1991,Janis1992}, or the cavity method \cite{Georges96}. 

In the following, we briefly review a diagrammatic argument which shows that for $d\to\infty$ DMFT reproduces the 
Feynman diagrams for $\Sigma$ to all orders in perturbation theory. For this purpose one considers the self-energy $\Sigma[G]$ 
as a functional of the interacting Green's function $G$. In terms of Feynman diagrams, $\Sigma[G]$ is the sum of all 
self-energy diagrams where internal lines have no self-energy insertions (skeleton diagrams), but represent $G$ instead 
of $G_0$. A power counting argument then shows that for $d\to\infty$ 
contributions from nonlocal diagrams should vanish for $\Sigma$: 
Each pair of vertices in a skeleton diagram for $\Sigma$ (with space indices $j$ and $l$) is at least 
connected by three independent path of Green's function lines. If one vertex is an internal vertex, summation over its 
space index contributes $\sim d^{ | l-j |}$ terms with distance $|l-j|$ between $l$ and $j$, while the factor due to the 
three $G$-lines scales as $\sim d^{-3|l-j|/2}$ for $d\to\infty$ due to Eq.~(\ref{dmft_formalism::dinf}). Hence only the term 
$l=j$ survives for $d\to\infty$. This shows that the functional relation between $G$ and $\Sigma$ is 
the same as the one for a general single-impurity Anderson model (\ref{dmft_formalism::imp-action-1}),
\begin{equation}
\label{dmft skeleton}
\Sigma_{ii}[G]=\Sigma_{SIAM}[G_{ii}].
\end{equation} 
By choosing the auxiliary quantity $\Delta_i$ such that Eq.~(\ref{dmft_formalism::imp-g-1}) yields a given $G_{ii}$, one 
thus ensures that the self-energy obtained from Eq.~(\ref{dmft_formalism::imp-dyson-1}) gives the correct value of the 
functional $\Sigma_{ii}[G]$. In this form, the argument was first stated for the Falicov-Kimball Model 
\cite{Brandt1989a,Brandt1990a,Brandt1991a}. Because the skeleton expansion is the derivative of the Luttinger-Ward 
functional \cite{Luttinger1960},  $\Sigma[G] = \frac{\delta \Phi[G]}{ \delta G }$, DMFT can be rephrased by stating that 
$\Phi[G] =\sum_i \Phi_{SIAM}[G_{ii}]$. This statement provides a suitable starting point for the formulation of  nonequilibrium 
DMFT.

\subsubsection{Nonequilibrium DMFT formalism}
\label{dmft_formalism::sec-noneq}

Quite generally, the mean-field concept can also be used to describe the time evolution of lattice systems, e.g.,  based on time-dependent Hartree 
or Gross-Pitaevskii equations. In a direct extension of the static Weiss mean-field theory for spins, the effective Hamiltonian would simply determine 
the time-evolution of ${\bm S}$, and not only its statistical average. This would result in a nonlinear initial value problem,
\begin{equation}
\label{dmft_formalism::static_mf-2}
\partial_t  \langle {\bm S}_i(t)\rangle =  
\Big[
{\bm h}_i 
+
\sum_{j} J_{ij} \langle {\bm S}_j(t)\rangle
\Big]
\times \langle {\bm S}_i(t)\rangle,
\end{equation}
where the term in brackets is the time-dependent mean-field. The initial value is obtained from the static mean-field theory. 
Similar to the equilibrium case discussed at the beginning of Sec.~\ref{dmft_formalism::sec-eq},
the time-dependent generalization of static (Hartree) mean-field
theory would fail to correctly describe the dynamics of correlated electrons. Instead, one
must make the dynamical mean-field $\Delta(\omega)$ time-dependent, thus going from a
Weiss field $\Delta(\omega)$ which captures particle fluctuations in equilibrium
to a time-dependent ``Weiss field'' $\Delta(t,t')$ which depends on two times.
This defines an effective model for the two-time Green's
function $G(t,t')$ in terms of an effective action that involves only
local degrees of freedom coupled to the Weiss field $\Delta(t,t')$, which in turn depends  on $G$ at earlier times.

The precise set of equations can again be obtained from the limit $d\to\infty$ in the Hubbard model. As long as the total 
length of the contour is finite, the power counting arguments based on the rescaling (\ref{dmft_formalism::dinf}) remain 
valid for contour-ordered Green's functions \cite{Schmidt2002}. Hence one may conclude that for $d\to\infty$
the (contour-ordered) self-energy is local in space,
\begin{equation}
\label{dmft_formalism::local-sigma-2}
\Sigma_{ij}(t,t') = \delta_{ij} \,\Sigma_{i}(t,t'),
\end{equation}
(spin indices are suppressed to simplify the notation), such that contour-ordered lattice Green's functions 
$G_{ij}(t,t') = -i \langle \mathcal{T}_\CC c_{i\sigma}(t) c_{j\sigma}^\dagger(t')\rangle$ can be obtained from a  Dyson equation
(Sec.~\ref{quantum boltzmann})
\begin{equation}
\label{dmft_formalism::latt-dyson-2}
(G^{-1})_{ij}(t,t') =  
\big[\delta_{ij}(i\partial_t + \mu) - v_{ij}(t) \big]\delta_\CC(t,t')  - \delta_{ij}\Sigma_{ii}(t,t'),
\end{equation}
where the first term $(G^{-1}_0)_{ij}(t,t') = \big[\delta_{ij}(i\partial_t + \mu) - v_{ij}(t) \big]\delta_\CC(t,t')$ is the inverse of the noninteracting 
lattice Green's function. Furthermore, Eq.~(\ref{dmft skeleton}) still holds, i.e., in order to evaluate 
the correct functional $\Sigma_{ii}[G]$ in $d\to\infty$ it is sufficient to solve a general local model with action
\begin{align}
 \label{dmft_formalism::imp-action-2}
 \CS_{i} 
 = 
 -i\intC \!dt \,H_{\text{loc}}(t) 
 -i 
 \sum_{\sigma}
 \intC \!dt 
 dt'\,
 c^\dagger_{\sigma}(t)
 \Delta_{i}(t,t') 
 c_{\sigma}(t'),
\end{align}
where the auxiliary field $\Delta(t,t')$ is chosen such that  
\begin{align}
\label{dmft_formalism::imp-g-2}
&G_{ii}(t,t') 
= 
-i
\langle\mathcal{T}_\CC c(t)c^\dagger(t')\rangle_{\mathcal{S}_i},
\end{align}
and $\Sigma$ is implicitly defined via the Dyson equation
\begin{align}
\label{dmft_formalism::imp-dyson-2}
&G_{ii}^{-1}(t,t')
=
(i \partial_t + \mu)\delta_\CC(t,t') - \Sigma_{ii}(t,t')   - \Delta_i(t,t').
\end{align}
Equations (\ref{dmft_formalism::local-sigma-2})-(\ref{dmft_formalism::imp-dyson-2}) provide the closed set of equations
for nonequilibrium DMFT. Although these equations look formally identical to 
Eqs.~(\ref{dmft_formalism::local-sigma-1})-(\ref{dmft_formalism::latt-dyson-1}) 
for equilibrium DMFT, they are conceptually very different: As discussed in Sections
\ref{quantum boltzmann} and \ref{KB-numerics}, Eq.~(\ref{dmft_formalism::imp-dyson-2}) can be viewed as a non-Markovian
equation of motion for $G_{ii}$, and through Eq.~(\ref{dmft skeleton}) the memory kernel $\Sigma$ has a non-linear 
dependence on $G_{ii}$. Like Eq.~(\ref{dmft_formalism::static_mf-2}), nonequilibrium DMFT (on the $L$-shaped 
contour $\CC$) is in essence a nonlinear initial value problem for $G_{ii}(t,t')$. 

Let us conclude this section with a brief remark on the limit of $d\to\infty$ for nonequilibrium. In general, small 
terms in the Hamiltonian or action can completely modify the long-time behavior of a system, even when they 
are irrelevant for the equilibrium properties. For example, one may imagine an ideal Fermi gas which is initially 
excited, such that its momentum distribution $n(\kk,t=0)$ deviates from a Fermi distribution. For a noninteracting gas, 
$n(\kk,t)$ is given by $n(\kk,0)$ for all times, but for an arbitrary weak interaction $n(\kk,t)$ is supposed to 
eventually reach a thermal equilibrium distribution. Hence the limit $t\to\infty$ is necessarily non-perturbative 
in terms of the interaction. 
In the same way, one cannot generally expect the limit of $t\to\infty$ and $d\to\infty$ to commute. In the Weiss mean-field theory 
(which is exact for $d=\infty$ with the scaling $J\sim J_*/d$ \cite{Brout1960}), e.g., spin-flip  terms 
are absent in a collinear antiferromagnet, but in a real system such terms may completely modify the long-time 
limit. The importance of nonlocal effects on the long-time dynamics of strongly correlated systems is an open 
question that can finally only be addressed by extensions of DMFT (some of which are mentioned in Sec.~\ref{cluster dmft}). 

\subsubsection{Models}

\paragraph{Overview}
Nonequilibrium DMFT can be applied to a large class of problems, including arbitrary electromagnetic driving 
fields, dissipative and non-dissipative systems, and many different types of local interaction terms. The general 
lattice Hamiltonian for the relevant (non-dissipative) models is of the form 
\begin{align}
\label{h-general}
H &=
\sum_{ij,\alpha\alpha'} 
v_{i\alpha j\alpha'}(t)
c_{i\alpha}^\dagger
c_{j\alpha'}
+
\sum_{i} 
H_{\text{loc}}^{(i)},
\end{align}
where $i$ and $\alpha$ denote site and orbital/spin labels.
The second term in this equation is a sum of {\rm local} interaction and single-particle terms. Apart from the Hubbard model 
[Eq.~(\ref{dmft_formalism::hint-hubbard})], important examples include, (i) the periodic Anderson model
\begin{equation}
\label{dmft_formalism::PAM-interaction}
H_{\text{loc}} = 
U 
f_\uparrow^\dagger f_\uparrow f_\downarrow^\dagger f_\downarrow
+E_f 
\sum_\sigma f_\sigma^\dagger f_\sigma + \sum_\sigma (V f_\sigma^\dagger c_\sigma + {\rm h.c.}),
\end{equation}
where conduction electrons hybridize with localized $f$-orbitals, (ii) the Kondo lattice model, 
\begin{equation}
\label{dmft_formalism::KLM-interaction}
H_{\text{loc}} = 
J \sum_{\sigma,\sigma'}
c_\sigma^\dagger {\bm \tau}_{\sigma\sigma'} c_{\sigma'} \cdot{\bm S},
\end{equation}
which describes electrons (with spin $\bm\tau$) 
that interact with a local spin ${\bm S}$, (iii),
the Holstein model, 
\begin{equation}
\label{dmft holstein interaction}
H_{\text{loc}} = 
 \omega_0(b^\dagger b + \tfrac12)
 +
g(b^\dagger + b) (n_\uparrow + n_\downarrow -1),
\end{equation}
for electron-phonon coupled systems, and (iv) the Hubbard-Holstein model
\begin{equation}
\label{dmft hubbard holstein interaction}
H_{\text{loc}} = U n_\uparrow n_\downarrow+
 \omega_0(b^\dagger b + \tfrac12)
 +
g(b^\dagger + b) (n_\uparrow + n_\downarrow -1).
\end{equation}
The first two models play an important role in the theory of heavy Fermion materials \cite{Coleman2007},
and all four have been studied intensively within equilibrium DMFT  
\cite{Meyer2002,DeLeo2008,Otsuki2009,Werner2009holstein}.

The Falicov-Kimball model, 
  \begin{align}%
    H
    &=
    \sum_{ij} v_{ij}(t)
    c_i^\mydagger
    c_j^\phdagger
    +
    E_f
    \sum_i
    f_i^\mydagger
    f_i^\phdagger
    +
    U
    \sum_i
    f_i^\mydagger
    f_i^\phdagger
    c_i^\mydagger
    c_i^\phdagger
    \,,\label{eq:FK-model}%
  \end{align}%
  describes itinerant $c$ electrons and
  immobile $f$ electrons on a lattice, interacting via a repulsive
  local interaction $U$ \cite{Falicov1969a,Brandt1989a,Freericks2003a}. 

\paragraph{Time-dependent electric fields}
\label{peierls}
The first term in Eq.~(\ref{h-general}), with arbitrary hoppings $v_{i\alpha j\alpha'}(t)$, can include time-dependent electromagnetic 
fields (Sec.~\ref{subsec:electricfields}). 
For a single-band model, the Peierls substitution \cite{Peierls1933a,Luttinger1951a,Kohn1959} introduces the vector potential 
$\VecA(\Vecr,t)$ as a phase factor in the hopping matrix elements,
\begin{equation}
\label{peierls-t}
v_{ij}(t) = v_{ij}
\exp\!\left(
-\frac{ie}{\hbar}
\int_{\VecR_i}^{\VecR_j} \!d\Vecr\, \cdot \VecA(\Vecr,t)\right),
\end{equation}
and adds a scalar potential term $e\sum_{i\sigma} \Phi(\VecR_i,t) c_{i\sigma}^\dagger c_{i\sigma}$ to the Hamiltonian
($e$ is the charge of an electron, and we use units in which the speed 
of light is set to $c=1$).
The Peierls substitution can apparently not describe inter-band Zener tunneling in multi-band systems, because 
dipole matrix elements between bands of different symmetry are neglected \cite{Foreman2002}. We will not discuss inter-band tunneling in 
this review. On the technical level, however, the DMFT formalism will not be modified when those dipole matrix elements 
are incorporated into the single-particle part of the general model (Eq.~(\ref{h-general})).  

The Peierls substitution is derived from the requirement that the Hamiltonian is invariant under the gauge transformation
\begin{subequations}
\label{gauge trafo}
\begin{align}
c_{j\sigma} & \to c_{j\sigma} \exp\Big({\frac{ie}{\hbar} \,\chi({\bm R}_j,t)}\Big),
\\
{\bm A}({\bm r},t)  &\to {\bm A}({\bm r},t) + {\bm \nabla \chi({\bm r},t)},
\\
\Phi({\bm r},t)  &\to \Phi({\bm r},t) -\frac{\partial \chi({\bm r},t)}{\partial t}.
\end{align}
\end{subequations}
The (gauge-invariant) current operator can be obtained from the derivative ${\bm j}({\bm r})$ $=$ $-\delta H / 
{\delta A}({\bm r})$ \cite{Scalapino1992a}, 
such that it satisfies the continuity equation 
for the charge density $\rho(\bm r) = e\sum_{i\sigma}\delta(\bm r-\bm R_i) c_{i\sigma}^{\dagger}c_{i\sigma}^{\phantom{\dagger}}$.
Usually one is concerned with situations in which the applied field varies only slowly on the atomic scale, 
which is the case
even for optical frequencies. When the $\bm r$-dependence of $\bm A$ is neglected, the Peierls substitution 
leads to a time-dependent dispersion
\begin{equation}
\epsilon_{\kk}(t) = \epsilon\Big(\kk-\frac{e}{\hbar}{\bm A(t)}\Big),
\end{equation}
where $\epsilon(\kk)$ is the dispersion for zero field, and $a$ is the lattice spacing, so that the hopping part of the 
Hamiltonian reads $H=\sum_{\kk\sigma} \epsilon_\kk(t)\,c_{\kk\sigma}^\dagger c_{\kk\sigma}$.
Correspondingly, the current operator in the limit of long (``optical'') wavelengths becomes 
\begin{align}
{\bm j}(t)
&=
\frac{e}{V}
\sum_{\kk\sigma}{\bm v}_{\kk}(t) n_{\kk\sigma},
\label{dmft current}
\end{align}%
where $V$ is the volume, and ${\bm v}_{\kk}$ is the  group velocity of the Bloch electrons,
\begin{align}%
\label{jvertex}
{\bm v}_{\kk}(t)
=
\frac{1}{\hbar}\partial _\kk \epsilon_{\kk}(t)
=
\frac{1}{\hbar}\partial _\kk \epsilon\Big(\kk-\frac{e}{\hbar}{\bm A}(t)\Big).
\end{align}%
Note that although the right-hand side of Eq.~\eqref{dmft current} involves gauge
dependent quantities, the current itself is of course gauge independent (see also
the later discussion in Sec.~\ref{dmft observables a}).

\paragraph{Dissipative systems}

To describe dissipation of energy to other degrees of freedom, the Hamiltonian (\ref{h-general}) may be coupled to some environment,
like in Eq.~(\ref{h with environment}). When the environment is traced out, one obtains an effective description of the system 
with an additional self-energy contribution $\Sigma_\text{bath}$, 
\begin{align}
\label{sigma diss def}
\Sigma = \Sigma_\text{bath}+\Sigma_\text{loc}.
\end{align}
Here $\Sigma_\text{loc}$ contains all diagrams due the local interaction $H_\text{loc}$. In the spirit of DMFT, it is local and will 
be evaluated by the solution of the impurity model, i.e., the functional relation (\ref{dmft skeleton}) is not modified by the dissipation. For a general 
Hamiltonian (\ref{h with environment}), this decoupling into bath and interaction self-energies is not exact, as it neglects vertex corrections. 
However, when discussing ``dissipation'' one is interested in a regime in which results become universal with respect to the type of 
dissipation. This is usually the regime of weak system-bath coupling. The decoupling (\ref{sigma diss def}) fails in the opposite
limit of strong system-bath coupling, where the dynamics is clearly no longer universal and the ``bath'' should rather be considered 
as an integral part of the system.

Two dissipation mechanisms have so far been considered within DMFT: 
(i) The free-fermion bath (B\"uttiker model, Sec.~\ref{buttiker}), in which
one additional reservoir of particles is coupled to each lattice site  \cite{Tsuji09,Amaricci2012,Aron12,Werner2012Kondo,Han2012}, 
and (ii) the coupling to a bath of harmonic 
oscillator modes \cite{Eckstein2012c,Eckstein2012d, Kemper2013}. For the B\"{u}ttiker model, the system-bath coupling is bilinear in the fermion 
creation and annihilation operators, such that  additional degrees of freedom can be integrated out exactly. Equation (\ref{sigma diss def}) 
is then exact with $\Sigma_\text{bath}$ given by Eq.~(\ref{sigma_diss}). In spite of its apparent simplicity and the absence of momentum 
scattering, the model can lead to a physically meaningful steady state \cite{Amaricci2012,Han2012}. 
For the bosonic bath, one considers an infinite 
set of phonons, coupled via the Holstein coupling (\ref{dmft holstein interaction}). In this case the decoupling in bath and interaction 
self-energies holds only at weak coupling. Consequently we take Eq.~(\ref{sigma diss def}) with a first-order electron phonon diagram
\begin{subequations}
\label{sigma diss phonon}
\begin{align}
&\Sigma_\text{bath}[G]
=
g^2 G(t,t') D(t,t'),
\\
&D(t,t')
=
-i
\int \!d\omega \,e^{-i\omega(t-t')}\, \Gamma(\omega) [\theta_\CC(t,t')+ b(\omega)],
\end{align}
\end{subequations}
where $D$ is the propagator for free bosons with a density of states $\Gamma(\omega)$,
$b(\omega)=1/(e^{\beta\omega}-1)$ is the bosonic occupation function,
and $g$  measures the coupling strength. In general, the precise mechanism of dissipation should be irrelevant 
for the physics. The fermion bath has the conceptual advantage of treating the system-bath coupling exactly, while the 
phonon bath is particle-number conserving by construction.

In pump-probe spectroscopy experiments on solids, the lattice acts as a heat bath for the electrons, but in many situations a quasi-equilibrium 
description such as the two-temperature model \cite{Allen1987a} is not applicable. It is therefore important to develop a formalism which can treat the quantum mechanical
time-evolution of electron-phonon coupled systems, such as Eqs.~(\ref{dmft holstein interaction}) and (\ref{dmft hubbard holstein interaction}).
Lattice perturbation theory has been used in \cite{Sentef2013} to study the time evolution of a photoexcited electron-phonon system in the weak correlation regime.  
A DMFT formalism for the Holstein-Hubbard model, which captures the effect of the nonequilibrium state of the electrons on the evolution of the phonons, 
and the feedback of the phonons on the electronic relaxation, has recently been presented in \cite{Werner2013phonon}.

\subsubsection{Implementation of the self-consistency}
\label{subsubsection:self-consistency}

\paragraph{Stable time propagation scheme}
Typically, the self-consistent solution of the  DMFT equations is achieved by some kind of iterative procedure 
(Fig.~\ref{intro:dmft}): Starting 
from a guess for $\Delta$, one must, (i) compute $G$ from Eq.~(\ref{dmft_formalism::imp-g-2}), (ii) solve Eq.~(\ref{dmft_formalism::imp-dyson-2}) 
for $\Sigma$, (iii) solve Eq.~(\ref{dmft_formalism::latt-dyson-2}) for $G_{ii}$, (iv) solve Eq.~(\ref{dmft_formalism::imp-dyson-2}) with 
the new $G$ to get a new  $\Delta$, and iterate steps (i)-(iv) until convergence is reached. In nonequilibrium DMFT, this procedure 
can be implemented as a time-propagation scheme: If self-consistent solutions $G(t_1,t_2)$ and $\Delta(t_1,t_2)$ have been obtained for 
$t_1,t_2 \le t$, one can extrapolate $\Delta$ to the next timestep ($t_1= t+\Delta t$ or $t_2 = t+\Delta t$) and again converge steps (i)-(iv), 
thereby updating only the data at $t_1= t+\Delta t$ or $t_2 = t+\Delta t$. Matsubara functions are obtained from a separate 
equilibrium DMFT calculation, i.e., by iterating steps (i)-(iv) for Green's functions on the imaginary branch of $\CC$.

In this section we discuss the ``self-consistency'' part in the above procedure [steps (ii)-(iv)], while the ``impurity solver'' [step (i)]  
is deferred to Sec.~\ref{impurity solver}. In contrast to equilibrium DMFT, the self-consistency can be numerically costly, as one 
must manipulate contour-ordered Green's functions that depend on two time variables. For an arbitrary spatially inhomogeneous system 
with $L$ inequivalent lattice sites, the memory to store a Green's function $G_{ij}$ scales like $N^2 L^2$ (where $N$ 
is the number of time-discretization steps), which can reach a terabyte for $N\approx1000$ and $L\approx 100$. In this section 
we restrict the discussion to translationally invariant systems. The inhomogeneous problem can be handled for layered systems, 
as discussed in Sec.~\ref{sec real space}.

The problem of computing $\Sigma$ [step (ii) above] or $\Delta$ [step (iv)] from 
Eq.~(\ref{dmft_formalism::imp-dyson-2}) is not a
contour equation of the type (\ref{ky-cinv-v01}) or (\ref{ky-vie2}), whose numerical solution was discussed in Sec. \ref{KB-numerics}.
In contrast, it seems more closely related to the numerically less favorable Volterra equations of the first kind (\ref{ky-vie1}). One can 
solve these equations in time-discretized form on the full contour \cite{Freericks2006,Freericks2008}, but with a slight reformulation 
it is also possible to take full advantage of the Kadanoff Baym propagation scheme discussed in Sec. \ref{KB-numerics}. For this purpose we introduce the isolated impurity Green's function $g$, which is defined via the impurity Dyson equation for $\Delta=0$,
\begin{equation}
g^{-1}(t,t') = (i\partial_t + \mu)\delta_\CC(t,t') - \Sigma(t,t').
\end{equation}
(For simplicity, the following equations are first stated for the single-band case without spin dependence.) In order to compute 
$g$ one can reformulate Eq.~(\ref{dmft_formalism::imp-dyson-2}) in an integral form, 
\begin{subequations}
\label{imp dyson 001}
\begin{align}
\label{imp dyson 002}
G 
&=  g + g\convz\Delta\convz G
\\
\label{imp dyson 002 cc}
&= g +  G\convz\Delta\convz g,
\end{align}
\end{subequations}
which leads to a Kadanoff-Baym equation of the type (\ref{ky-vie2}),
\begin{align}
\label{imp get g}
[1+F] \convz g =Q,\,\,\,
%F =  -G\convz\Delta,\,\,\,
F =  G\convz\Delta,\,\,\,
Q=G.
\end{align}
[If the impurity solver gives $\Sigma$, as is the case for the weak-coupling solver (Sec.~\ref{weak-coupling perturbation}) and 
CTQMC (Sec.~\ref{ctqmc}), solving for $g$ is a standard Kadanoff-Baym equation (\ref{ky-cinv-v01}).]

Next one computes momentum-resolved Green's functions $G_\kk$ from the Dyson equation (\ref{dmft_formalism::latt-dyson-2})
in the momentum representation,
\begin{equation}
\label{latt dyson kk}
G_\kk^{-1}= (i\partial_t+\mu)\delta_\CC(t,t') - \Sigma(t,t') - \epsilon_\kk(t,t')  \equiv g^{-1} - \epsilon_\kk,
\end{equation}
with $\epsilon_\kk(t,t') = \delta_\CC(t,t') \epsilon_\kk(t)$. I reads in the integral form,
\begin{equation}
\label{kk dyson 01}
G_\kk =  g + g\convz \epsilon_\kk \convz G_\kk.
\end{equation}
Hence, $G_\kk$ can again be obtained by solving Kadanoff-Baym equations of the type (\ref{ky-vie2}),
\begin{align}
\label{get gk}
[1+F_\kk] \convz G_\kk =Q_\kk,\,\,
F_\kk =  -g\convz \epsilon_\kk,\,\,
Q_\kk=g.
\end{align}
To compute the updated $\Delta$ we start by summing Eq.~(\ref{kk dyson 01}) over $\kk$,
\begin{equation}
G =  g + g\convz \sum_\kk (\epsilon_\kk \convz G_\kk ),
\end{equation}
where we used the normalization $\sum_\kk=1$, and the relation $\sum_\kk G_\kk = G$.
Comparison with Eq.~(\ref{imp dyson 002}) gives
\begin{subequations}
\label{compute delta}
\begin{equation}
\label{summed dyson 1}
 \Delta \convz G= \sum_\kk  \epsilon_\kk \convz G_\kk \equiv G_1.
\end{equation}
Solving this integral equation for $\Delta$ would still be an integral equation of the less stable type (\ref{ky-vie1}). 
However, after inserting the conjugate of Eq.~(\ref{kk dyson 01}) and Eq.~(\ref{imp dyson 002 cc}) into the 
r.h.s. and l.h.s. of Eq.~(\ref{summed dyson 1}), respectively, one finds
\begin{equation}
\label{summed dyson 2}
 \Delta + \Delta \convz G \convz \Delta = \sum_\kk \big( \epsilon_\kk + \epsilon_\kk \convz G_\kk \convz \epsilon_\kk\big)
 \equiv G_2,
\end{equation}
such that also $\Delta$ can be obtained from a Kadanoff-Baym equation (\ref{ky-vie2}),
\begin{align}
\label{delta from g1 g2}
&[1+ G_1]  
\convz \Delta = G_2.
\end{align}
\end{subequations}
In an implementation of nonequilibrium DMFT that uses this scheme, steps (ii), (iii) above are replaced by the solution of Eq.~(\ref{imp get g})
and (\ref{get gk}), respectively. Instead of step (iv), one computes the $\kk$ sums (\ref{summed dyson 1}) and (\ref{summed dyson 2}) and 
solves Eq.~(\ref{delta from g1 g2}) for a new $\Delta$ \cite{Eckstein11}.

If one adopts an impurity solver based on the weak-coupling expansion 
(Sec.~\ref{weak-coupling perturbation}), one can skip step (ii),
as $\Sigma$ is directly given by a functional of the Weiss Green's function $\mathcal{G}_0$, which is defined by
\begin{align}
\mathcal{G}_0^{-1}(t,t')=(i\partial_t+\mu)\delta_{\CC}(t,t')-\Delta(t,t').
\end{align}
In this case, one can take $\mathcal{G}_0$ to represent the dynamical mean field, and $\Delta$ does not explicitly appear
in the self-consistency calculation.
In step (iii), one solves the Kadanoff-Baym Eq.~(\ref{dmft_formalism::latt-dyson-2}) for $G$ with the given $\Sigma$.
A new $\mathcal{G}_0$ is derived from the impurity Dyson equation of the form
\begin{align}
[1+F]\ast \mathcal{G}_0=Q,\,\,
F=G\ast\Sigma,\,\,
Q=G.
\end{align}

For dissipative systems, it is convenient to include the dissipative self-energy (\ref{sigma diss def}) into the definition of $g$, 
\begin{equation}
g^{-1}(t,t') = (i\partial_t + \mu)\delta_\CC(t,t') - \Sigma(t,t')-\Sigma_\text{bath}(t,t'),
\end{equation}
and define a corresponding ``lattice hybridization function'' $\Delta_\text{lat} =  \Delta-\Sigma_\text{bath}$. 
Then Eqs.~(\ref{imp dyson 001}) - (\ref{compute delta}) hold with the replacement $\Delta \to \Delta_\text{lat}$,
and the only additional step in the DMFT self-consistency is to compute the hybridization function $\Delta$ from 
$\Delta=\Delta_\text{lat}x+\Sigma_\text{bath}$ \cite{Eckstein2012c}.   

\paragraph{Momentum summations}
The momentum summations appearing the DMFT self-consistency [e.g., Eqs.~(\ref{summed dyson 1}) and (\ref{summed dyson 2})] can be simplified in special situations. First, without external electromagnetic fields $\epsilon_\kk$ is time independent, such that 
$\kk$-dependent quantities depend on $\kk$ only via $\epsilon_\kk$, and momentum sums can be reduced to integrals 
over a one-dimensional density of states,
\begin{equation}
\sum_\kk g(\epsilon_\kk)= \int \!d\epsilon\,g(\epsilon) \sum_\kk \delta(\epsilon-\epsilon_\kk) \equiv 
\int \!d\epsilon\, D(\epsilon) g(\epsilon).
\end{equation}
In particular, for a semielliptic density of states (\ref{eq:semielliptic}) one can collapse the whole self-consistency 
into a single equation $\Delta(t,t')=v_\ast^2G(t,t')$ \cite{epjst2010a},
like for equilibrium [Eq.~(\ref{bethe selfconsistency})]. If the
hopping is such that the (positive)
bandwidth ($=$ $4v_*$) depends on time, this equation generalizes to
\cite{Eckstein2010ramps}
\begin{align}
  \Delta(t,t')=v_*(t)G(t,t')v_*(t').
  \label{eq:selfcons-timedep}
\end{align}

The situation is more involved in the presence of electromagnetic fields, where one might have to do the $\kk$-sum 
explicitly. A simplification is possible for a hypercubic lattice with bare dispersion 
$\epsilon_\kk=-2v_*/\sqrt{2d}\sum_{\alpha} \cos(k_\alpha)$
and gaussian density of states
\begin{align}
D(\epsilon)=\frac{1}{\sqrt{2\pi}v_\ast} e^{-\epsilon^2/2v_\ast^2},
\end{align}
when a homogeneous field 
${\bm A}=A(t)(1,1,1,...)$ points along the body diagonal of the unit cell  \cite{Turkowski2005a}.
One has
\begin{align}
\epsilon(\kk -{\bm A}(t)) 
&= -2v \sum_{\alpha} \cos(k_\alpha - A(t))
\nonumber
\\
&=   \cos(A(t)) \epsilon_\kk + \sin(A(t)) \bar \epsilon_\kk,
\end{align}
with $\bar \epsilon_\kk=-2v\sum_{\alpha} \sin(k_\alpha)$ (taking $a=e=\hbar=1$). Momentum summations 
then reduce to integrals over a two-dimensional joint density of states $D(\epsilon,\bar\epsilon)=\sum_\kk \delta(\epsilon-\epsilon_\kk)
\delta(\bar\epsilon-\bar\epsilon_\kk)$,
\begin{equation}
\sum_\kk g(\epsilon(\kk-\bm A)) = 
\int \!d\epsilon d\bar\epsilon\, D(\epsilon,\bar \epsilon) \,g(\epsilon\cos(A) + \bar \epsilon\sin(A)).
\end{equation}
For the hypercubic lattice, one has $D(\epsilon,\bar\epsilon)=D(\epsilon)D(\bar\epsilon)$ \cite{Turkowski2005a}.
For other examples of an infinite-dimensional lattice structure, see \onlinecite{Tsuji08}.

\subsubsection{Observables and conservation laws}
\label{dmft observables}

\paragraph{      Equal-time observables}
\label{dmft observables a}
Equal-time observables (\ref{expectation value}) of the lattice model can directly be computed from the lattice 
Green's functions. Before we define the observables, we make
an important remark that in the presence of external electromagnetic fields
the Green's functions are not a priori gauge invariant. Under the gauge transformation (\ref{gauge trafo}),
the Green's function $G_{ij}(t,t')=-i\langle\TC c_i(t)c_j^\dagger(t')\rangle$ transforms as
\begin{align}
G_{ij}(t,t')\to G_{ij}(t,t')\exp\left(\frac{ie}{\hbar}[\chi(\bm R_i,t)-\chi(\bm R_j,t')]\right).
\label{gauge transformation of G}
\end{align}
Since physical observables should not depend on the choice of the gauge, the gauge dependent Green's function cannot 
be generally used in the present form. A widely adopted prescription is to put an additional phase factor (``string'')
to the Green's function \cite{Boulware1966,Davies1988,Bertoncini1991},
\begin{align}
\label{davies G}
\tilde{G}_{ij}(t,t')=\exp\left(\frac{-i e}{\hbar}\int_{(\bm R_j,t')}^{(\bm R_i,t)}
\left[d\bar{\bm r}\cdot \bm A(\bar{\bm r},\bar{t})-d\bar{t} \Phi(\bar{\bm r},\bar{t})\right]\right)G_{ij}(t,t').
\end{align}
The phase factor cancels the change of the phase of $G$ (\ref{gauge transformation of G}),
hence $\tilde{G}$ remains gauge invariant. However, $\tilde{G}$ depends on the path of the line integral
in the exponential. A standard convention is to take a straight line in the four dimensional spacetime
connecting $(\bm R_i,t)$ and $(\bm R_j,t')$ \cite{Boulware1966}.

For a uniform electric field which is often studied with the nonequilibrium DMFT (see Sec.~\ref{subsec:electricfields}),
one can take the temporal gauge [$\Phi=0$ and $\bm A=\bm A(t)$]. In this case (which is focused on in the following), 
the local Green's function is gauge invariant,
$G_{ii}(t,t')=\tilde G_{ii}(t,t')$. The equal-time Green's function, on the other hand, becomes gauge invariant 
if one shifts the momentum, $G_{\bm k+\bm A(t)}(t,t)=\tilde{G}_{\bm k}(t,t)$.

Using the gauge invariant Green's function, one can safely construct ``physical'' observables.
For example, the number of particles on site $i$ with spin $\sigma$ is
\begin{align}
 \label{density}
 n_{i,\sigma}(t) 
 &= \expval{ c_{i\sigma}^\dagger(t) c_{i\sigma}(t)}
= 
-i G_{ii,\sigma}^<(t,t)
=
-i \tilde{G}_{ii,\sigma}^<(t,t),
\end{align}
and the current (\ref{dmft current}) is
\begin{align}
{\bm j}(t)
&=
-\frac{ie}{V}
\sum_{\kk\sigma}{\bm v}_{\kk-\bm A(t)}
G_{\kk,\sigma}^<(t,t)
=
-\frac{ie}{V}
\sum_{\kk\sigma}{\bm v}_{\kk}
G_{\kk+\bm A(t),\sigma}^<(t,t)
\nonumber
\\
&=
-\frac{ie}{V}
\sum_{\kk\sigma}{\bm v}_{\kk}
\tilde{G}_{\kk,\sigma}^<(t,t),
\label{dmft current expval}
\end{align}%
both of which are explicitly gauge invariant.
The momentum occupation defined by $n(\kk,t)=-iG_{\kk,\sigma}^<(t,t)$ is apparently not gauge invariant.
Instead, one can take a co-moving wave vector $\tilde{\bm k}=\bm k+\bm A(t)$ \cite{Davies1988}, with which
\begin{equation}
\label{neps}
n(\tilde{\kk},t) = -i G_{\kk+{\bm A(t)},\sigma}^<(t,t)
= -i \tilde{G}_{\kk,\sigma}^<(t,t)
\end{equation}
becomes gauge invariant, and hence can be interpreted as a physically meaningful observable.

Energies are also calculated from the Green's functions.
The kinetic energy per spin and site is (denoting the number of sites by $L$)
\label{ekin}
\begin{equation}
E_\text{kin}(t)
= 
\frac{-i}{L} \sum_{jl} v_{jl}(t) G_{lj,\sigma}^<(t,t)
=
\frac{-i}{L} \sum_{j} [\Delta_j\convz G_{jj,\sigma}]^<(t,t),
\end{equation}
where the second equation follows from a comparison of 
Eqs.~(\ref{dmft_formalism::latt-dyson-2}) and (\ref{dmft_formalism::imp-dyson-2}).
The interaction energy can be computed from the self-energy, using equations of motion. 
In general, comparison of the equation of motion and the lattice Dyson equation gives
\begin{equation}
\expval{ c_{i\alpha}^\dagger [H_\text{loc},c_{i\alpha'}]} 
= i\sum_{\alpha''} [\Sigma_{i\alpha,i\alpha''} \convz G_{i\alpha'',i\alpha'}]^<(t,t).
\end{equation}
For the Hubbard interaction  (\ref{dmft_formalism::hint-hubbard}) and a homogeneous 
state, one has 
\begin{equation}
U(t)\langle n_{i\sigma}(t)[n_{i\bar\sigma}(t)-\tfrac{1}{2}]\rangle
     = -i [\Sigma\convz G_{ii,\sigma}]^<(t,t),
\end{equation}
which allows to compute the double occupancy $d(t)=\langle n_{i\uparrow}(t)n_{i\downarrow}(t) \rangle$.

Finally let us remark that DMFT (with an exact impurity solver) is a conserving approximation in the sense of Baym and Kadanoff 
\cite{Baym1961,Baym1962a}, because the self-energy is related to the Green's function as a functional 
derivative of a Luttinger-Ward functional. The latter can be used to prove, along the lines of  
\cite{Baym1962a}, particle number conservation and energy conservation,
\begin{equation}
\label{energy conservation with jj}
\frac{d}{dt} \expval{H(t)} = {\bm j(t)} \cdot {\bm E(t)},
\end{equation}
where $\bm E$ is the electric field.

\paragraph{Photoemission spectrum}
Time and angular-resolved photoemission spectroscopy (ARPES) is the most direct experimental tool 
to probe the time evolution of both the electronic spectrum and the occupation on ultrafast time scales. 
Static photoemission spectroscopy in equilibrium is often analyzed in terms of the momentum-resolved 
spectral function, 
\begin{equation}
\label{pes equilibrium}
I(\kk_f,E) =
\sum_\kk |M_\kk|^2 \delta_{\kk_\pp+\qq_\pp,\kk_{f\pp}} \,
N_{\kk}(E-\hbar\omega_q-W),
\end{equation}
where  $N_\kk(\omega)=f(\omega)A_\kk(\omega)$ is the occupied density of states at momentum $\kk$
[cf.~Eq.~\ref{occupied}], $\qq$ is the momentum of the incoming photon, and $I(\kk_f,E)$ is the photoemission 
intensity at final momentum $\kk_f$ and energy $E$, and $W$ is the work function.
 The delta function accounts for  momentum conservation 
parallel to the surface, and the $M_\kk$ denote matrix elements, which are often taken as $\kk$-independent
as a first approximation. The most important approximation entering this expression is the so-called sudden 
approximation \cite{Hedin2002a}, which neglects interactions between the outgoing electron and the bulk and thus 
allows to express the photoelectron current in terms of single-particle properties of the sample. 

In time-resolved ARPES one probes the state of a system with a short pulse with center-frequency $\Omega$, 
and counts the  total number of electrons emitted with a certain momentum $\kk_f$ and energy $E$. The electric 
field of the probe pulse is of the form $\cos[\Omega (t-t_p+\phi) ] S(t-t_p)$, where $S(t)$ is the probe envelope, 
$t_p$ is the probing time, and $\phi$ is the carrier-envelope phase. The system can be in an arbitrary nonequilibrium 
state due to an earlier pump excitation. Equation (\ref{pes equilibrium}) can directly be generalized to this situation 
\cite{FreericksKrishnamurthyPruschke2009}: In the sudden approximation, the electric probe field couples the 
electronic orbitals in the sample to the outgoing electron states $| \kk_f\rangle$ via some dipole matrix element. 
If we again disregard for a moment the $\kk$-dependence of this matrix element, straightforward second-order 
time-dependent perturbation theory gives (after averaging over the carrier-envelope phase $\phi$ )
\begin{align}
\label{arpes-2d}
&I({\kk}_f,E;t_p) 
\propto 
\sum_\kk \delta_{\kk_\pp+\qq_\pp,\kk_{f\pp}} 
I_\kk( E-\hbar\omega_q-W;t_p),
  \\
  \label{arpes-simple}
  &I_\kk(\omega;t_p)
  = -i
  \int \!\!dtdt'\,S(t)S(t')
  e^{i\omega(t'-t)}
 \tilde G^<_\kk(t+t_p,t'+t_p),
\end{align}
where $\tilde G_\kk$ is the Fourier transform of the gauge invariant Green's function 
(\ref{davies G}).
This expression provides a convenient starting point to analyze time-resolved ARPES in terms of the contour Green's 
functions determined in DMFT. (In contrast, the imaginary part of a partial Fourier transform 
$G^<(\omega,t)=\int d\bar{t} e^{i\omega\bar{t}}G^<(t+\bar{t}/2,t-\bar{t}/2)$ is not always positive.) 
The equation contains the fundamental frequency-time uncertainty \cite{Eckstein2008c}: 
When the probe pulse is very short, $S(t)=\delta(t)$, one measures instantaneous occupations, 
$I_\kk(\omega;t_p)=n_\kk(t_p)$
but all energy resolution is lost. In the limiting case of a stationary state, Eq.~(\ref{arpes-simple}) reduces to a 
convolution of the equilibrium result (\ref{pes equilibrium}) with the spectral density $|\tilde S(\omega)|^2$ of 
the probe pulse. For a quasi-stationary state, when $G(t,t')$ can be approximated as translationally invariant in 
time during the probe-pulse, $I_\kk(\omega;t_p)$ is given by a corresponding convolution of the partial Fourier 
transform $G^<(\omega,t_p)$. Applications of this formula are discussed in Sec.~(\ref{subsubsec:photodoping}).

\paragraph{Optical conductivity}

Another very powerful tool to probe the time evolution of strongly correlated systems is the electromagnetic 
response, ranging from terahertz, which is suitable for the analysis of carrier mobilities or phonons, to optical frequencies, 
which can be used to study, e.g., inter-band or charge transfer excitations. In this frequency range, the electromagnetic
response is described by the conductivity in the limit $\qq\to0$, i.e., the response of a translationally invariant 
current ${\bm j}$ to a translationally invariant electric probe field $\delta {\bm E}$,
\begin{align}
\label{sigma}
\sigma_{\alpha\beta}(t,t')=\frac{\delta \langle j_\alpha(t) \rangle}{\delta E_\beta(t')},
\end{align}
[$\delta j_\alpha(t) =\int_{-\infty}^t d\bar t\, \sigma_{\alpha\beta}(t,\bar t)\,\delta E_\beta(\bar t)$ in integral notation.]
This response function of the nonequilibrium state can be computed from the contour-ordered current-current 
correlation function (Kubo relation), which can be evaluated in nonequilibrium DMFT \cite{Eckstein2008b,
Tsuji09} in analogy to equilibrium \cite{Pruschke1993a}. For this purpose it is convenient to define the susceptibility
\begin{align}
    \label{chi}
\chi_{\alpha\beta}(t,t')=\frac{\delta \langle j_\alpha(t) \rangle}{\delta A_\beta(t')},
\end{align}
which is related to the optical conductivity $\sigma_{\alpha\beta}(t,t')$ via 
\begin{align}
    \label{chi2sigma}
    \sigma_{\alpha\beta}(t,t')
    &=
    -\int_{t'}^t d\tb \,\chi_{\alpha\beta}(t,\tb) \quad (t\ge t'),
  \end{align}
within the chosen gauge ${\bm E}(t)$ $=$ $-\partial_t {\bm A}(t)$.
The susceptibility (\ref{chi}) can be obtained by taking the derivative  of the expression (\ref{dmft current expval}), 
where the vector potential enters both in the vertex ${\bm v}_\kk(t)$ (leading to the diamagnetic contribution to $\chi$) and in 
the Green's function $G_{\kk\sigma}^<(t,t)$ (leading to the paramagnetic contribution)
\cite{Eckstein2008b},
  \begin{subequations}%
    \begin{align}%
      \chi_{\alpha\beta}(t,t')
      &=
      \chi^{{\text{dia}}}_{\alpha\beta}(t,t')
      +
      \chi^{{\text{pm}}}_{\alpha\beta}(t,t')
      \,,
      \\
      \chi^{{\text{dia}}}_{\alpha\beta}(t,t')
      &=
      -\frac{ie}{V}
      \sum_{\kk\sigma}\frac{\delta v^\alpha_{\kk}(t)}{\delta A_\beta(t')}\,
      G_{\kk\sigma}^<(t,t)
      \,,\label{chi-dia1}
      \\
      \chi^{{\text{pm}}}_{\alpha\beta}(t,t')
      &=
      -\frac{ie}{V}
      \sum_{\kk\sigma} v^\alpha_{\kk}(t)\,
      \frac{\delta G_{\kk\sigma}^<(t,t)}{\delta A_\beta(t')}
      \,.\label{chi-para1}
    \end{align}%
  \end{subequations}%
  The paramagnetic contribution can be found from a variation of the lattice Dyson equation (\ref{latt dyson kk})
  \begin{align}
  \label{deltagk}
  \delta G_{\kk\sigma} =
  -
  G_{\kk\sigma} \convz [-{\delta \epsilon_\kk} - \delta\Sigma_{\sigma}]
  \convz G_{\kk\sigma},
  \end{align}
where the term in brackets is $\delta G_\kk^{-1}$.
In equilibrium DMFT, the contribution from $\delta\Sigma$,  which is related to the vertex $\delta \Sigma/\delta G$ 
of the impurity model, vanishes in Eq.~(\ref{chi-para1}) because (i) $\Sigma$ is $\kk$-independent, and (ii)
$G_{\kk\sigma}G_{\kk\sigma}$ and ${\bm v}_{\kk\sigma}$ are even and odd with respect to $\kk$, respectively
\cite{Khurana1990}.  This symmetry argument leads to the drastic simplification that the conductivity can be 
expressed in terms of the product of two Green's functions (bubble diagram), without local vertex corrections. 
In nonequilibrium, conditions (ii) can be violated in several ways: The vertex (\ref{jvertex}) is no longer antisymmetric 
when an electric field is present in addition to the probe field, e.g., when the system is constantly driven by an ac field
\cite{Tsuji09}, or when the pump and probe fields overlap in time. Furthermore, the isotropy $G_{\kk\sigma}
=G_{-\kk\sigma}$ is lost whenever the system is driven out of equilibrium by a polarized pump pulse 
(the system is clearly not isotropic as long as a current is flowing). However, in this case the isotropy of $G_\kk$ is often
restored on a much faster time scale than other interesting relaxation processes, such that one can still neglect the vertex 
$\delta \Sigma$ for most of the time. Otherwise, one would have to compute a Bethe-Salpeter equation for the vertex on 
the Keldysh contour, which has been done for the ac-field-driven Falicov-Kimball model \cite{Tsuji09}. In the most 
general case, solving the Bethe-Salpeter equation for a vertex that depends on four different time arguments is a formidable 
numerical task, so that it might be easier to compute the induced current $j$ for small probe fields directly from 
Eq.~(\ref{dmft current expval}), and take the numerical derivative.

Neglecting the vertex, one obtains the following expressions \cite{Eckstein2008b}
for the susceptibility
  \begin{subequations}%
    \label{chifinal}%
    \begin{align}%
      \label{chi-para}%
      \chi^{{\text{pm}}}_{\alpha\beta}(t,t') &= 
      -2\chi_0
      \sum_{\kk\sigma}
      v_{\kk}^\alpha(t)
      v_{\kk}^\beta(t')
      \text{Im}[G^R_{\kk\sigma}(t,t')G^<_{\kk\sigma}(t',t)],
\\
      \label{chi-dia}%
      \chi^{{\text{dia}}}_{\alpha\beta}(t,t') &= 
      -\chi_0
      \delta(t-t')
      \sum_{\kk\sigma}
      \frac{\partial_{\Vk_\alpha}\partial_{\Vk_\beta}\epsilon_{\kk}(t)}{\hbar}
      {\rm Im} G^<_{\kk\sigma}(t,t),
    \end{align}%
  \end{subequations}%
with $\chi_0$ $=$ $e^2/(V\hbar)$. 
Together with (\ref{chi2sigma}), 
they constitute the final DMFT expressions for the optical conductivity. 
In equilibrium, these equations reduce to the familiar relation \cite{Pruschke1993a}
\begin{multline}
\label{sigma-eq}
\text{Re}\, \sigma_{\alpha\beta}(\omega) 
\stackrel{\omega>0}{=}
\pi \chi_0
\sum_{\kk\sigma}
v_{\kk}^\alpha
v_{\kk}^\beta
\\
\times
\int\limits_{-\infty}^{\infty} \!d\omega'\,
\frac{A_{\kk\sigma}(\omega')A_{\kk\sigma}(\omega+\omega')
[f(\omega')-f(\omega+\omega')]}{\omega}.
\end{multline}
Even out of equilibrium, one can generally show from analytic properties of 
$\sigma(\omega,t)=\int d\bar{t} e^{i\omega\bar{t}}\sigma(t+\bar{t}/2,t-\bar{t}/2)$ that 
 the (nonequilibrium) $f$-sum rule \cite{TsujiPhD,Shimizu2011},
\begin{equation}
\int_0^\infty \!d\omega \,
\text{Re}\, \sigma_{\alpha\beta}(\omega,t) 
=  \frac{\pi e^2}{2V}
      \sum_{\kk\sigma}
      \frac{\partial_{\Vk_\alpha}\!
      \partial_{\Vk_\beta}
      \epsilon_{\kk}(t)}{\hbar^2}
      n_{\kk\sigma}(t)
      \equiv \frac{\pi}{2}\varepsilon_0\omega_p^2(t)
\end{equation}
 holds, which gives the familiar result for the plasma frequency $\omega_p^2 = n e^2/\varepsilon_0 m^\ast$ of electrons
near the bottom of a parabolic band $\epsilon_{\bm k}=\hbar^2\bm k^2/2m^\ast$.

\subsection{Real-time impurity solvers}
%%%%%%%%%%%%%%%%%%%%%%%%%%%%%%%%%%%%%%%%%%%%%%%%%
\label{impurity solver}

\subsubsection{General remarks}
\label{sec-def}

In this section we discuss methods to solve the nonequilibrium impurity models which play a central role in the DMFT formalism (Sec.~\ref{Sec_DMFT}). A general impurity action defined on the L-shaped contour $\CC$ (Fig.~\ref{L-shaped contour}) has the form
\begin{subequations}
\begin{align}
\CS_\text{imp} 
&= 
\CS_\text{loc} + \CS_\text{hyb},
\\
\label{sloc}
\CS_\text{loc} 
&= 
-i\!
\intC \!\!dt\, H_\text{loc}[d_{p}^\dagger(t),d_{p}(t),t],
\\
\label{shyb}
\CS_\text{hyb}
&=
-i
\!
\intC \!\!dt_1 dt_2\,   
\!\!
\sum_{p_1,p_2}\!
d_{p_1}^\dagger\!(t_1) \,\Delta_{p_1,p_2}(t_1,t_2)\, d_{p_2}\!(t_2). 
\end{align}
\label{impurity action}
\end{subequations}  
Here $d_{p}$ and $d_{p}^\dagger$ denote, respectively, annihilation and creation operators for an electron 
in the impurity level $p$ ($p$ labels spin and orbital degrees of freedom), 
and $H_\text{loc}$ is the local Hamiltonian of the impurity site, which can be 
interacting and time-dependent in general. The hybridization function 
$\Delta_{p_1,p_2}(t_1,t_2)$ gives the amplitude for the hopping of an 
electron from the $p_2$-orbital into the bath at time $t_2$, its 
propagation within the bath, and the hopping back into the impurity 
orbital  $p_1$ at time $t_1$. 
It is related to the Weiss Green's function ${\mathcal G}_{0,p_1,p_2}(t,t')$ in DMFT by
\begin{align}
{\mathcal G}_{0,p_1,p_2}^{-1}(t,t')=(i\partial_t+\mu)\delta_\CC(t,t')\delta_{p_1,p_2}-\Delta_{p_1,p_2}(t,t').
\label{bath_GF}
\end{align}

The action (\ref{impurity action}) can be 
derived from an impurity Hamiltonian with a time-dependent coupling between the impurity and the bath, 
\begin{subequations}
\begin{align}
H_\text{imp}(t) 
&= H_\text{loc}(t) + H_\text{bath}(t) + H_\text{hyb}(t),  \label{H_impurity} \\
H_\text{bath}(t) &=  \sum_\nu \epsilon_\nu(t) c_\nu^\dagger c_\nu, \\
H_\text{hyb}(t) &= \sum_{p,\nu} [V_{p,\nu}(t) \,d_p^\dagger c_\nu + {\rm h.c.}] ,
\end{align}
\label{impurity Hamiltonian}
\end{subequations}
by tracing out the bath degrees of freedom $c_\nu$. The hybridization function is given by
\begin{align}
\Delta_{p_1,p_2}(t,t')=\sum_{\nu} V_{p_1,\nu}(t)g_\nu(t,t')V_{p_2,\nu}(t')^\ast
\end{align}
with 
\begin{align}
g_\nu(t,t')
=i[f(\epsilon_\nu(0))-\theta_\CC(t,t')]e^{-i\int_{t'}^{t}d\bar{t} \epsilon_\nu(\bar{t})}
\end{align}
being the noninteracting bath Green's function without coupling to the impurity site.

By ``solving the impurity problem", we essentially mean computing the 
single-particle Green's function of the impurity model (\ref{impurity action}),  
\begin{equation}
  \label{impg}
    G_{p,p'}(t,t') = -i \expval{\TC\, d_{p}(t)d_{p'}^\dagger (t')}_{\CS_\text{imp} }.
\end{equation}
The nonequilibrium impurity solvers may be classified into two classes:
diagrammatic approaches and a Hamiltonian-based approaches.
The former treat the impurity action (\ref{impurity action}) using diagrammatic techniques
without direct reference to a given Hamiltonian formulation,
while the latter solve the time-dependent Hamiltonian (\ref{impurity Hamiltonian}) directly. 
In nonequilibrium DMFT calculations, the Hamiltonian approach requires one to reconstruct a given hybridization function 
$\Delta(t,t')$ by optimizing the (time-dependent) bath parameters $\epsilon_\nu(t)$ and $V_{p,\nu}(t)$, which is a nontrivial problem. 
The Hamiltonian approach has been used in combination with
the numerical renormalization group \cite{Joura08}, and very recently
with the exact-diagonalization-based method \cite{Arrigoni2013,Gramsch2013}.
In the case of a steady state, the effective Matsubara method \cite{Han07}
based on Hershfield's expression \cite{Hershfield93}
for a steady-state density matrix has been tested as an impurity solver
 \cite{Aron13}. 
In the following, we focus on the diagrammatic approaches, which have been used in various types of applications.

\subsubsection{Falicov-Kimball model}
\label{sec:FKequations}

The Falicov-Kimball (FK) model [Eq.~(\ref{eq:FK-model})] may be regarded as a 
simplified version of the Hubbard model, because only 
one electron species ($c$, $c^\dagger$)
can hop between lattice sites. 
The localized electrons ($f$, $f^\dagger$)
provide an annealed disorder 
potential, i.e., their equilibrium
distribution is not governed by a fixed probability
distribution (which would correspond to quenched disorder),
but it 
is determined by statistical mechanics,
which assumes that all 
$f$ configurations 
are in
principle accessible and can contribute to the partition function.
In nonequilibrium, however,  the FK model is special
because each $f_i^\mydagger f_i^\phdagger$ is a constant of motion and therefore
the $f$ electrons will maintain their initial distribution.
The FK model is a useful starting point for DMFT studies, because the 
effective local DMFT action for the $c$ particles becomes quadratic. Thus the 
$c$-electron Green's function can be calculated exactly \cite{Brandt1989a,vanDongen1990a,vanDongen1992a}. 
%In thermodynamic equilibrium this yields correlation-induced transitions 
In thermodynamic equilibrium, the model exhibits correlation-induced transitions 
between metallic, 
insulating, and charge-ordered phases~\cite{Freericks2003a}. 
 
In nonequilibrium DMFT, the single-site action 
and the $c$-electron Green's function
for the homogeneous phase are given
by~\cite{Turkowski2005a,Freericks2006}
  \begin{subequations}\label{eq:local-g}%
    \begin{align}%
      &\mathcal{S}
      = -i\int_\CC\!\!dt dt' \,
        c^\mydagger(t)\Delta(t,t')c(t')
        \!
        -i\!\!\int_\CC\!\!dt\, U(t) n_c(t)n_f(t),
\label{fkm action}
\\
      &G(t,t') 
      = -i\frac{ \TR_{c,f} [e^{-\beta H_0} \TC e^{\mathcal{S}}
        c(t)c^\mydagger(t')] }{ \TR_{c,f} [e^{-\beta H_0} \TC e^{\mathcal{S}}]}.
    \end{align}%
  \end{subequations}%
  Here $\CC$ is the L-shaped contour in Fig.~\ref{L-shaped contour},
  $n_c=c^\dagger c$,   $n_f=f^\dagger f$,
   and the  operators are in the interaction representation with respect to
  $H_0$ $=$ $(E_f-\mu)f^\mydagger f-\mu c^\mydagger c$.  The $f$
  electrons 
  can thus be traced out 
  and the $c$ Green's function is obtained as
  \begin{subequations}\label{eq:g-from-lambda}%
    \begin{align}%
      G(t,t')
      &=
      w_0 \Gwnull(t,t')
      +
      w_1 \Gweins(t,t')
      \,.\label{eq:g=b+d}
    \end{align}%
    Here $w_1$ $=$ $\langle f^\mydagger f\rangle$ $=$ $1-w_0$, and
    $\Gwnull(t,t')$ and $\Gweins(t,t')$ are defined 
    as $G(t,t')$ in Eq.~(\ref{eq:local-g}b) but without $\TR_{f}$ and with
    $f^\mydagger(\bar{t})f(\bar{t})$ replaced by 0 and 1,
    respectively,
    i.e., they are determined by the equations of motion 
    \begin{align}%
      {}[i\partial_{t}\! + \mu]
      \Gwnull(t,t')
      -
      (\Delta \myast \Gwnull)(t,t')
      &=
      \delta_\CC(t,t')
      \,,\label{eq:eqm-d}
      \\
      {}[i\partial_{t}\! + \mu - U(t)]
      \Gweins(t,t')
      -
      (\Delta \myast \Gweins)(t,t')
      &=
      \delta_\CC(t,t')\,,
      \label{eq:eqm-b}
    \end{align}%
  \end{subequations}%
  with an anti-periodic boundary condition (see Sec.~\ref{nonequilibrium green}).
  These equations must be solved together with the self-consistency condition (see Sec.~\ref{Sec_DMFT}).
  The solution for abrupt and slow interaction changes is discussed
  in Secs.~\ref{subsubsec:quenches} and~\ref{subsubsec:ramps}.

\subsubsection{Continuous-time quantum Monte Carlo algorithms}
\label{ctqmc}

\paragraph{General remarks}

A general strategy for evaluating expectation values such as Eq.~(\ref{impg}) is to write $H_\text{imp}$ as a sum of two terms: one, $H_1$, for which the time evolution can be computed exactly and another, $H_2$, which is treated by a formal perturbative expansion. The expansion in $H_2$ generates a series of diagrams which are sampled stochastically, using an importance sampling which accepts or rejects proposed diagrams on the basis of their contributions to the partition function of the initial state.  We discuss two types of expansions: In the ``weak coupling'' method, $H_\text{imp}$ is split into a quadratic part $H_0$ and an interacting part $H_\text{int}$, and the expansion is performed in terms of $H_\text{int}$. In the ``strong coupling'' approach, $H_\text{loc}$ and $H_\text{bath}$ are treated exactly, while $H_\text{hyb}$ is treated as a perturbation.  

These so-called continuous-time quantum Monte Carlo (CTQMC) algorithms provide very efficient and flexible solvers for equilibrium quantum impurity problems at temperature $T>0$ 
(for a review, see \onlinecite{Gull2011}).
In equilibrium, the expansion can be formulated on the imaginary time branch, and at least for single-site impurity problems with diagonal bath, the fermionic sign problem can be avoided. 
In the nonequilibrium extension of these methods \cite{Muehlbacher2008, Werner2009}, the expansion must also be performed on the real-time branches of the contour $\mathcal C$, where the convergence of the perturbation theory is oscillatory rather than monotonic. The weights of the diagrams become complex, and the resulting dynamical sign problem limits the time range over which accurate results can be obtained.

For nonequilibrium DMFT applications, the times which can be reached with the strong-coupling CTQMC approach are usually too short. We introduce this formalism mainly to set the stage for more useful approximate strong-coupling methods (see Sec.~\ref{strong-coupling perturbation}). The weak-coupling CTQMC method on the other hand, thanks to a simplification of the diagrammatic structure in half-filled single-band systems, has proven useful in situations where the interesting phenomena happen on a fast time scale and where a numerically exact treatment, e.g., of the intermediate correlation regime, is essential. This method has enabled some pioneering nonequilibrium DMFT studies of the Hubbard model \cite{Eckstein09, Tsuji11}.\\

\paragraph{Weak-coupling CTQMC}

We will first discuss the implementation of the weak-coupling continuous-time Monte Carlo algorithm 
\cite{Rubtsov05, Gull08_ctaux, Werner2009, Werner2010} for the Anderson impurity model
\begin{equation}
H_\text{loc}(t)=U(t)(n_\uparrow-\alpha_\uparrow)(n_\downarrow-\alpha_\downarrow)-\sum_\sigma[\mu-U(t)\alpha_{\bar\sigma}]n_\sigma,
\label{H_loc_weakcoupling}
\end{equation}
where a spin-dependent chemical potential shift [$U(t)\alpha_{\bar\sigma}$] has been introduced to avoid a trivial sign problem in the repulsively interacting case \cite{Rubtsov05} with ${\bar\sigma}=\downarrow,\uparrow$ for $\sigma=\uparrow,\downarrow$. The impurity action (\ref{impurity action}) is split into an interaction term, 
\begin{equation}
\CS_\text{int}=-i\int_\CC dt H_\text{int}(t)=-i\int_\CC dt U(t)(n_\uparrow-\alpha_\uparrow)(n_\downarrow-\alpha_\downarrow),
\label{S_int_weakcoupling}
\end{equation}
and the rest, $\CS_\text{0}=\CS_\text{imp}-\CS_\text{int}$, which is quadratic in the fermionic operators. Expanding the partition function $Z=\Tr[\TC e^{\CS_{\rm imp}}]=\Tr[\TC e^{\CS_0+\CS_\text{int}}]$ in powers of $\CS_\text{int}$ leads to an expression 
\begin{align}
&Z=\sum_{n=0}^\infty \frac{(-i)^n}{n!} \int_\CC{dt_1}\ldots \int_\CC{dt_n}\TR\Bigg[\TC e^{\CS_0} H_\text{int}(t_1)\ldots H_\text{int}(t_n)\Bigg].
\end{align}
After separating the trace into spin-up and spin-down factors and using the definition (\ref{cntr-expval}), we obtain
\begin{align}
&\frac{Z}{Z_0}=\sum_{n=0}^\infty \frac{(-i)^n}{n!} \int_\CC{dt_1}\ldots \int_\CC{dt_n}U(t_1)\ldots U(t_n)\nonumber\\
&\hspace{1.5cm}\times \prod_\sigma \big\langle (n_\sigma(t_1)-\alpha_{\sigma}) \ldots (n_\sigma(t_n)-\alpha_{\sigma}) \big\rangle_{\CS_0},
\label{Z_weak}
\end{align}
with $Z_0=\TR[\TC e^{\CS_0}]$. Wick's theorem allows to express the expectation values in (\ref{Z_weak}) as the determinant of an $n\times n$ matrix $M_\sigma^{-1}$, with elements
\begin{align}
& (M_\sigma^{-1})_{i,j}=-i\mathcal{G}_{0,\sigma}(t_i,t_j)-\alpha_{\sigma}\delta_{i,j},\\
& \mathcal{G}_{0,\sigma}(t,t')=-i\big\langle \TC d_\sigma(t)d_\sigma^\dagger(t')\big\rangle_{\CS_0},\label{G_bath}
\end{align}
and the convention $\mathcal{G}_{0,\sigma}(t,t)\equiv \mathcal{G}_{0,\sigma}^<(t,t)$ (note that these functions depend on the choice of $\alpha_\sigma$). Equation (\ref{Z_weak}) therefore expresses
$Z/Z_0=\sum_c w(c)$ 
as a sum over {\it configurations} $c=\{ t_1\prec t_2 \prec \cdots \prec t_n\}$, which are collections of time-points on the contour $\CC$, with a weight 
\begin{equation}
w(c)=(-i)^n (U(t_1)dt_1)\ldots (U(t_n)dt_n)\prod_\sigma \det M_\sigma^{-1}.
\label{w_weak}
\end{equation}
Here, $dt_i=dt$ for $t_i$ on the forward branch $\CC_1$, $-dt$ for $t_i$ on the backward branch $\CC_2$, and $-id\tau$ for $t_i$ on the imaginary-time branch $\CC_3$ (Fig.~\ref{L-shaped contour}). 

\begin{figure}[t]
\begin{center}
\includegraphics[angle=0, width=0.8\columnwidth]{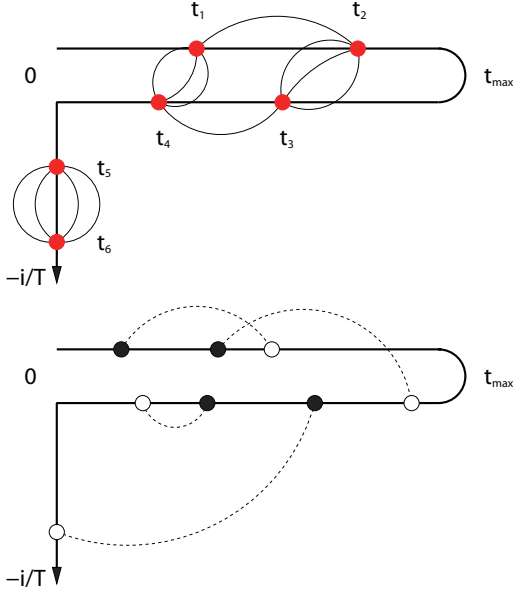}
\caption{Illustration of the CTQMC method. 
Top panel:
Example of a sixth order diagram appearing in the weak-coupling expansion. Interaction vertices (red) are linked by bath Green's functions $\mathcal{G}_0$ (black lines). %Right panel: 
Bottom panel: 
Example of a fourth order diagram appearing in the-strong coupling expansion: full (empty) dots represent impurity creation (annihilation) operators. Pairs of creation and annihilation operators are linked by hybridization functions $\Delta$ (dashed lines). In both methods, all the diagrams obtained by linking a given collection of operators by Green's functions or hybridization lines are summed up into a determinant.  
}
\label{contour_hyb}
\end{center}
\end{figure}

A Monte Carlo sampling over all configurations $c$ can then be implemented on the basis of these weights \cite{Werner2009}. The determinants in Eq.~(\ref{w_weak}) sum up all the $n!$ connected and disconnected bare diagrams for a given set of $n$ interaction vertices. An example of such a diagram is shown in the top panel of Fig.~\ref{contour_hyb}.
While this summation (for a proper choice of the $\alpha_\sigma$) absorbs the sign cancellations originating from Fermi statistics, the weights $w(c)$ are in general complex, so the sampling suffers from a ``phase problem" (dynamical sign problem). This phase problem grows exponentially with the number of interaction vertices on the real-time branches, and hence the length of the real-time contour (i.e., $\expval{\rm phase}\sim Ce^{-\alpha \tmax}$). 

In a practical implementation, one generates a Markov chain of configurations $c_1\to c_2\to c_3\to\cdots$ in such a way that every configuration can be visited from any other within a finite number of steps (ergodicity) and that the probability for the configuration to change from $c_i$ to $c_j$ [$p(c_i\to c_j)$] satisfies the detailed balance condition,
\begin{align}
p(c_i\to c_j)|w(c_i)|=p(c_j\to c_i)|w(c_j)|.
\label{detailed balance}
\end{align}
This guarantees that the configuration $c$ is realized with a probability $\propto |w(c)|$. To sample the configurations, one usually adopts the Metropolis-Hastings algorithm \cite{Metropolis1953,Hastings1970}. 
One proposes to insert an $n$th interaction vertex at $t_n\in \CC$ with probability $p^{\rm prop}(n-1\to n)=|dt_n|/(2\tmax+\beta)$, or to remove the $n$th vertex with probability $p^{\rm prop}(n\to n-1)=1/n$. 
Then one accepts the propoed insertion (removal) with the probability $\min[1,R^{(n)}]$ ($\min[1,R^{(n)-1}]$), where
\begin{align}
R^{(n)}&=\frac{p^{\rm acc}(n-1\to n)}{p^{\rm acc}(n\to n-1)}=
\frac{p^{\rm prop}(n\to n-1)|w(n)|}{p^{\rm prop}(n-1\to n)|w(n-1)|}
\nonumber
\\
&=
|U(t_n)|\frac{2t_{\rm max}+\beta}{n}\prod_\sigma\frac{|\det (M_\sigma^{(n)})^{-1}|}{|\det (M_\sigma^{(n-1)})^{-1}|}.
\label{acc prob}
\end{align}
The ratio of two determinants in Eq.~(\ref{acc prob}) can be obtained by a fast-update algorithm with a computational cost of $O(n^2)$ \cite{Gull2011}.

To measure an observable by the Monte Carlo sampling, $\langle O(t)\rangle_{MC}$, 
we perform a similar expansion in the presence of the operator $O$ at time $t$. The measurement formula becomes
\begin{align}
\langle O(t)\rangle_{MC}
&=
\frac{\sum_c w_c \frac{O_c}{w_c}}{\sum_c w_c}
=
\frac{\sum_c |w_c| {\rm phase}_c \frac{O_c}{w_c}}{ \sum_c |w_c|{\rm phase}_c},
\end{align}
where 
$O_c$ is the ``weight" corresponding to the collection $c$ of operators in the presence of $O(t)$,
and ${\rm phase}_c = w_c/|w_c|$.
In the case of the Green's function (\ref{impg}), the ratio $O_c/w_c$ (or rather the analogue for two-time operators) becomes
$\mathcal{G}_{0,\sigma}(t,t')+ i \sum_{i,j=1}^n \mathcal{G}_{0,\sigma}(t,t_i)(M_\sigma)_{i,j}\mathcal{G}_{0,\sigma}(t_j,t')$, and the measurement formula reads
\begin{align}
G_\sigma(t,t')&=
\mathcal{G}_{0,\sigma}(t,t')+\int_\mathcal{C} ds_1 \int_\mathcal{C} ds_2 
\mathcal{G}_{0,\sigma}(t,s_1)\mathcal{G}_{0,\sigma}(s_2,t')\nonumber
\\
&\quad\times 
\Bigg \langle i \sum_{i,j=1}^n \delta_\mathcal{C}(s_1,t_i)(M_\sigma)_{i,j}\delta_\mathcal{C}(s_2,t_j)\Bigg\rangle_{MC}.
\label{X}
\end{align}
It is therefore sufficient to accumulate the quantity (improper self-energy) \cite{Gull08_ctaux, Werner2010}
\begin{equation}
X_\sigma(s_1, s_2)=\Bigg\langle i \sum_{i,j=1}^n \delta_\mathcal{C}(s_1,t_i)
(M_\sigma)_{i,j}\delta_\mathcal{C}(s_2,t_j)\Bigg\rangle_{MC}.
\label{Xonly}
\end{equation}
Comparison of Eq.~(\ref{X}) to the Dyson equation shows that $X$ is
related to the (proper) self-energy $\Sigma$ by
$X \convz \mathcal{G}_0=\Sigma \convz G$,
so the measurement of $X$ allows to extract $\Sigma$. 
This equation 
is a Volterra integral equation of the first kind (\ref{ky-vie1}), for which numerical solutions are known to be unstable. 
We may however combine it with 
the Dyson equation $G_{\sigma}=\mathcal{G}_{0,\sigma}+G_{\sigma} \convz \Sigma_{\sigma} \convz \mathcal{G}_{0,\sigma}$ to find 
\begin{equation}
\label{dmft-getsigma}
(1 + X_{\sigma} \convz \mathcal{G}_{0,\sigma}) \convz \Sigma_{\sigma}
=
X_{\sigma}.
\end{equation}
This equation is a Volterra equation of the second kind (\ref{ky-vie2}) with unknown 
$G=\Sigma$ and kernel $F=X_{\sigma} \convz \mathcal{G}_{0,\sigma}$,
which can be solved in the same way as the Dyson equation (Sec.~\ref{dmft_formalism::sec-noneq}). 

Let us briefly mention the relationship of the above algorithm to the continuous-time auxiliary field method \cite{Gull08_ctaux, Werner2009}. In practice, it is advantageous to symmetrize the $\alpha_\sigma$, which amounts to introducing an Ising spin degree of freedom at every vertex position. For example, up to irrelevant constants we may write the interaction term 
$Un_\uparrow n_\downarrow$   
in the form \cite{Assaad2007}
\begin{equation} 
\frac{U(t)}{2}\sum_{s=\pm 1}(n_\uparrow -\tfrac12 -s\delta)(n_\downarrow -\tfrac12+s\delta)+\frac{U(t)}{2}(n_\uparrow+n_\downarrow),
\label{int_spin}
\end{equation}
and absorb the last (quadratic term) into $\mathcal{G}_0$. The configuration space then becomes the space of all Ising spin configurations on the contour $\CC$: $c=\{ (t_{1}, s_1),(t_{2},s_2), \ldots (t_{n}, s_n) \}$. To avoid a fermionic sign problem in the repulsively interacting case (away from half-filling) one has to choose 
$\delta>1/2$.
At half-filling, one should use $\delta=0$, i.e., the symmetric form $H_\text{int}=U(t)(n_\uparrow-\tfrac12)(n_\downarrow-\tfrac12)$. In this case, all odd-order diagrams vanish, which leads to a less severe sign problem. 
The interaction expansion based on (\ref{int_spin}) is equivalent to the auxiliary field algorithms described in \cite{Gull08_ctaux, Werner2009}, as was shown in \cite{Mikelsons2009}.

\noindent
\paragraph{Strong-coupling CTQMC}
\label{strong coupling ctqmc}

A complementary diagrammatic Monte Carlo algorithm can be obtained by performing an expansion in powers of the hybridization term $H_\text{hyb}$ \cite{Muehlbacher2008, Werner2009, Schiro2009}. We will sketch here the derivation for the general impurity model defined in Eq.~(\ref{H_impurity}). 
In the hybridization expansion approach the time evolution of the operators is given by $H_\text{loc}+H_\text{bath}$, and the starting point is the identity
\begin{equation}
Z=\TR\big[\TC e^{-i\int_\CC ds (H_\text{loc}(s)+H_\text{bath}(s))-i\int_\CC ds H_\text{hyb}(s)}\big]. 
\end{equation}
Expanding the contour-ordered exponential into a power series yields
\begin{align}
&Z=\TR\Bigg[\TC e^{-i\int_\CC ds (H_\text{loc}(s)+H_\text{bath}(s))}\sum_{n=0}^\infty \frac{(-i)^n}{n!} \int_\CC{dt_1}\ldots \int_\CC{dt_n}\nonumber\\
&\hspace{4cm}\times H_\text{hyb}(t_1)\ldots H_\text{hyb}(t_n)\Bigg].
\label{strong coupling expansion hyb}
\end{align}
We may now proceed in exactly the same way as in equilibrium \cite{Werner06, Werner06Kondo}, namely separate the $H_\text{hyb}$ factors into  impurity creation terms (time arguments $t'_i$) and impurity annihilation terms (time arguments $t_j$). Then, because the bath is noninteracting, we can evaluate the trace $\TR_\text{bath}[\ldots]$ over the bath Hilbert space exactly. Wick's theorem yields a determinant $\det {N}^{-1}$, with the size of the matrix ${N}^{-1}$ given by the number of impurity creation (or annihilation) operators on the contour $\mathcal C$, and with matrix elements given by the hybridization functions $\Delta$:
\begin{equation}
({N}^{-1})_{i,j}=\Delta_{p'_i,p_j}(t'_i,t_j).
\label{M_inv}
\end{equation}
Here, $t'_i$ denotes the position of the $i$th creation operator and $t_j$ the position of the $j$th annihilation operator, with flavor $p_i$ and $p_j$, respectively. The weight of a Monte Carlo configuration with $n$ creation and $n$ annihilation operators on the contour then becomes
\begin{align}
&w(\{(t'_1, p'_1),\ldots,(t'_n, p'_n); (t_1,p_1)\ldots (t_n,p_n)\})=\nonumber\\
&\hspace{5mm}\TR_\text{loc}\Big[\TC e^{-i\int_\CC ds H_\text{loc}(s)} d^\dagger_{p'_1}(t'_1)d_{p_1}(t_1)\ldots d^\dagger_{p'_n}(t'_n)d_{p_n}(t_n) \Big]\nonumber\\
&\hspace{5mm}\times dt_1\ldots dt'_n \frac{(-1)^n}{(n!)^2}\text{det} {N}^{-1},
\label{weight_hyb}
\end{align}
where the $dt_i$ again contains factors $+1$, $-1$ or $-i$, depending on the branch corresponding to $t_i$. 
The trace over the impurity states, $\TR_\text{loc}[\dots]$, is calculated explicitly, for example by expressing the impurity creation and annihilation operators, as well as the time-evolution operators $U(t_2,t_1)=\TC e^{-i\int_{t_1}^{t_2}H_\text{loc}(s)ds}$ as matrices in the eigenbasis of the Hamiltonian $H_\text{loc}(0)$. This is particularly simple if the eigenbasis of $H_\text{loc}$ is the occupation number basis, and if the eigenbasis does not change in time (as is the case, for example, for the Anderson impurity model with a time-dependent interaction term). Monte Carlo configurations are then generated stochastically by insertion and removal of pairs of impurity creation and annihilation operators, based on the absolute value of the weight (\ref{weight_hyb}).

The interpretation of the determinant of the $n\times n$ hybridization matrix $N^{-1}$ is that it sums all the strong-coupling diagrams which can be obtained for a given collection of $n$ creation and $n$ annihilation operators, by connecting pairs of these operators by hybridization functions $\Delta$, see bottom panel of Fig.~\ref{contour_hyb}.

The Green's function can be measured by removing a single hybridization line, in complete analogy to the equilibrium case \cite{Werner06}. This leads to a measurement procedure which accumulates delta-functions with weights given by the matrix elements of $({N})_{j,i}$ at times $t'_i$ and $t_j$. 
Particularly simple is the measurement of local observables $O(t)$, such as the density or double occupancy. They can be measured by inserting the corresponding operator into the trace factor $\TR_\text{loc}[\dots]$.

\subsubsection{Weak-coupling perturbation theory}
\label{weak-coupling perturbation}

The phase problem of CTQMC techniques can be avoided if subclasses of diagrams are summed up analytically using Dyson's equation (\ref{dyson}). In this approach, the object of interest is the self-energy, whose diagrammatic expansion is truncated at some given order. We will discuss here the weak-coupling perturbation theory for the Anderson impurity model with the local Hamiltonian defined in Eq.~(\ref{H_loc_weakcoupling}). 

Perturbation theory for nonequilibrium impurity problems 
\cite{HershfieldDaviesWilkins1991,HershfieldDaviesWilkins1992,FujiiUeda2003} is a straightforward generalization
of the equilibrium perturbation theory formulated on the Matsubara branch 
\cite{AbrikosovGorkovDzyaloshinskiBook,MahanBook,YosidaYamada1970,Yamada1975,YosidaYamada1975}.
The weak-coupling perturbation theory has been used for a long time as an impurity solver in equilibrium DMFT calculations 
\cite{Georges1992a,ZhangRozenbergKotliar1993,Georges96,Freericks1994,FreericksJarrell1994}, and in the nonequilibrium DMFT context, 
it has enabled a range of studies of the Hubbard model 
\cite{Schmidt2002,Heary2009,Eckstein2010,Eckstein2011bloch,Tsuji2012,TsujiOkaAokiWerner2012,Tsuji2013,Aron11,Amaricci2012}
and of the Falicov-Kimball model \cite{TurkowskiFreericks2007b}.

We again split the impurity action into a quadratic term $\CS_0$ and an interaction term (\ref{S_int_weakcoupling}), and expand $\exp(\CS_{\rm imp})=\exp(\CS_0+\CS_\text{int})$ in Eq.~(\ref{impg}) into a Taylor series with respect to the interaction term,
\begin{align}
&G_\sigma(t,t')
=
(-i)\frac{1}{Z}\sum_{n=0}^\infty \frac{(-i)^n}{n!} \int_{\mathcal{C}} dt_1 \cdots dt_n 
\nonumber
\\
&\quad\times
\TR\left[ \TC e^{\CS_0} H_\text{int}(t_1)\cdots H_\text{int}(t_n) 
d_\sigma(t) d_\sigma^\dagger(t') \right].
\label{expanded Green's function}
\end{align}
The linked cluster theorem ensures that all the disconnected diagrams that contribute to Eq.~(\ref{expanded Green's function}) 
can be factorized to give a proportionality constant $Z/Z_0$
with $Z_0=\TR\left[\TC e^{\CS_0}\right]$.
As a result, the expansion of the Green's function may be written as
\begin{align}
G_\sigma(t,t')
&=
(-i)\sum_{n=0}^\infty (-i)^n \int_{\mathcal{C}, t_1\prec\cdots\prec t_n} dt_1 \cdots dt_n 
\nonumber
\\
&\quad\times
\langle \TC H_\text{int}(t_1)\cdots H_\text{int}(t_n) 
d_\sigma(t) d_\sigma^\dagger(t') \rangle_{\CS_0}^\text{conn.}.
\label{connected Green's function}
\end{align}
where `conn.' means that we only consider the connected diagrams. The factor $n!$ has been cancelled by specifying the contour ordering as $t_1\prec\cdots\prec t_n$. Owing to Wick's theorem, one can evaluate each term in Eq.~(\ref{connected Green's function})
using the bath Green's function (\ref{G_bath}).

\begin{figure}[t]
\includegraphics[width=5cm]{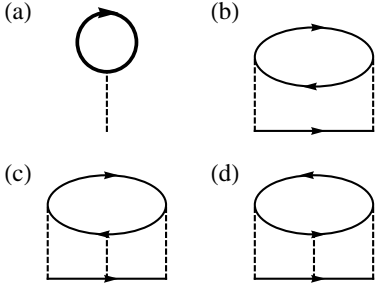}
\caption{The self-energy diagrams up to third order. The thin lines represent $\mathcal{G}_{0\sigma}(t,t')$, the bold line $G_\sigma(t,t')$,
and the dashed lines the interaction vertices.}
\label{third-order diagrams}
\end{figure}

In the weak-coupling perturbation theory, one usually considers an expansion of the self-energy $\Sigma_\sigma(t,t')$ instead of the Green's function. 
This allows to generate an infinite series of diagrams for the Green's function by solving the Dyson equation (\ref{dyson}).
The self-energy consists of the one-particle irreducible parts of the expansion (\ref{connected Green's function}). We show topologically distinct Feynman diagrams
of the self-energy up to third order in Fig.~\ref{third-order diagrams}. Due to the chemical potential shift 
$\mu\to\mu-U(t)\alpha_{\bar\sigma}$ in Eq.~(\ref{H_loc_weakcoupling}), a tadpole diagram with a bold  or bare line
amounts to $n_\sigma(t)-\alpha_{\bar \sigma}$ or $n_{0\sigma}(t)-\alpha_{\bar\sigma}$, where $n_\sigma(t)=-iG_\sigma^<(t,t)$ and $n_{0\sigma}(t)=-i\mathcal{G}_{0\sigma}^<(t,t)$. 

The Feynman rules to calculate the self-energy diagrams on the contour $\mathcal{C}$ are:
(1) Draw topologically distinct one-particle irreducible diagrams. 
(2) Associate each solid line [bold line] with the Weiss Green's function $(-i)\mathcal{G}_{0\sigma}(t,t')$ [the interacting Green's function $(-i)G_\sigma(t,t')$].
(3) Multiply $(-i)U(t)$ for each interaction vertex (dashed line).
(4) Multiply $n_\sigma(t)-\alpha_{\sigma}$ [$n_{0\sigma}(t)-\alpha_{\sigma}$] for each bold [bare] tadpole diagram.
(5) Multiply $(-1)$ for each Fermion loop.
(6) Multiply an additional factor $(-i)$, coming from the definition of the Green's function (\ref{impg}).
(7) Carry out a contour integral along $\mathcal{C}$ for each internal vertex.
For example, the first-order diagram [the Hartree term, Fig.~\ref{third-order diagrams}(a)] is given by
\begin{align}
\Sigma_\sigma^{(1)}(t,t')
&=
U(t)(n_{\bar\sigma}(t)-\alpha_{\bar\sigma})\delta_{\mathcal{C}}(t,t'),
\end{align}
and the second-order diagram [Fig.~\ref{third-order diagrams}(b)] by
\begin{align}
\Sigma_\sigma^{(2)}(t,t')
&=
U(t)U(t')\mathcal{G}_{0\sigma}(t,t')\mathcal{G}_{0\bar\sigma}(t',t)\mathcal{G}_{0\bar\sigma}(t,t').
\end{align}

At half filling in the paramagnetic phase, it is natural to choose $\alpha_\sigma=\frac{1}{2}$.
This cancels all the tadpole diagrams since $n_\sigma-\alpha_{\sigma}=n_{0\sigma}-\alpha_{\sigma}=0$, and due to the particle-hole symmetry 
[i.e., $\mathcal{G}_{0\sigma}(t,t')=-\mathcal{G}_{0\sigma}(t',t)$], all the odd-order diagrams vanish as well.
On the other hand, away from half filling or in a spin-polarized phase (whenever $n_\uparrow\neq n_\downarrow$), it is nontrivial how to deal with
the Hartree and other tadpole diagrams. One may take bold diagrams [such as Fig.~\ref{third-order diagrams}(a)] with the interacting density $n_\sigma$, 
which is self-consistently determined, or bare diagrams with the noninteracting density $n_{0\sigma}$.
One can also explicitly expand the Hartree diagram with respect to the interaction up to a given order. 
For the antiferromagnetic phase at half filling, the best accuracy is obtained when one takes $\alpha_\sigma=\frac{1}{2}$ and expands all the bold loops in the tadpole diagrams up to the same order as the rest of the self-energy \cite{Tsuji2012, Tsuji2013}. 

One issue with the expansion of $\Sigma$ into bare diagrams is that the resulting perturbation theory is not a conserving approximation. As a consequence, the total energy of the system will drift with increasing time, even if the Hamiltonian is time-independent. In DMFT simulations of the Hubbard model, one finds that this drift is very small for weak-to-intermediate coupling ($U\lesssim U_c/2$, with $U_c$ the critical value for the Mott transition in the paramagnetic phase). The drift ``saturates'' at a certain time scale, so that the bare perturbation theory can be trusted up to very long time for these $U$ \cite{Tsuji2013}. A similar behavior is found
when second order perturbation theory is used to study the Hubbard  model driven by an electric field:  For small $U$, energy conservation in the form of 
Eq.~(\ref{energy conservation with jj}) is satisfied for long times \cite{Eckstein2011bloch}. 

\begin{figure}[t]
\includegraphics[width=8cm]{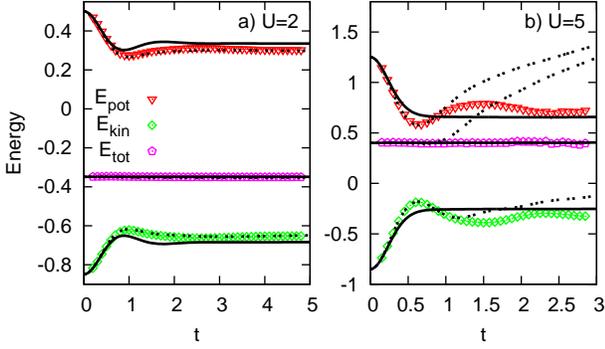}
\caption{Nonequilibrium DMFT results for the kinetic, potential and total energy of the Hubbard model after an interaction quench from $U=0$ to $U=2$ (left panel) and $U=5$ (right panel). Symbols show the weak-coupling CTQMC result, dashed lines the result from bare second order perturbation theory, and the solid lines the result from self-consistent second order perturbation theory. (Adapted from \onlinecite{Eckstein2010}.)}
\label{fig_energy_perturbation}
\end{figure}

For stronger coupling,  the perturbation theory can however fail rapidly and quite abruptly. This contrasts with the equilibrium case, where 
the second-order perturbation theory accidentally reproduces the correct strong-coupling limit at half-filling, gives a reasonable interpolation 
between the weak- and the strong-coupling regime, and even captures the Mott transition \cite{ZhangRozenbergKotliar1993,Georges96}. 
A simple way to fix the problem with the drifting energy might seem to switch to self-consistent perturbation theory. Here, one replaces the bare propagators $\mathcal{G}_0$ in the self-energy diagrams by bold propagators $G$, and considers only the diagrams which are two-particle irreducible. However, even though this approximation is conserving, 
it reproduces the time evolution of the system very poorly \cite{Eckstein2010, Tsuji2013}.
Figure~\ref{fig_energy_perturbation} shows the kinetic and potential energy, as well as the total energy after a quench in the Hubbard model (semi-circular density of states, bandwidth 4), from a noninteracting initial state to $U=2$ (left panel) and $U=5$ (right panel). Self-consistent perturbation theory conserves the total energy, but gives wrong values for the kinetic and potential energy, already after a very short time. While bare second-order perturbation theory fails for the quench to $U=5$, it at least reproduces the short-time dynamics correctly.  
%%%
For driven steady states \cite{Amaricci2012}, the accuracy of the weak-coupling approach has yet to be tested in detail. 
Here, non-conserving nature of the approximation becomes apparent when the energy current to the reservoir differs from 
the power injected by the field [Eq.~(\ref{energy conservation with jj})]. 

\subsubsection{Strong-coupling perturbation theory}
\label{strong-coupling perturbation}
%\input{nca.tex}
%%%%%%%%%%%%%%%%%%%%%%%%%%%%%%%%%%
\newcommand{\pseudoG}{{\mathcal{G}}}
\newcommand{\pseudog}{{g}}
\newcommand{\pseudoS}{{\mathfrak{S}}}
\newcommand{\pseudoF}{{\mathfrak{F}}}
\newcommand{\cconv}{{\hspace{2pt}*\hspace{-7.2pt}\raisebox{-1pt}{$\circlearrowright$}\hspace{1pt}}}

\paragraph{General remarks}

The strong-coupling CTQMC approach (Sec.~\ref{strong coupling ctqmc}) sums all diagrams generated by a 
hybridization expansion on the contour $\mathcal{C}$. While the summation into a determinant allows to absorb 
some of the sign cancellations between these diagrams, the Monte Carlo weights are complex, and the resulting 
phase problem restricts simulations to relatively short times. To avoid this sign problem, one can analytically 
sum certain subsets of the strong-coupling diagrams using a Dyson equation. This approach is expected to 
work well in the strong-coupling regime, in particular the Mott insulating phase, where the hybridization can be 
treated as a perturbation. 

The lowest order perturbative strong-coupling method is called the ``non-crossing approximation" (NCA), 
because it sums all diagrams without crossing hybridization lines. It was originally proposed 
as an approximate solution for the Anderson impurity model \cite{Keiter1971,Grewe1981,Kuramoto1983},
where it gives a reasonable description of the physics down to the Kondo temperature $T_K$, but fails in 
the Fermi liquid regime for $T\ll T_K$ \cite{MuellerHartmann1984}. The deficiencies of NCA are partly 
cured by higher order summations \cite{Pruschke1989,Haule2001a}. In particular, the simplest 
extension of NCA, the so called one-crossing approximation (OCA) \cite{Pruschke1989}, largely corrects 
the underestimation of $T_K$ of the NCA at $U<\infty$. From early on \cite{Pruschke1993a} until today 
(see, e.g., \onlinecite{Shim07Pu,Shim07Hf}), both the NCA and the OCA have been used as impurity 
solvers for DMFT. In the context of nonequilibrium DMFT, an attractive feature of the perturbative 
strong-coupling expansion is its conserving nature, and its good convergence properties with 
increasing  order of the approximation in the Mott phase \cite{Eckstein2010nca}.

Within any approximation to the strong-coupling expansion, one must re-sum parts of the series to 
infinite order to avoid severe artifacts and make the theory conserving. There exist 
various derivations of a resummed strong-coupling expansion
\cite{Keiter1971,Grewe1981,Kuramoto1983,Barnes1976,Bickers1987,
Coleman1984a,Bickers1987b}, that all circumvent the problem that Wick's theorem does not apply, because 
the unperturbed action is not quadratic.  A detailed derivation of the strong-coupling equations on the Keldysh 
contour in the context of nonequilibrium DMFT \cite{Eckstein2010nca} employed the 
pseudo-particle technique \cite{Coleman1984a} (for earlier real-time formulations and applications of the NCA, see 
\cite{Nordlander1999,Okamoto2008layer}). 
Here, we provide an alternative derivation of the same equations, which builds on the strong-coupling CTQMC formalism introduced in Sec.~\ref{strong coupling ctqmc}.

\paragraph{Self-consistent strong-coupling equations}

The starting point for NCA and its extensions is a Taylor expansion of the action (\ref{impurity action}) in terms of the 
hybridization $\Delta$, analogous to Eq.~(\ref{strong coupling expansion hyb}). To re-sum terms of this expression 
to infinite order, one must decouple trace terms like $\text{Tr}[\mathcal{T}_\CC e^{\mathcal{S}_\text{loc}} 
d_{p_1}^\dagger(t_1) d_{p_2}(t_2) \cdots ]$. Wick's theorem does not apply, but for any given 
collection of the times  
$t_1,t_2,...$
along $\CC$, one can insert a complete set of states of the impurity Hilbert space, $\sum_n|n\rangle\langle n|$, between 
consecutive operators, and thus factor the trace into a matrix-product of impurity propagators $\pseudog$ and hybridization 
vertices $F^p$ and $\bar F^p\equiv(F^p)^\dagger$,
\begin{align}
&\pseudog_{nm}(t,t')
\stackrel{t\succ  t'}{=}
-i\langle
n
|
\mathcal{T}_\CC e^{-i\int_{t'}^{t}\! d\bar t\, H_\text{loc}(\bar t) }
|
m
\rangle,
\label{bare level ret}
\\
\label{hyb vertex}
&F_{nm}^p 
= \langle n |  d_p | m  \rangle, 
\text{~~~}
\bar F_{nm}^p = \langle n |  d_p^\dagger | m  \rangle.
\end{align}
The factor $-i$ in $\pseudog$ is inserted for convenience.  For example, 
$\text{Tr}[\mathcal{T}_\CC e^{\mathcal{S}_\text{loc}}  d^\dagger _p(t)d_{q}(t')]$
$=$ $-i\,\text{Tr}[\, \pseudog(-i\beta,t) \bar F^{p} \pseudog(t,t')  F^{q}  \pseudog(t',0)]$ for $t\succ t'$,
and 
$i\,\text{Tr}[\, \pseudog(-i\beta,t') F^{q} \pseudog(t',t)  \bar F^{p}  \pseudog(t,0)]$ for $t'\succ t$.
With a suitable graphical representation of vertices $F$ and propagators $\pseudog$ 
(Fig.~\ref{fig-nca-new}a), one can represent the Taylor expansion of $Z[\Delta]$
as a sum of diagrams constructed according to the following rules: {\em (i)} The $n$-th order 
contribution to $Z[\Delta]$ is given by all topologically inequivalent diagrams consisting of one 
sequence of $2n+1$ directed lines $\pseudog$ (``backbone"), separated by $2n$ three-leg vertices ($n$ 
annihilation events $F$, $n$ creation events $\bar F$), which are connected by $n$ (directed) 
hybridization lines in all possible ways. {\em (ii)} External time arguments ($0$ and $-i\beta$)
and internal times are contour-ordered, $-i\beta \succ t_{2n} \succ \cdots \succ t_1 \succ 0$. Perform 
the trace over the product of $\pseudog$ and $F$-factors, sum over all internal flavor indices, and 
integrate over the internal times (keeping the contour ordering). {\em (iii)} The sign of the diagram 
is $(-1)^{s+f}$, where $s$ is the number of crossings of hybridization lines, and $f$ is the number of 
hybridization lines that point opposite to the direction of the backbone. {\em (iv)} An overall 
factor $i^{n+1}$ is added. For example, the expression for the simplest diagram in 
Fig.~\ref{fig-nca-new}b is
\begin{equation}
\nonumber 
\sum_{p_1,p_2}
\hspace*{-11mm}\int\limits _{\hspace*{12mm}-i\beta \succ t_2 \succ t_1 \succ 0}
\hspace*{-12mm}\!dt_2 dt_1\hspace{0mm}
 \text{Tr} \big[\pseudog(-i\beta,t_2) F^{p_2} \pseudog(t_2,t_1)  \bar F^{p_1}  \pseudog(t_1,0) \big]\Delta_{p_1p_2}(t_1,t_2),
\end{equation}
where the sign is $-(-i)^2=+1$. 

To re-sum the series, one can define an impurity level self-energy $\pseudoS_{nm}(t,t')$ as the sum 
of all parts of the above diagrams that cannot be separated into two by cutting a single $\pseudog$-line, without 
the outer trace and the outer two $\pseudog$ factors, and with an overall factor $i^n$ instead of $i^{n+1}$. 
For example, the self-energy part in the first diagram in Fig.~\ref{fig-nca-new}b is $\pseudoS(t,t')=-i\sum_{p,q} 
F^{p} \pseudog(t,t')  \,\bar F^{q} \Delta_{qp}(t',t)$. One can then introduce renormalized propagators $\pseudoG$ 
via the Dyson equation,
\begin{equation}
\label{pseudo dyson}
\pseudoG
=\pseudog + \pseudog \cconv \pseudoS \cconv \pseudoG
=\pseudog +  \pseudoG\cconv\pseudoS \cconv \pseudog ,
\end{equation}
where, for $t\succ t'$,  $[a\cconv b](t,t')$  denotes the convolution $\int _\CC\!d\bar t \,a(t,\bar t)b(\bar t,t')$ restricted to contour-ordered time 
arguments $t\succ \bar t \succ t'$. Since there are no symmetry factors associated with the diagrams, the 
partition function is given by 
\begin{equation}
\label{Z from gret}
Z
=
i\text{Tr} \big[ \pseudoG(-i\beta,0^+) \big].
\end{equation} 

Observables in the initial state can simply be evaluated by computing $\expval{\mathcal{O}(0)} = i\text{Tr} 
[\mathcal{O}\pseudoG(-i\beta,0)]/Z$, where $\mathcal{O}$ is the matrix 
with elements 
$\langle n |\mathcal{O}| m\rangle$.
For time $t>0$, however, a diagram for  $\expval{\mathcal{O}(t)}$ in general contains ``initial state vertex 
corrections'' (shaded part of Fig.~\ref{fig-nca-new}c), i.e., it is not proportional to $\text{Tr} [\pseudoG
(-i\beta,t)\mathcal{O}\pseudoG(t,0)]$. To obtain a consistent description of the self-energy $\pseudoS$ and 
these initial state corrections, one can show that the latter can be viewed as the ``lesser'' part of $\pseudoS$, 
i.e., a self-energy $\pseudoS(t,t')$ with $t\prec t'$. Since spectral functions and occupations are 
no longer related for conventional nonequilibrium Green's functions, one must separately compute the 
``lesser'' ($t\prec t'$) and ``retarded'' ($t\succ t'$) components of $\pseudoS(t,t')$ and $\pseudoG(t,t')$ 
within the nonequilibrium strong-coupling expansion. To be precise, we first define the lesser component 
of the bare propagators (\ref{bare level ret}),
\begin{equation}
\pseudog(t,t')
=
-i
\xi 
\pseudog(t,0^+)\pseudog(-i\beta,t') 
\text{~~for~~} t\prec  t',
\label{bare level les}
\end{equation}
i.e., the time-evolution is performed in ``clockwise order'' along $\CC$, from $t'$ to $-i\beta$ and then from $0$ to $t$. 
For later convenience a sign $\xi$ is included in the definition: $\xi$ is a diagonal matrix, with $\xi_{m}=-1$ ($+1$) 
when the number of particles in $|m\rangle$ is odd and even, respectively ($[\xi,\pseudog]=[\xi,\pseudoG]=0$, because 
$H_\text{loc}$ conserves the particle number). One can then define the lesser self-energy by the same diagram 
rules as before, and $\pseudoG$ by the Dyson equation (\ref{pseudo dyson}), only extending the contour-ordering 
constraint in the time integrals to clock-wise order,
\begin{multline}
[a\cconv b](t,t') = 
\\
=\left\{
\begin{array}{ll}
\int_{\CC,t'}^t d\bar t \,a(t,\bar t)b(\bar t,t') 
&
t\succ t' 
\\
\int_{\CC,t'}^{-i\beta} d\bar t \,a(t,\bar t)b(\bar t,t') 
+ 
\int_{\CC,0^+}^t d\bar t \,a(t,\bar t)b(\bar t,t') 
&
t\prec t' 
\end{array}
\right..
\end{multline}  
These definitions for $\pseudog$, $\pseudoG$, and $\pseudoS$ are actually equivalent to the ``projected 
pseudo particle'' propagators used in the alternative formulation \cite{Eckstein2010nca}. 
We can now write the partition function   
in terms of the lesser propagators, 
\begin{align}
\label{Z from glesser}
Z &= i \text{Tr} \big[ \xi \pseudoG(t^+,t^-) \big ],
\\
\expval{\mathcal{O}(t) }&= i \frac{1}{Z} 
\text{Tr} \big[ \mathcal{O} \xi \pseudoG(t^+,t^-)\big],
\label{obs from glesser}
\end{align}
where $t^\pm$ is a time on the upper (lower) branch of $\CC$. The equality of Eqs.~(\ref{Z from glesser})
and (\ref{Z from gret}) can be verified most easily order by order: A given term in the expansion of $\pseudoG(t^+,t^-)$ 
can be mapped onto a term of the expansion of $\pseudoG(-i\beta,0)$ by using Eq.~(\ref{bare level les}) 
and cyclically permuting operators under the trace. The change of the sign of the diagram, associated with 
the number of $\Delta$-lines which are flipped with respect to the $\pseudog$-lines, is accounted for by the 
$\xi$ factors in Eqs.~(\ref{bare level les}) and (\ref{Z from glesser}).

Approximations for $\pseudoG$, $Z$, and any observable are obtained by truncating the series for $\pseudoS$. 
Since there are no symmetry 
factors in  the diagrams for $\pseudoS$, one can formally reproduce all terms in the series by replacing 
$\pseudog$-lines in a diagram by fully renormalized $\pseudoG$-lines, and in turn omitting all diagrams 
in which the lines have self-energy insertions (skeleton diagrams). All skeleton diagrams up to third order 
are shown in Fig.~\ref{fig-nca-new}d. The truncation of this skeleton series at a given order defines a conserving 
approximation. More generally, an approximation is conserving when $\pseudoS[\pseudoG,\Delta]$ is 
a derivative 
$\delta \Phi /\delta \pseudoG(t',t)$ 
of some functional $\Phi$, where the 
exact $\Phi$ is the Luttinger-Ward functional. 
While it is not a priori clear that conserving approximations are necessarily better than non-conserving ones, 
the perturbative strong-coupling impurity solvers which have been tested so far are self-consistent approximations
which are conserving and satisfy $\sum_n | n \rangle\langle n  | =1$ at any time.

The most imporant observable for DMFT is the Green's function
\begin{equation}
G_{pp'}(t,t') = -\frac{1}{Z} \frac{\delta Z[\Delta] }{\delta \Delta_{p'p}(t',t)}.
\end{equation}
One can derive diagrammatic rules for $G$ and reformulate them in terms of fully interacting 
propagators $\pseudoG$ similar to time-local observables (\ref{obs from glesser}). Again, it is 
convenient not to deal with initial state correlations but instead ``close'' the contour at $-i\beta$ 
and $0^+$ and include lesser components. The following rules for $G$ result: 
(i) The 
$n$-th order contribution consists of all topologically inequivalent loops of 2$n$ $\pseudoG$-lines 
which connect $n$ $F$-vertices and $n$ $\bar F$-vertices, where one of each type is an external 
vertex. The internal vertices are connected by $\Delta$ lines such that no $\pseudoG$-line has 
a self-energy insertion. (ii) Sum over all internal flavor indices, and integrate over times, 
respecting their clockwise order. (iii) To determine the sign of a diagram, reinsert the 
$\Delta$-line between the external vertices, open the loop at any point, insert the sign $\xi$, and 
add the factor $(-1)^{s+f}$, where $s$ is the number of crossings of hybridization lines, and $f$ is 
the number of hybridization lines that point opposite to the $\pseudog$ lines. {(iv) Add a 
pre-factor $i^n$. 

It can be shown that $G$ is the derivative of the Luttinger-Ward functional,
$G(t,t')=-\frac{1}{Z} \frac{\delta \Phi[\Delta,\pseudoG] }{\delta \Delta(t',t)}$
(at fixed $\pseudoG$). Hence consistent self-consistent diagrammatic expansions for $\pseudoS$ 
and $G$ are obtained by choosing one approximation to the Luttinger-Ward functional. 
All skeleton diagrams for $\pseudoS$ and $G$ up to third order are shown in Fig.~\ref{fig-nca-new}d 
and Fig.~\ref{fig-nca-new}e.

  \begin{figure}
    \centerline{\includegraphics[width=0.8\columnwidth]{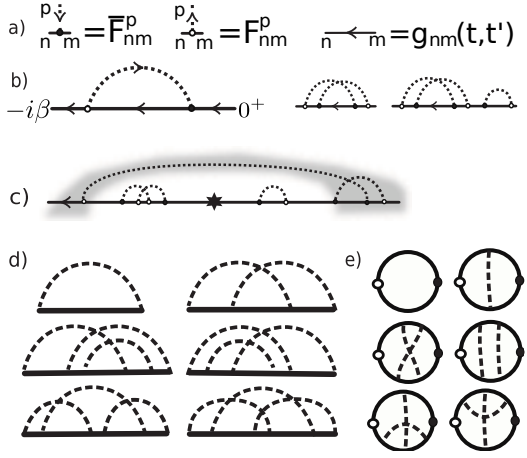}}
    \caption{
    Elements of the strong coupling expansion:
    (a) Vertices [Eq.~(\ref{hyb vertex})] and lines [Eq.~(\ref{bare level ret})].
    (b) Some diagrams for $Z$, in open-contour representation (from $0^+$ on the 
    upper branch of $\CC$, to $-i\beta$).
    (c) Diagram for $\langle \mathcal{O}(t)\rangle$ (star) in the open contour representation. 
    The shaded part can be viewed as an initial state correction, or lesser self-energy.
    (d) All skeleton diagrams $\pseudoS[\pseudoG,\Delta]$ up to third order. The first 
    two diagrams define the NCA and OCA, respectively. 
    (e) All skeleton diagrams $G[\pseudoG,\Delta]$ up to third order.
    }
    \label{fig-nca-new}
  \end{figure}

\paragraph{Numerical implementation}
\label{sec-implementation}

 Within DMFT, one numerically computes the integrals for $\pseudoS$ (lesser and greater components),
solves the Dyson equation (\ref{pseudo dyson}), computes the diagams for $G$, and feeds the result 
into the self-consistency (\ref{subsubsection:self-consistency}) to update $\Delta$.  
The numerical effort is mainly determined by the contour integrals over the internal vertices in the 
evaluation of the diagrams, and scales like $N^3$,  $N^4$,  and $N^5$ for the first, second and third 
order, respectively (where $N$ is the number of timesteps \cite{Eckstein2010nca}). For solving the 
Dyson equation (\ref{pseudo dyson}) one can switch to a differential notation, starting from the 
equation of motion for bare propagators (\ref{bare level ret}) and (\ref{bare level les}),
\begin{align}
[i\partial_t-H_\text{loc}(t)] \pseudog(t,t') = 0,\\
 \pseudog(t,t')[-i\overleftarrow{\partial}_{t'}-H_\text{loc}(t')] = 0.
\end{align}
which must be solved with the initial condition $g(t^-,t^+)=-i$. Applying the term in brackets 
to Eq.~(\ref{pseudo dyson}), one obtains two equivalent integral-differential equations for $\pseudoG$,
\begin{align}
[i\partial_t-H_\text{loc}(t)] \pseudoG(t,t') - [\pseudoS \cconv \pseudoG ](t,t')  = 0,\\
\pseudoG(t,t')[-i\overleftarrow{\partial}_{t'}-H_\text{loc}(t')] - [ \pseudoG \cconv \pseudoS](t,t')  = 0,
\end{align}
which must be solved for $t\ne t'$ with the initial condition, $\pseudoG(t^-,t^+)=-i$. Although this is 
an initial value problem instead of a boundary value problem (even on the Matsubara branch), 
and integrals are time-ordered, this equation is a causal contour equation like Eqs.~(\ref{ky-cinv-v01})
(more details are given in \cite{Eckstein2010nca}). Hence, the whole set of nonequilibrium DMFT 
equations can again be implemented in the form of a time-propagation scheme.

Results up to third order, for the time-evolution of the double occupancy after an interaction quench in 
the Hubbard model (DMFT calculation for a semi-circular density of states with bandwidth $4$), are 
shown in Fig.~\ref{fig_d_nca}. For not too low temperatures (which is often the relevant regime 
in experiments, such as pump-probe experiments which strongly excite electrons), one finds a good convergence of the results with increasing order.  (Note 
that in the correlated metal phase ($U_0=3$), NCA cannot provide an adequate description of 
the initial equilibrium phase.)

\begin{figure}[t]
\includegraphics[width=8cm]{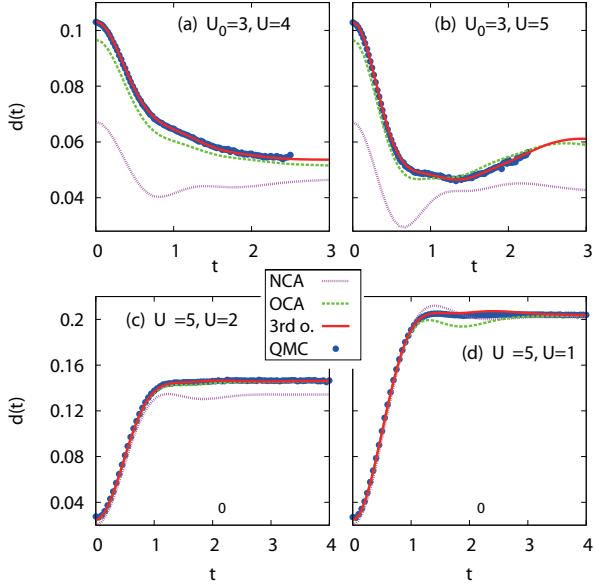}
\caption{Nonequilibrium DMFT results for the time-evolution of the double occupancy in the Hubbard model after a quench from $U(t=0)=U_0$ to $U$. The symbols show weak-coupling CTQMC results, and the lines results from strong-coupling perturbation theory. (From \cite{Eckstein2010nca}.)}
\label{fig_d_nca}
\end{figure}

\paragraph{Monte Carlo sampling around NCA}

A numerically exact Monte Carlo method based on NCA (or OCA) propagators has been described in \cite{Gull2010}. The idea here is to stochastically 
sample all the strong-coupling diagrams using a worm-type Monte Carlo algorithm \cite{Prokofev1998}. Since the building blocks of the diagrams are NCA propagators, 
the number of diagrams is reduced with respect to the strong-coupling CTQMC method, and longer times can be reached. The method is particularly suitable for the study of insulating phases, 
where NCA is a good starting point and the corrections from crossing diagrams are small. In the metallic regime, where diagrams 
with complicated topologies become relevant, the sampling of individual bold-line diagrams (instead of summing up collections of bare 
diagrams into a determinant) leads to a fermionic sign problem, in addition to the dynamical sign problem. In practice, all the contributing bold-line diagrams are summed up at factorial cost to reduce this problem.

Applications of this method to nonequilibrium quantum impurity models have been presented by \cite{Gull2011bold}. While it has not yet been 
used in the context of nonequilibrium DMFT, such an application, in appropriate parameter regimes, seems promising. The Monte Carlo 
sampling around NCA or higher-order approximations could also become a useful tool to estimate the errors accumulated in perturbative 
strong-coupling calculations. 

The method has also been used to evaluate the memory function of the Nakajima-Zwanzig-Mori equation \cite{Nakajima1958,Zwanzig1960,Mori1965}, 
a quantum master equation for the reduced density matrix of the impurity problem, which enables one to compute the time evolution
of the reduced density matrix up to longer times \cite{Cohen2011,Cohen2012}.

\subsection{Floquet formalism for periodically driven systems}
\label{floquet intro}
%\input{floquet.tex}
%%%%%%%%%%%%%%%%%%%%%%%%%%%%%%%%%%%%%%%%%%%%%%%%
When a quantum system is continuously driven by a time-periodic external force,
it may enter a nonequilibrium steady state in which the overall time-dependence of the 
system is periodic.
For example, the pump-pulse in ultrafast pump-probe experiments
may be viewed (during irradiation) as a time-periodic ac electric field 
if the laser pulse contains many 
oscillation cycles.
A theoretical approach to treat periodically driven states is the Floquet method 
\cite{Shirley1965, Zeldovich1966, Ritus1966, Sambe1973, GrifoniHanggi1998, Dittrich1998}.
It originates from Floquet's theorem \cite{Floquet1883, Hill1886, MagnusWinklerBook}, 
a temporal analogue of Bloch's theorem for a spatially periodic system. Due to the periodicity 
of external fields, the time-dependent problem can be mapped onto a {\it time-independent}
eigenvalue problem. Recently, the Floquet method has been employed in combination 
with DMFT to study nonequilibrium steady states of periodically driven correlated systems 
\cite{Tsuji08, Tsuji09, Joura08, Schmidt2002, LubatschKroha2009, FreericksJouraBook}.
An advantage of the Floquet DMFT is that one does not have to calculate the full time evolution until a 
nonequilibrium steady state is reached. It suffices to solve small-size matrix equations for nonequilibrium 
Green's functions represented in frequency space, which greatly reduces the computational cost.

The Floquet formalism has been
used in the study of Floquet topological insulators
\cite{OkaPHE09,Kitagawa11Haldane,Lindner11}.
The topology of quantum systems can be controlled by 
external time-periodic perturbations;
For example, by applying circularly polarized light
to graphene (or other many-band systems), 
one can change a trivial state into a 
quantum Hall insulator \cite{OkaPHE09,Kitagawa11Haldane}.
The methods which we describe here can be applied to such problems.

\subsubsection{Overview of Floquet's theorem}
\label{floquet theorem}

Floquet's theorem \cite{Floquet1883, Hill1886, MagnusWinklerBook}
is a general statement about the solution of an ordinary differential equation
$dx(t)/dt=C(t)x(t)$ with time-periodic coefficients $C(t)$. 
Here we apply the theorem to the time-dependent Schr\"{o}dinger equation, 
\begin{align}
i\frac{d}{dt} \Psi(t)
&=
H(t) \Psi(t),
\label{schrodinger}
\end{align}
where the Hamiltonian $H(t)$ is assumed to be periodic in time with period $\mathscr{T}$,
$H(t+\mathscr{T})=H(t)$.
Floquet's theorem states that there exists a solution of Eq.~(\ref{schrodinger}) of the form 
\begin{align}
\Psi_\alpha(t)
&=
e^{-i\varepsilon_\alpha t} u_\alpha(t),
\label{floquet_state}
\end{align}
where $u_\alpha(t)=u_\alpha(t+\mathscr{T})$ is a periodic function of $t$, 
and the real number $\varepsilon_\alpha$ is called the {\it quasienergy}, which is
unique up to integer multiples of $\Omega$ $=2\pi/\mathscr{T}$.
To prove this, let us write the formal solution of Eq.~(\ref{schrodinger}) as
\begin{align}
\Psi(t)
&=
\mathcal{U}(t,t_0)\Psi(t_0),
\label{formal solution}
\end{align}
with the time-evolution operator 
$\mathcal{U}(t,t_0)=\mathcal{T}e^{-i\int_{t_0}^{t}d\bar{t}\,H(\bar{t})}$.
Then we consider the operator
\begin{align}
\mathcal{U}(t+\mathscr{T},t_0)
&=
\mathcal{U}(t+\mathscr{T},t_0+\mathscr{T})\,\mathcal{U}(t_0+\mathscr{T},t_0),
\label{U divided}
\end{align}
which is split into two via the chain rule. The first part is
$\mathcal{U}(t+\mathscr{T},t_0+\mathscr{T})=\mathcal{U}(t,t_0)$
due to the periodicity of the 
Hamiltonian. 
The second part $\mathcal{U}(t_0+\mathscr{T},t_0)$
is called the Floquet operator, 
which we can write in terms of a Hermitian operator $Q(t_0)$,
\begin{align}
  e^{-iQ(t_0)\mathscr{T}}
    &\equiv
      \mathcal{U}(t_0+\mathscr{T},t_0).
\end{align}
Multiplying both sides of 
Eq.~(\ref{U divided}) 
with $e^{iQ(t_0)(t+\mathscr{T})}$ from the right, we 
thus have
\begin{align}
  \mathcal{U}(t+\mathscr{T},t_0)e^{iQ(t_0)(t+\mathscr{T})}
    &=
      \mathcal{U}(t,t_0)e^{iQ(t_0)t},
  \label{U periodic2}
\end{align}
which shows that the unitary operator $P(t,t_0)\equiv \mathcal{U}(t,t_0)e^{iQ(t_0)t}$ 
is periodic in $t$, i.e.,  $P(t+\mathscr{T},t_0)=P(t,t_0)$. The solution 
(\ref{formal solution}) of the Sch\"{o}dinger equation (\ref{schrodinger}) has thus been 
written in terms of a ($t$-independent) Hermitian operator $Q(t_0)$ and a unitary time-periodic 
operator $P(t,t_0)$,
\begin{align}
  \Psi(t)
    &=
      P(t,t_0)e^{-iQ(t_0)t}\,\Psi(t_0).
  \label{floquet state}
\end{align}
When the initial state $\Psi(t_0)$ is an eigenstate of $Q(t_0)$ [denoted by $\Psi_\alpha(t_0)$] with eigenvalue $\varepsilon_\alpha$,
the solution is given by Eq.~(\ref{floquet_state})
with $u_\alpha(t)=P(t,t_0)\Psi_\alpha(t_0)$.
One can see that $u_\alpha(t)$ is a periodic function [$u_\alpha(t+\mathscr{T})=u_\alpha(t)$],
and that the 
quasienergy spectrum $\varepsilon_\alpha$ 
does not depend on $t_0$. 

To determine $\varepsilon_\alpha$, 
one can Fourier expand $u_\alpha(t)$ as $u_\alpha(t)=\sum_n e^{-in\Omega t}\, u_\alpha^n$
(with $\Omega=2\pi/\mathscr{T}$), where $u_\alpha^n$
is called the $n$th Floquet mode of the Floquet state (\ref{floquet_state}). 
Then Eq.~(\ref{schrodinger}) gives
\begin{align}
  \sum_n (H_{mn}-n\Omega\delta_{mn}) u_\alpha^n
    &=
      \varepsilon_\alpha u_\alpha^m,
  \label{floquet}
\end{align}
where 
\begin{align}
  H_{mn}
    &\equiv 
      \frac{1}{\mathscr{T}}\int_{0}^{\mathscr{T}} dt \;
      e^{i(m-n)\Omega t} H(t)
  \label{floquet_hamiltonian}
\end{align}
is the Floquet matrix form of the Hamiltonian. 
Thus the quasienergies $\varepsilon_\alpha$ are the eigenvalues 
of the infinite dimensional Floquet matrix $H_{mn}-n\Omega\delta_{mn}$.
Note that if $\varepsilon_\alpha$ is an eigenvalue of $H_{mn}-n\Omega\delta_{mn}$, 
the same holds for $\varepsilon_\alpha+n\Omega$, for arbitrary integer $n$. To avoid 
the redundancy in $\varepsilon_\alpha$, we impose the condition that 
$-\frac{\Omega}{2}<\varepsilon_\alpha\le\frac{\Omega}{2}$.
As a consequence of the Floquet theorem,
the time-dependent differential Eq.~(\ref{schrodinger})
has been transformed into a {\it time-independent} eigenvalue problem (\ref{floquet}), 
which can be solved by simple linear algebra (if one truncates the matrix size).

\subsubsection{Floquet Green's function method}
\label{floquet-green}

\paragraph{General formulation}

We now apply the Floquet method 
to the nonequilibrium Green's function formalism (Sec.~\ref{steady-states}),
as introduced earlier in several papers \cite{Faisal1989, Althorpe1997, BrandesRobinson2002, Martinez2003, Martinez2005, MartinezMolina2006}.
In general, Green's functions $G^{R,A,K}(t,t')$  in the Keldysh formalism (Sec.~\ref{steady-states}) 
have two independent time arguments, $t$, $t'$ $\in (-\infty,+\infty)$  (two-time representation) , or equivalently, relative time 
$t_{\rm rel}\equiv t-t'$ and averaged time $t_{\rm av}\equiv (t+t')/2$ \cite{Wigner1932}.
We will say that a periodically driven system has reached a nonequilibrium steady state (NESS) when 
its Green's functions become periodic as a function of $t_{\rm av}$, i.e.,
\begin{align}
G^{R,A,K}(t+\mathscr{T},t'+\mathscr{T})
&=
G^{R,A,K}(t,t').
\label{G periodicity}
\end{align}
The Floquet Green's function formalism (and Floquet DMFT) can describe such NESSs not only for the unitary evolution of isolated systems as considered in Sec.~\ref{floquet theorem}
but also for the dissipative evolution of open systems coupled to environment.
For the latter, one can determine the Floquet Green's function without considering 
the earlier transient dynamics. Mathematically, it is hard to prove that a driven system approaches a NESS 
in the long-time limit (in the same way as it is difficult to prove ``thermalization'' for an isolated system). However, 
for a dissipative system that is continuously driven by a time-periodic perturbation, NESS solutions obtained by 
the Floquet method usually exist and are unique, so that one may simply assume that this state is indeed established 
after the dependence on the initial condition is wiped out in the presence of dissipation. 

For a Green's function $G(t,t')$ that satisfies the periodicity condition (\ref{G periodicity}), 
we first define the Wigner representation \cite{Wigner1932} by a Fourier-transformation with respect to relative time,
\begin{align}
G(\omega,t_{\rm av})
&=
\int_{-\infty}^{\infty} dt_{\rm rel}\, e^{i\omega t_{\rm rel}} \,G(t,t').
\label{wigner_rep}
\end{align}
From this we introduce the Floquet matrix form of $G$,
\begin{align}
\bm G_{mn}(\omega)
&\equiv
\frac{1}{\mathscr{T}}\int_0^{\mathscr{T}} \!\!dt_{\rm av}\, e^{i(m-n)\Omega t_{\rm av}}
G\left(\omega+\frac{m+n}{2}\Omega,t_{\rm av}\right).
\label{floquet_rep}
\end{align}
In this Floquet representation, we will 
use the reduced zone scheme to avoid degeneracies, i.e., 
the range of $\omega$ is restricted to a ``Brillouin zone'' 
$-\Omega/2 < \omega \leq \Omega/2$.
To reconstruct $\bm G_{mn}(\omega)$ outside the Brillouin zone, 
one can shift the frequency with an integer $l$ such that $-\Omega/2<\omega-l\Omega\le \Omega/2$,
using the symmetry relation $\bm G_{m,n}(\omega)=\bm G_{m+l,n+l}(\omega-l\Omega)$. 
In Fig.~\ref{floquet-green diagram}, we show a diagrammatic representation of $\bm G_{mn}(\omega)$,
a fermion propagator with multiple photon absorption/emission (external wavy lines).
As a fermion propagates in the presence of the driving field, 
its energy changes from $\omega+m\Omega$ to $\omega+n\Omega$.
\begin{figure}[t]
\begin{center}
\includegraphics[width=7cm]{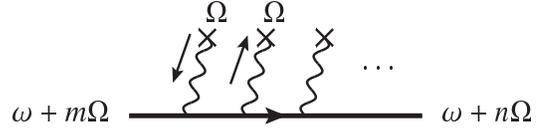}
\caption{Diagrammatic representation of the Floquet Green's function $\bm G_{mn}(\omega)$. 
The wavy lines denote photon propagators of the pump light, and the arrows indicate the energy flow.}
\label{floquet-green diagram}
\end{center}
\end{figure}

For time-periodic NESSs, the two-time representation, the Floquet representation, and the 
Wigner representation for Green's functions are equivalent. The relations among these
three representations  are summarized in Table~\ref{three rep}.
The advantage of the Floquet representation is that a convolution in the two-time representation
(or a Moyal product in the Wigner representation) can be translated into a simple matrix product
in the Floquet representation.
This greatly simplifies the solution of the Dyson equation (\ref{dyson}),
since the problem is reduced to a matrix inversion in the Floquet representation,
and one can usually truncate the Floquet matrix to a relatively small size
as long as higher order nonlinear processes (multi-photon absorption/emission) are suppressed.
\begin{table*}[htbp]
\begin{center}
\begin{tabular}{|c|c|c|}
\hline
Two-time representation & Wigner representation & Floquet representation \\
\hline
\hline
\rule{0cm}{0.5cm}
$\displaystyle G(t,t')=\int_{-\infty}^\infty \frac{d\omega}{2\pi} e^{-i\omega t_{\rm rel}}G(\omega,t_{\rm av})$
& 
$\displaystyle G(\omega,t_{\rm av})=\int_{-\infty}^\infty dt_{\rm rel} e^{i\omega t_{\rm rel}} G(t,t')$
& 
\rule{0cm}{0.6cm}
$\displaystyle \bm G_{mn}(\omega)=\frac{1}{\mathscr{T}}\int_0^{\mathscr{T}} dt_{\rm av} e^{i(m-n)\Omega t_{\rm av}}
G\left(\omega+\frac{m+n}{2}\Omega, t_{\rm av}\right)$
\\
\rule{0cm}{0.5cm}
$\displaystyle =\sum_{mn}\int_{-\Omega/2}^{\Omega/2} \frac{d\omega}{2\pi} 
e^{-i(\omega+m\Omega)t+i(\omega+n\Omega)t'}\bm G_{mn}(\omega)$
&
$\displaystyle =\sum_{m-n} e^{-i(m-n)\Omega t_{\rm av}} \bm G_{mn}\left(\omega-\frac{m+n}{2}\Omega\right)$
&
\rule{0cm}{0.6cm}
$\displaystyle =\frac{1}{\mathscr{T}}\int_0^{\mathscr{T}} dt_{\rm av} \int_{-\infty}^\infty dt_{\rm rel} 
e^{i(\omega+m\Omega)t-i(\omega+n\Omega)t'} G(t,t')$
\\
\hline
\rule{0cm}{0.4cm}
convolution: & 
Moyal product: $(A\star B)(\omega,t_{\rm av})=$ & 
matrix product: \\
\rule{0cm}{0.5cm}
$\displaystyle (A\ast B)(t,t')=\int_{-\infty}^\infty d\bar{t} A(t,\bar{t})B(\bar{t},t')$
& 
$\displaystyle A(\omega,t_{\rm av})\exp 
\left(\frac{i}{2}[\overleftarrow{{\partial}_{\omega}}\overrightarrow{\partial_{t_{\rm av}}}
-\overleftarrow{\partial_{t_{\rm av}}}\overrightarrow{\partial_{\omega}}]\right)
B(\omega,t_{\rm av})$
&
$\displaystyle (\bm A\cdot \bm B)_{mn}(\omega)=\sum_l \bm A_{ml}(\omega) \bm B_{ln}(\omega)$ \\
\hline
\end{tabular}
\caption{Three representations of nonequilibrium Green's functions.}
\label{three rep}
\end{center}
\end{table*}

To further establish the formalism, we start from the retarded Green's function
for a noninteracting  system, which satisfies the Dyson equation 
(or equation of motion),
\begin{align}
[i\partial_t-H(t)]G^R(t,t')
&=
\delta(t-t'),
\label{kernel}
\end{align}
where $H$ is the single-particle Hamiltonian, and all objects are regarded as matrices in orbital indices.
A solution of Eq.~(\ref{kernel}) is explicitly given in terms of the Floquet wave functions 
$\Psi_\alpha(t)=e^{-i\varepsilon_\alpha t}u_\alpha(t)$ (\ref{floquet_state}) as
\begin{align}
G_\alpha^R(t,t')
&=
-i\theta(t-t')\Psi_\alpha(t)\Psi_\alpha^\dagger(t').
\label{psi psi}
\end{align}
Since $\Psi_\alpha(t+\mathscr{T})=e^{-i\varepsilon_\alpha\mathscr{T}}\Psi_\alpha(t)$,
the Green's function (\ref{psi psi}) manifestly satisfies the periodicity condition (\ref{G periodicity}).
Its Floquet representation is given by \cite{Tsuji08}, 
\begin{align}
\bm G_{\alpha}^R(\omega)
&=
\bm U_{\alpha}\cdot \bm Q_{\alpha}(\omega)\cdot \bm U_{\alpha}^\dagger,
\label{lambda_q_lambda}
\end{align}
with a unitary matrix 
\begin{align}
(\bm U_{\alpha})_{mn}
&=
\frac{1}{\mathscr{T}}\int_{0}^{\mathscr{T}} dt\, e^{i(m-n)\Omega t} u_\alpha(t),
\label{lambda}
\end{align}
and a diagonal matrix
\begin{align}
(\bm Q_{\alpha})_{mn}(\omega)
&=
\frac{1}{\omega+n\Omega+\mu-\varepsilon_\alpha+i\eta}\,\delta_{mn}.
\label{q}
\end{align}
Thus, $\bm G_{\alpha}^R(\omega)$ has poles at the 
Floquet quasienergies $\omega=\varepsilon_{\alpha}-n\Omega$
($n=0,\pm1,\pm2,\dots$).
For a noninteracting single-band system described by the Hamiltonian 
$H_0(t)=\sum_{\bm k} \epsilon_{\bm k}(t) c_{\bm k}^\dagger c_{\bm k}$, 
$\varepsilon_\alpha$ is given by the time-averaged band dispersion 
$\langle\!\langle\epsilon_{\bm k}\rangle\!\rangle \equiv\frac{1}{\mathscr{T}}\int_0^{\mathscr{T}} dt\,\epsilon_{\bm k}(t)$, 
and the Floquet wave function is $\Psi_{\bm k}(t)=e^{-i\langle\!\langle\epsilon_{\bm k}\rangle\!\rangle t}u_{\bm k}(t)$
with $u_{\bm k}(t)=e^{-i\int_0^t d\bar{t}[\epsilon_{\bm k}(\bar{t})-\langle\!\langle\epsilon_{\bm k}\rangle\!\rangle]}$ \cite{Tsuji08}.
Effectively, the band dispersion in the periodically driven system is renormalized from the original $\epsilon_{\bm k}$ 
to the time-averaged $\langle\!\langle\epsilon_{\bm k}\rangle\!\rangle$, and the renormalized band splits into
replicas with a spacing $\Omega$.

In practice, it is convenient to use the inverse $\bm G_{0}^{R-1}(\bm k,\omega)$
rather than $\bm G_{0}^R(\bm k,\omega)$
when one solves the Dyson equation (\ref{dyson with dissipation}).
Using the relation (\ref{lambda_q_lambda}) and the unitarity of $\bm U_{\bm k}$,
one has $\bm G_{0}^{R-1}(\bm k,\omega)=\bm U_{\bm k}\,
\bm Q_{\bm k}^{-1}(\omega)\, \bm U_{\bm k}^\dagger$, which reads
\begin{align}
(\bm G_0^{R-1})_{mn}(\bm k,\omega)
&=
(\omega+n\Omega+\mu+i\eta)\delta_{mn}-(\bm \epsilon_{\bm k})_{mn}.
\label{inv_g}
\end{align}
Here $\bm\epsilon_{\bm k}$ is the Floquet matrix defined by
\begin{align}
(\bm \epsilon_{\bm k})_{mn}
&=
\frac{1}{\mathscr{T}}\int_{0}^{\mathscr{T}} dt\,
e^{i(m-n)\Omega t} \epsilon_{\bm k}(t).
\label{e_flq}
\end{align}
The Keldysh component of the noninteracting Green's function is not uniquely determined 
by Eq.~(\ref{kernel}). One way to state this fact is to say that $(\bm G_0^{-1})_{mn}^K(\bm k,\omega)\equiv 
-(\bm G_0^{R-1}\bm G_{0}^K\,\bm G_{0}^{A-1})_{mn}(\bm k,\omega)=2i\eta F(\epsilon_{\bm k}-\mu)\delta_{mn}$ 
is proportional to the infinitesimal $\eta$ and an arbitrary distribution function. The latter is usually fixed by the 
equilibrium distribution $F(\omega)=\tanh(\beta\omega/2)$ [Eq.~(\ref{coth, tanh})] for fermions. However, any  other nonzero 
term in the Keldysh self-energy (e.g., a bath self-energy $\Sigma_{\rm bath}^K$) dominates $(\bm G_{0}^{-1})^K$ 
and thus completely determines the steady state distribution.

\paragraph{Simple example}

To see how the Floquet Green's function technique works, we consider a one-dimensional 
electric-field-driven tight-binding model \cite{Han2012} coupled to free-fermion baths 
(Sec.~\ref{buttiker}). In the temporal gauge, the Hamiltonian is
\begin{align}
H_{\rm s}(t)&=
-2\gamma\sum_k \cos(k-A(t))c_k^\dagger c_k,
\label{1d tight binding}
\end{align}
where the dc electric field is introduced as the Peierls phase $A(t)=-\Omega t$, with the Bloch-oscillation 
frequency $\Omega=eEa$. Although $A(t)$ is not periodic in time, the Hamiltonian (\ref{1d tight binding}) 
has the periodicity with period $\mathscr{T}=2\pi/\Omega$, and we can apply the Floquet method to this dc-electric-field 
problem. In the present case, we have $\epsilon_k(t)=-2\gamma\cos(k+\Omega t)$,
which leads to $\langle\!\langle\epsilon_k\rangle\!\rangle=0$ and
\begin{align}
(\bm U_k)_{mn}
&=e^{-i(m-n)k-2i\frac{\gamma}{\Omega}\sin k}\mathcal{J}_{n-m}\left(\frac{2\gamma}{\Omega}\right),
\end{align}
where $\mathcal{J}_{n}$ is the $n$th order Bessel function.  
The Floquet Green's function is derived from the Dyson equation (\ref{dyson with dissipation}), 
\begin{align}
\begin{pmatrix}
\bm G^R & \bm G^K \\
O & \bm G^A
\end{pmatrix}^{-1}
&\!=\!
\begin{pmatrix}
\bm G_0^{R-1} & (\bm G_0^{-1})^K \\
O & \bm G_0^{A-1}
\end{pmatrix}
\!-\!
\begin{pmatrix}
\bm \Sigma_{\rm bath}^R & \bm \Sigma_{\rm bath}^K\\
O & \bm \Sigma_{\rm bath}^A
\end{pmatrix},
\end{align}
where
\begin{align}
\label{fermion bath floquet}
\begin{pmatrix}
\bm \Sigma_{\rm bath}^R(\omega) & \bm \Sigma_{\rm bath}^K(\omega)\\
O & \bm \Sigma_{\rm bath}^A(\omega)
\end{pmatrix}
&=
\begin{pmatrix}
-i\Gamma\mathbbm{1} & -2i\Gamma\bm F(\omega) \\
O & i\Gamma\mathbbm{1}
\end{pmatrix}
\end{align}
is the Floquet representation of the bath self-energy (\ref{sigma bath}), with 
\begin{align}
\bm F_{mn}(\omega)
&\equiv
\tanh\left(\frac{\omega+n\Omega}{2T}\right)\delta_{mn}
\end{align}
being the Floquet representation of the Fermi distribution function for fermions
with the bath temperature $T$.
With the decomposition (\ref{lambda_q_lambda}),
we have the retarded Floquet Green's function
$\bm G^R=\bm U_k\cdot [\bm Q_k^{-1}(\omega)+i\Gamma\mathbbm{1}]^{-1}\cdot\bm U_k^\dagger$,
which reads
\begin{align}
(\bm G^R)_{mn}(k,\omega)
&=
e^{-i(m-n)k}\sum_l \frac{\mathcal{J}_{l-m}\left(\frac{2\gamma}{\Omega}\right)
\mathcal{J}_{n-l}\left(\frac{2\gamma}{\Omega}\right)}
{\omega+l\Omega+i\Gamma}.
\end{align}
The lesser Floquet Green's function is given by
$\bm G^<=2i\Gamma \bm G^R\cdot \bm f\cdot \bm G^A$ with 
the Fermi distribution function $\bm f_{mn}=f(\omega+n\Omega)\delta_{mn}$.
Using this, we obtain the momentum distribution function 
$n(\bm k,t)=-i\int_{-\Omega/2}^{\Omega/2}\frac{d\omega}{2\pi}
\sum_{mn}e^{-i(m-n)\Omega t}\bm G_{mn}^<(\bm k,\omega)$,
which can be evaluated analytically for temperature $T=0$ \cite{Han2012},
\begin{align}
n(k,t)
&=
\frac{\Gamma}{\pi}\sum_{m,n} e^{i(m-n)(k+\Omega t)}
\frac{\mathcal{J}_m\left(\frac{2\gamma}{\Omega}\right)
\mathcal{J}_n\left(\frac{2\gamma}{\Omega}\right)}{(n-m)\Omega+2i\Gamma}
\ln\left(\frac{m\Omega-i\Gamma}{n\Omega+i\Gamma}\right).
\end{align}
The function $n(k,t)$ defined in this way is not gauge invariant. To obtain a physical observable, 
we must evaluate it at a co-moving wave vector $\tilde{k}=k-\Omega t$
(see Sec.~\ref{dmft observables}). Hence the gauge-invariant
momentum distribution function 
$\tilde n (k,t) \equiv n(\tilde k,t) = n(k-\Omega t,t)$
is time-independent.
In the weak-field limit ($\Omega\ll \gamma, \Gamma$), we have
\begin{align}
n(\tilde{k})
&=
\frac{1}{2}+\frac{1}{\pi}\arctan\left(\frac{2\gamma\cos(\tilde{k}-\delta k)}{\Gamma}\right)
+O(\Omega^2)
\end{align}
with the momentum shift
\begin{align}
\delta k=\frac{\Omega}{\Gamma}\left[1+\left(\frac{2\gamma\cos \tilde{k}}{\Gamma}\right)^2\right]^{-1}.
\end{align}
At zero field, $n(k)=\int d\omega\,\frac{\Gamma/\pi}{(\omega-\epsilon_k)^2+\Gamma^2}f(\omega)
=\frac{1}{2}+\frac{1}{\pi}\arctan\left(\frac{-\epsilon_k}{\Gamma}\right)$.
When the field is turned on, the momentum distribution shifts in the field direction
by $\delta k\sim \Omega/\Gamma$ near the Fermi surface, which is expected from
Boltzmann's semiclassical transport theory with the relaxation time approximation $(\tau\sim\Gamma^{-1})$.
The current is obtained from $j=\int \frac{dk}{2\pi}\, v(k-A(t))n(k,t)=
\int \frac{d k}{2\pi}\, v(k)\tilde n(k)$ with the  group velocity 
$v(k)=\partial \epsilon(k)/\partial k$.
It is time independent (no Bloch oscillations), and consistent with the linear-response result
in the weak-field limit,
\begin{align}
j&=
\frac{2\gamma^2}{\pi\Gamma\sqrt{4\gamma^2+\Gamma^2}}\Omega
+O(\Omega^2).
\end{align}
When $\Gamma\ll \gamma$,
it reproduces the Drude formula,
\begin{align}
j\sim \frac{\gamma\Omega}{\pi\Gamma}
\propto \frac{E\tau}{m^\ast},
\end{align}
with $\gamma\sim1/m^\ast$ ($m^\ast$: effective mass) and $\Gamma\sim 1/\tau$ ($\tau$: relaxation time).
This shows that  that although the free-fermion bath model (Sec.~\ref{buttiker})
is somewhat artificial in the sense that it only includes single-particle processes,
it correctly reproduces the conventional semiclassical transport picture 
without momentum scattering, so that it serves as a minimal model for dissipation mechanisms.

\subsubsection{Floquet dynamical mean-field theory}
\label{floquet dmft}

\paragraph{General formalism}

In this section, we describe the application of Floquet theory to nonequilibrium DMFT,
to study periodically driven states of strongly correlated systems.
The original idea goes back to the pioneering work of \onlinecite{Schmidt2002}.
The formalism was further developed by \onlinecite{Joura08,FreericksJouraBook}, and 
\onlinecite{Tsuji08, Tsuji09}. 
In the Floquet DMFT formalism, one considers a dissipative system continuously driven by a time-periodic 
perturbation.
It is assumed that a time-periodic NESS exists in the long-time limit, after all memory on the 
initial condition has been wiped out by the dissipation. 
(These assumptions have been numerically tested for the driven Hubbard model coupled to a heat bath by \onlinecite{Amaricci2012}.)
Floquet DMFT can directly access this time-periodic NESS, without computing the full time evolution from the initial 
state,
by mapping 
the time-periodic NESS of the lattice model to the corresponding time-periodic NESS of 
an effective single-site impurity model. 

As an example, let us take the Hubbard model,
for which the action $S_{\rm imp}^{\rm NESS}$ of the effective impurity problem 
is the same as Eq.~(\ref{dmft_formalism::imp-action-2}),
but on the Keldysh contour $\mathcal{C}_K$ (Fig.~\ref{keldysh contour})
instead of the $L$-shaped contour, and with a hybridization function 
$\Delta_\sigma(t,t')$, that has the time periodicity 
$\Delta_\sigma(t+\mathscr{T},t'+\mathscr{T})=\Delta_\sigma(t,t')$.
Due to the periodicity in time, the impurity Green's function defined by
\begin{align}
\label{ness g imp}
G_\sigma^{\rm imp}(t,t')=
-i\langle \mathcal{ T}_{\mathcal{C}_K} c_\sigma(t)c_\sigma^\dagger(t')\rangle_{S_{\rm imp}^{\rm NESS}}
\end{align}
also satisfies $G_\sigma^{\rm imp}(t+\mathscr{T},t'+\mathscr{T})=G_\sigma^{\rm imp}(t,t')$.
As described in Sec.~\ref{floquet-green}, the time-periodic Green's function can thus be represented
by a Floquet Green's function $\bm G_\sigma^{\rm imp}(\omega)$ in frequency space. 
Each Green's function
or self-energy in the Floquet representation 
then has a supermatrix structure of the Larkin-Ovchinnikov form,
\begin{align}
\bm G
=
\begin{pmatrix}
\bm G^R & \bm G^K \\
O & \bm G^A
\end{pmatrix}.
\end{align}
The mapping 
from the lattice to the impurity model is defined such that
the local part of the lattice Floquet Green's function is reproduced by the impurity Floquet Green's function,
\begin{align}
\bm G^{\rm imp}_{\sigma}(\omega)
&=
\sum_{\bm k} \bm G^{\rm lat}_{\sigma}(\bm k, \omega),
\label{Floquet DMFT1}
\end{align}
and one makes the approximation that the lattice Floquet self-energy $\bm \Sigma_\sigma^{\rm lat}$ 
is local in space 
(or independent of $\kk$)
and can be identified with the Floquet self-energy of the impurity,
\begin{align}
\bm \Sigma^{\rm lat}_{\sigma}(\bm k, \omega)
&=
\bm \Sigma^{\rm imp}_{\sigma}(\omega).
\label{Floquet DMFT2}
\end{align}
The lattice Floquet Green's function satisfies the Dyson equation for the lattice model,
\begin{align}
\bm G^{\rm lat}_\sigma(\bm k,\omega)
&=
\left[
\bm G_{0\sigma}^{-1}(\bm k,\omega)-\bm \Sigma^{\rm bath}(\omega)-\bm \Sigma^{\rm lat}_\sigma(\bm k,\omega)
\right]^{-1},
\label{Floquet lattice Dyson}
\end{align}
where $\bm G_{0\sigma}(\bm k,\omega)$ is the noninteracting Floquet Green's function, 
and $\bm \Sigma^{\rm bath}(\omega)$ is a dissipation term coming from the coupling to
the external heat bath. Usually $\bm \Sigma^{\rm bath}(\omega)$ is a local function,
such as the Floquet representation (\ref{fermion bath floquet}) of Eq.~(\ref{sigma_diss}) for the free-fermion bath.
The impurity Floquet Green's function satisfies the Dyson equation for the impurity model,
\begin{align}
\bm G_\sigma^{\rm imp}(\omega)
&=
\left[
\bm \omega+\mu \mathbbm{1}-\bm\Delta_\sigma(\omega)-\bm\Sigma_\sigma^{\rm imp}(\omega)
\right]^{-1},
\label{Floquet impurity Dyson}
\end{align}
with $\bm\omega_{mn}=(\omega+n\Omega)\delta_{mn}$. 
The self-consistency condition of the Floquet DMFT consists of
Eqs.~(\ref{Floquet DMFT1}), (\ref{Floquet DMFT2}), (\ref{Floquet lattice Dyson}), and (\ref{Floquet impurity Dyson})
in combination with the solution of the effective impurity problem (\ref{ness g imp}).

\paragraph{Impurity solver}

In practical implementations of the Floquet DMFT, one has to solve the time-periodic 
nonequilibrium impurity problem (\ref{ness g imp}). Since the length of the real-time axis
is infinite by construction, the available impurity solvers (Sec.~\ref{impurity solver}) are limited.
For example, quantum Monte Carlo techniques cannot be used  
due to the dynamical sign problem. \onlinecite{Joura08} applied
the numerical renormalization group technique \cite{Bulla08} to calculate the density of states
for the NESS of the driven Hubbard model based on the approximation of using the thermal density matrix.
\onlinecite{Tsuji09} studied the Falicov-Kimball model \cite{Freericks2003a} 
driven by an ac field with the Floquet DMFT,
for which an exact solution for the impurity problem out of equilibrium can be used.
\onlinecite{LubatschKroha2009} employed the iterated perturbation theory as an impurity solver
to study the Hubbard model driven by an ac field,
where the field was introduced to linear order by the $\bm j\cdot \bm A$ coupling. 
In general, one can directly apply diagrammatic techniques
such as the weak-coupling (Sec.~\ref{weak-coupling perturbation}) and strong-coupling 
(\ref{strong-coupling perturbation}) perturbation theories to the impurity problem.
In the presence of a heat bath, numerical simulations usually become stable, even for nonconserving approximations.

\subsection{Extensions of DMFT and alternative approaches}
\label{subsec:extensions}

Single-site dynamical mean field theory provides a qualitatively correct description of high-dimensional lattice models. 
Since nonequilibrium applications often involve highly excited systems, and nonlocal correlations tend to become less 
relevant in equilibrium models at high temperature, one might expect that the local self-energy approximation of single-site 
DMFT is even better in the nonequilibrium context than it is in equilibrium. On the other hand, as 
mentioned in Sec.~\ref{dmft_formalism::sec-noneq}, 
even small perturbations to the Hamiltonian can have a pronounced influence 
on the long-time behavior of a system out of equilibrium. It is thus not a priori obvious how severely the single-site DMFT approximation affects the time-evolution. To study the nonequilibrium properties of models with reduced 
dimensionality, it is important to develop methods which take spatial correlations into account. Extensions of the 
DMFT formalism are also needed to study inhomogeneous systems, such as hetero-structures or cold atoms in a trapping 
potential. 

\subsubsection{Cluster perturbation theory}
%\input{cpt.tex}
%%%%%%%%%%%%%%%%%%%%%%%%%%%%%%%%%%%%%%%
%%%%%%%%%%
\begin{figure}[t]
\includegraphics[width=7cm]{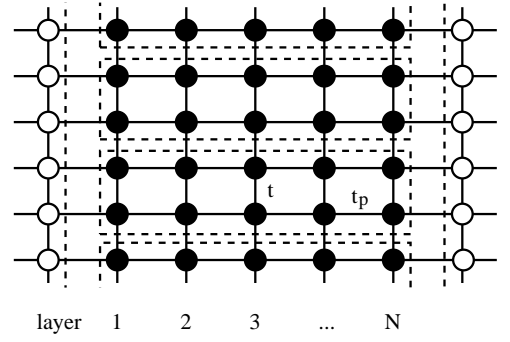}
\caption{Illustration of a system consisting of $N$ correlated layers (full dots) between noninteracting leads (open dots). The intra-layer hopping $t$, inter-layer hopping $t_p$, interaction $U$, chemical potential $\mu$, can be layer- and time-dependent. }
\label{fig_layer}
\end{figure}
%%%%%%%%%%%%%%%%%

A simple and computationally cheap method which allows to treat short-range correlations explicitly is the cluster perturbation theory 
(CPT) \cite{Gros1993, Senechal2000}. This method has recently been adapted to nonequilibrium systems \cite{Balzer2011,Knap2011}. 
The idea is to decompose the system into small clusters, whose dynamics can be computed exactly, for example using exact diagonalization
or Krylov-space methods \cite{Balzer2012}, and to treat the inter-cluster hopping as a perturbation. An example of such a decomposition is shown 
in Fig.~\ref{fig_layer}, for a system consisting of several layers with local interactions $U$, sandwiched between two noninteracting leads, where 
we chose clusters of length $L_x=N$ and width $L_y=2$ (dashed lines). If we denote the cluster part (local terms and intra-cluster hoppings) 
of the Hamiltonian by $h$ and the inter-cluster hopping terms by $T$, we obtain the decomposition
\begin{equation}
H=h+T. \label{decomp}
\end{equation}
Let us denote the Green's function and self-energy corresponding to $H$ by $G$ and $\Sigma$, and those corresponding to the unperturbed 
Hamiltonian $h$ by $g$ and $\Sigma_h$. 
When these functions are viewed as matrices in intra-cluster orbital and spin indices, the Dyson equation becomes
\begin{equation}
G=g+g(T+\delta\Sigma)G,
\end{equation}
with $\delta\Sigma=\Sigma-\Sigma_h$. In a nonequilibrium calculation, $G$ and $g$ are Keldysh 
Green's functions, and the product is some convolution (depending on whether one works within the 
Kadanoff-Baym formalism, the Floquet formalism, or the Keldysh formalism for steady states).
The CPT approximation neglects $\delta\Sigma$, i.e., we compute an approximate lattice Green's function 
$G_\text{CPT}$ using the Dyson equation 
\begin{equation}
\label{cpt-dyson}
G_\text{CPT}=g+gT G_\text{CPT}.
\end{equation}
Only $\Sigma_h$ is taken into account (in the exact calculation of $g$). 

In a setup with leads like in Fig.~\ref{fig_layer}, 
we can restrict the description to the correlated region, and denote 
the full Green's function and the Green's function 
of the cluster part by $G_c$ and $g_c$, respectively. 
The effect of the noninteracting leads is to add a 
lead selfenergy $\Sigma_\text{leads}=\Sigma_L+\Sigma_R$, with $\Sigma_{L}=T_{Lc}g_{L}T_{cL}$ [cf.~Eq.~(\ref{sigma_diss})] and similarly for $\Sigma_R$ 
(here, $g_\alpha$ denotes the Green's function for lead 
$\alpha$ and $T_{c\alpha}$ the hopping between the correlated region and lead $\alpha$). Hence, we have to solve 
the Dyson equation
\begin{equation}
G_c=g_c+g_c(T_c+\Sigma_\text{leads})G_c,
\end{equation}
for known $g_c$, which can be done using the techniques described in Sec.~\ref{Sec_DMFT}. 

It was shown in \cite{Balzer2011} by comparison to the exact solution that nonequilibrium CPT correctly reproduces the short-time dynamics 
in small lattice systems. The advantages of this approach are that it respects causality, treats correlations exactly within the cluster, and is 
moderate in terms of computational cost. Both the noninteracting limit ($U=0$) and the isolated cluster limit $T=0$ are recovered exactly. 
%%%%%%%%%%
\begin{figure}[t]
\includegraphics[width=7cm]{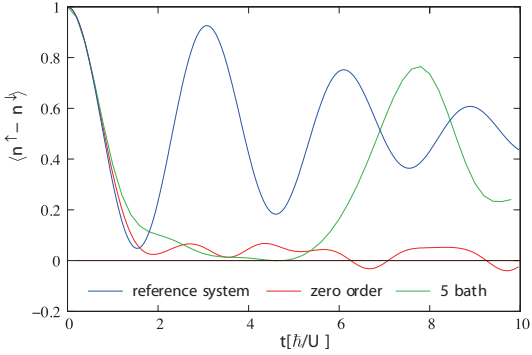}
\caption{Time evolution of the magnetization in a semi-infinite chain with interaction $U=1$ on the first (impurity) site. 
The hopping between the noninteracting bath sites is $t=1$ and the initial state is the impurity, occupied with a spin-up electron, 
decoupled from the bath. The time-evolution is triggered by the sudden switch-on of the hopping between the impurity and the bath. 
The curves show the time-evolution of the spin-polarization in the impurity, computed in a finite chain with five bath sites,
the CPT result for a reference cluster $C'$ that consists of only the impurity and one bath site (``zeroth order''), and for the 
isolated cluster $C'$ (``reference system''). 
(From \onlinecite{Jung2012}.)}
\label{fig_dual_cpt}
\end{figure}
Furthermore, although the dynamics of the exactly solved small cluster shows severe finite-size artifacts, such as persistent
beating oscillations after a perturbation, the Dyson equation (\ref{cpt-dyson}) can restore the relaxation that is characteristic for 
an infinite size system \cite{Jung2012}. This fact is illustrated in Fig.~\ref{fig_dual_cpt}. 

Nevertheless, a main limitation of the approach is the  
limited feed-back to the exactly solved sub-system.  A possibility to partly resolve this issue is to perform a self-consistent 
re-summation  of certain classes of correction terms to the isolated cluster (or single-site) self-energy \cite{Mikelsons2012}, 
in the spirit of a linked cluster expansion around the atomic limit \cite{Metzner1990}. 
As an alternative way to improve the method by including a feedback to the reference system, \onlinecite{Knap2011} proposed a 
nonequilibrium generalization of the variational cluster approach (VCA) \cite{Potthoff2003} for nonequilibrium steady states.  
This generalization exploits the fact that the decomposition (\ref{decomp}) of the Hamiltonian into a cluster contribution $h$ 
and a perturbation $T$ is not unique. We can add arbitrary single-particle terms $\delta h$ to $h$, provided that we subtract 
them from $T$, that is, we can write $H=\tilde h+\tilde T$ with $\tilde h=h+\delta h$ and $\tilde T=T-\delta h$. By optimizing
the parameters of these additional single-particle terms one can hope to achieve a better description of the system. The main 
question is how the optimization should be done in practice. 
 For nonequilibrium steady states, \onlinecite{Knap2011} proposed the following:
let us denote the variational parameters (intra-cluster hoppings and on-site energies) by $p$, and the operators coupled to these 
parameters by $O_p$ ($O_p=d\delta h/dp$). The self-consistency condition which fixes the parameter $p$ demands that the
expectation value of the operators $O_p$ are the same in the unperturbed and in the perturbed state.  
With an infinite number of variational parameters, corresponding to an infinite number of bath sites attached to the cluster, 
the above procedure allows to optimize the system in such a way that the cluster Green's function $g_c$ becomes identical 
to the cluster projection of the full Green's function $G_c$. This is precisely the self-consistency condition which determines 
the bath parameters in the cluster extension of DMFT (cellular DMFT).  
The extension of the corresponding variational principle to time-evolving systems, rather than steady states, has been proposed recently \cite{Hofmann2013}, but has not yet been implemented numerically.

\subsubsection{Cluster extension of nonequilibrium DMFT}
\label{cluster dmft}
%\input{cluster.tex}
%%%%%%%%%%%%%%%%%%%%%%%%%%%%%%%%%%%%%%%%%%
In cluster extensions of DMFT \cite{Maier2005}, one maps the lattice model to
a multi-site cluster embedded in a dynamical mean-field bath, which is self-consistently 
determined. When the number of cluster sites $N_c$ is 1, the formalism reduces to 
the original DMFT. By increasing $N_c$, one can systematically introduce the momentum 
dependence of the self-energy, which has been neglected in DMFT. This allows to address 
the role and importance of spatially nonlocal correlations in the nonequilibrium dynamics of correlated 
systems, especially in low dimensions.

The mapping to the cluster model is not unique unlike the single-site DMFT.
There are two well-established approaches to construct a cluster extension of DMFT, namely
the cellular DMFT \cite{Lichtenstein2000,Kotliar2001} and the dynamical cluster approximation (DCA) 
\cite{Hettler1998,Hettler2000}. They differ in the way the effective cluster problem is constructed. Since 
the nonequilibrium generalization of both methods is straightforward, we briefly review here, 
as an example, the nonequilibrium DCA for the Hubbard model \cite{Tsuji2013cluster}, which enforces translational 
symmetries. The action of the effective cluster problem for DCA is a functional of a hybridization 
function $\Delta_{\sigma}(\bm R; t,t')$,
\begin{align}
S_{\rm clust}[\Delta]
&=
\sum_{\bm R\bm R'\sigma} \int_{\mathcal C}dt\,
(v_{\bm R,\bm R'}-\mu\delta_{\bm R,\bm R'})d_{\bm R\sigma}^\dagger(t)d_{\bm R'\sigma}(t)
\nonumber
\\
&\quad
+
\sum_{\bm R\bm R'\sigma}
\int_{\mathcal{C}} dt \int_{\mathcal{C}} dt' d_{\bm R\sigma}^\dagger(t)
\Delta_{\sigma}(\bm R-\bm R'; t,t') d_{\bm R'\sigma}(t')
\nonumber
\\
&\quad
+U\sum_{\bm R}\int_{\mathcal{C}} dt\, \hat{n}_{\bm R\uparrow}(t)\hat{n}_{\bm R\downarrow}(t),
\end{align}
where $\mathcal{C}$ is the L-shaped contour contour (Sec.~\ref{noneq green function}), and the cluster sites 
are labeled by $\bm R$ and $\bm R'$ with a hopping amplitude $v_{\bm R,\bm R'}$. 
By solving the cluster problem, one obtains the cluster Green's function 
$G_{\sigma}^{\rm clust}(\bm R; t,t')=-i\langle \mathcal{T}_{\mathcal{C}} 
d_{\bm R\sigma}(t)d_{\bm 0\sigma}^\dagger(t')\rangle_{S_{\rm clust}}$.
After Fourier transformation, it is represented in momentum space as $G_{\sigma}^{\rm clust}(\bm K; t,t')$,
where $\bm K$ is a reciprocal vector of $\bm R$. The Brillouin zone is then divided into $N_c$ patches, centered 
at the wave vectors $\bm K$. In the lattice problem, an arbitrary wave vector $\bm k$ is represented by 
$\bm k=\bm K+\tilde{\bm k}$, where $\tilde{\bm k}$ is a wave vector of the superlattice defined by the clusters. 
The cluster problem is constructed such that the cluster Green's function corresponds to the lattice Green's  
function averaged over each patch, i.e., 
\begin{align}
G_{\sigma}^{\rm clust}(\bm K; t,t')
&=
\frac{N_c}{N}\sum_{\tilde{\bm k}} G_{\sigma}^{\rm lat}(\bm K+\tilde{\bm k}; t,t'),
\end{align}
with $N$ the total number of $k$ points in the Brillouin zone. The approximation of the method is to identify 
the lattice self-energy with the cluster self-energy,
\begin{align}
\Sigma_{\sigma}^{\rm lat}(\bm K+\tilde{\bm k}; t,t')
&=
\Sigma_{\sigma}^{\rm clust}(\bm K; t,t'),
\end{align}
i.e., the $\tilde{\bm k}$ dependence of $\Sigma_{\sigma}^{\rm lat}(\bm K+\tilde{\bm k}; t,t')$ is neglected.
With this, the equations for the Green's function and self-energy are closed, and one can obtain the self-consistency 
condition using the lattice Dyson equation,
\begin{align}
(i\partial_t+\mu-\epsilon_{\bm k})G_{\sigma}^{\rm lat}(\bm k)
-\Sigma_{\sigma}^{\rm lat}(\bm k)\ast G_{\sigma}^{\rm lat}(\bm k)
=
\delta_{\mathcal C}(t,t'),
\end{align}
and the cluster Dyson equation,
\begin{align}
&(i\partial_t+\mu-\bar{\epsilon}_{\bm K})G_{\sigma}^{\rm clust}(\bm K)
-\Delta_{\sigma}(\bm K)\ast G_{\sigma}^{\rm clust}(\bm K)
\nonumber
\\
&\quad
-\Sigma_{\sigma}^{\rm clust}(\bm K)\ast G_{\sigma}^{\rm clust}(\bm K)
=\delta_{\mathcal C}(t,t'),
\end{align}
with $\bar{\epsilon}_{\bm K}=(N_c/N)\sum_{\tilde{\bm k}}\epsilon_{\bm K+\tilde{\bm k}}$.

An open issue is how to solve the nonequilibrium cluster problem.
The impurity solvers used for the nonequilibrium DMFT (Sec.~\ref{impurity solver})
are in principle applicable to cluster problems. The CTQMC is a numerically exact method,
but its applicability for cluster problems is very limited since the dynamical sign problem is 
expected to become even more severe than for single-site impurity problems. In the weak-coupling 
regime, the weak-coupling perturbation theory is most promising, since it is easily generalized 
by assigning cluster-site labels $\bm R$ to each interaction vertex, and is computationally 
feasible for large clusters \cite{Tsuji2013cluster}. In the strong-coupling regime, the NCA-type expansion can be 
extended straightforwardly. However, the bare level-propagators $g$ then have a 
$D\times D$ orbital matrix structure, where $D$ is the dimension of the Hilbert space of the 
cluster (respecting symmetries). The memory requirements for $g$ are of the order 
of $\sim D^2 N_t^2$ complex numbers (where $N_t$ is the number of timesteps), such 
that NCA in nonequilibrium is restricted to rather small clusters 
(e.g., $D_{\rm max}= 12$ for $N_c=4$, using spatial symmetries).

\subsubsection{Dual fermions}
%\input{dual.tex}
%%%%%%%%%%%%%%%%%%%%%%%%%%%%%%%%%%%%%%%%%%%
A systematic diagrammatic extension of single-site DMFT, which is also related to the VCA, is the dual-fermion method \cite{Rubtsov2008}. 
The idea here is to represent the lattice model as a collection of impurity models (with the same local interactions as in the lattice), plus 
quadratic terms, and to formulate a systematic expansion in the quadratic terms which provide the coupling between the impurities. The 
extension of this method to nonequilibrium systems has been detailed in \cite{Jung2012}. 

At present, the dual-fermion method has been implemented and tested for the nonequilibrium dynamics of impurity systems, 
within the framework of ``superperturbation theory" \cite{Hafermann2009}. 
In this scheme, one chooses a reference impurity model with a small enough Hilbert space that the impurity Green's functions and vertex functions 
can be computed exactly. As in the VCA case, the parameters of the impurity model may be treated as parameters which can be  
optimized to achieve a better description within a low-order approximation. 
We briefly summarize this formalism, following \onlinecite{Jung2012}. The action of the impurity system is 
\begin{align}
S[d^\dagger, d]
&=
-i\int dt dt' \sum_{a,b} d^\dagger_{a}(t)\Delta_{ab}(t,t')d_b(t')+S_\text{loc}
\nonumber \\
&\equiv
-i\,d^\dagger_{1}\Delta_{12}d_2+S_\text{loc},
\end{align}
where $a$ denotes spin and orbital indices and $\Delta$ is the hybridization function, which is related to the bath Green's function 
$\mathcal G_0$ by (\ref{bath_GF}). In the second line, we introduce subscripts which represent combined spin, orbital and time indices, 
and assume a summation (contour-integration) over repeated indices. 
The idea is to introduce a reference impurity system with a finite number of bath levels, corresponding to the hybridization function $\tilde \Delta$, 
\begin{equation}
\tilde S[d^\dagger, d]=-i\, d^\dagger_{1}\tilde \Delta_{12}d_2+S_\text{loc},
\end{equation}
such that the original action is given by the sum of the action of the reference system and a 
quadratic correction term $i\,d^\dagger_{1}(\tilde \Delta_{12}-\Delta_{12})d_2.$
In order to formulate the perturbation expansion around the reference system, dual fermions $f^\dagger$ and $f$ are 
introduced via a Gaussian integral with coupling term $if^\dagger(t)g^{-1}(t,t')d(t')$, and $g$ the Green's function of the 
reference problem, leading to the action 
\begin{align}
S[d^\dagger, d, f^\dagger, f] =& \tilde S[d^\dagger, d]
+
S^c[d^\dagger, d, f^\dagger, f]\nonumber\\
&+i\,f^\dagger_1[g^{-1}(\tilde \Delta-\Delta)^{-1}g^{-1}]_{12}f_2,
\end{align}
with $S^c[d^\dagger, d, f^\dagger, f]=-if^\dagger_1 g^{-1}_{12}d_2-id^\dagger_1 g^{-1}_{12}f_2$. 
The last step is to integrate out the original $d$-fermions, which leads to the dual action 
\begin{equation}
S^d[f^\dagger, f]=-if_1^\dagger \Delta^d_{12} f_2
-i\frac{1}{4}\gamma^{4}_{1234}f^\dagger_1 f_2 f^\dagger_3 f_4 +\ldots.
\end{equation}
Here, $\gamma^{(4)}$ is the two-particle vertex of the {\it reference system}: 
\begin{equation}
\gamma^{4}_{1234}=g^{-1}_{11'}g^{-1}_{33'}(\chi_{1'2'3'4'}-\chi^0_{1'2'3'4'})g^{-1}_{2'2}g^{-1}_{4'4},
\end{equation}
with $\chi$ and $\chi^0$ the full and disconnected two-particle Green's function of the reference problem. The bare dual Green's function is defined as
$G_0^d=-g[g+(\tilde \Delta-\Delta)^{-1}]^{-1}g$. The partition function of the dual theory may now be expanded in powers of $\gamma$, which leads to the dual self-energy
\begin{equation}
\label{dual1}
\Sigma^d_{12} = -i\gamma^{4}_{1234}(G_0^{d})_{43}+\ldots. 
\end{equation}
Because of the complexity of calculating and storing higher-order vertices, practical implementations of the dual-fermion scheme will likely be restricted to the lowest-order diagram.

Once an approximate solution for the dual Green's function $G^d$ has been obtained, the Green's function $G$ of the original impurity problem is calculated using the relation
\begin{equation}
G=(\tilde \Delta-\Delta)^{-1}+[g(\tilde \Delta-\Delta)]^{-1} G^d [(\tilde \Delta - \Delta)g]^{-1}.
\end{equation}

\begin{figure}[t]
\includegraphics[width=\columnwidth]{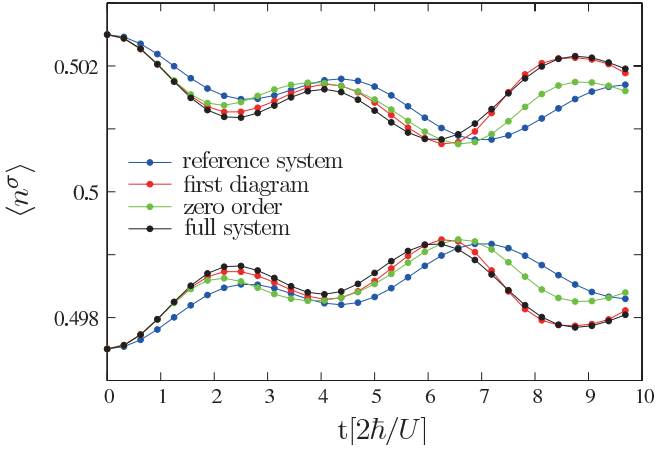}
\caption{Time evolution of the impurity occupation in a two-site system
(one impurity with interaction $U=2$, one noninteracting bath site, hopping $v=0.5$ switched on at $t=0$).
The exact solution is compared to various approximations within the dual fermion
method (see text). (From \onlinecite{Jung2012}.)}
\label{fig_dual}
\end{figure}

A test calculation which illustrates the potential of the dual-fermion method is shown in Fig.~\ref{fig_dual}. 
The impurity system itself is only a two-site problem, consisting of one impurity (interaction $U=2$) coupled
to one bath site. A hopping $v=0.5$ is switched on at time $t=0$. The reference system is the same model with a different hopping ($v'=0.4$, also switched on at $t=0$),
and an expansion is performed in the difference $v-v'$. Taking into account only the first dual diagram 
(\ref{dual1}) leads to a considerable improvement over the zeroth order ($\Sigma^d=0$, which is equivalent 
to VCA), or the solution of the reference problem alone.

\subsubsection{Inhomogeneous DMFT}
\label{sec real space}
%\input{realspace.tex}
%%%%%%%%%%%%%%%%%%%%%%%%%%%%%%%%%%%%%%%%%%%%%%%
An approximate treatment of inhomogeneous systems is possible with the ``inhomogeneous" or ``real-space" extension of  DMFT \cite{Potthoff1999, Freericks2004}, 
in which the self-energy is local, but site-dependent. This technique can be adapted to nonequilibrium situations \cite{Eckstein2013layer,Okamoto2007layer,Okamoto2008layer}
in order to describe, e.g., the nonlinear transport through correlated heterostructures,
the trapping potential in cold-atom experiments, or time-dependent surface phenomena in condensed matter systems, 
such as the propagation of excitations from the surface of a sample into the bulk \cite{Andre2012}. In the most general setup, the two space and 
two time arguments of the Green's function $G_{ij}(t,t')$ cannot be decoupled, neither by introducing momentum-dependent 
Green's functions $G_\kk(t,t')$ (as in homogeneous nonequilibrium DMFT), nor by using frequency-dependent  Green's functions  $G_{ij}(\omega)$ 
(as in inhomogeneous equilibrium DMFT). Inhomogeneous nonequilibrium DMFT simulations thus require a very large amount of memory 
in general. The problem turns out to be numerically tractable for a simpler layered geometry \cite{Potthoff1999, Freericks2004}, in which the properties 
depend on the lattice position in one direction, but are homogeneous in the $d-1$ other dimensions. 

To describe the approach, we consider the model illustrated in Fig.~\ref{fig_layer}, consisting of $N$ correlated layers 
and connected to uncorrelated leads, with inter-layer hopping $t_p$.
(The equations can easily be generalized to a time and layer-dependent inter-layer hopping \cite{Eckstein2013layer}.)
After Fourier transformation within the layers ($y$-direction), we have the following $N\times N$ matrix expression for the 
momentum-dependent Green's function $G_k$:
\begin{equation}
\label{realspace-dyson}
(G^{-1}_k)_{m,n}=(i\partial_t+\mu-\epsilon_{k,m}-\Sigma_m)\delta_{m,n}-t_p(\delta_{m,n+1}+\delta_{m+1,n}),
\end{equation}
where $\epsilon_{k,n}$ is the intra-layer dispersion of layer $n$. 
(Time arguments $t,t'$ on both sides of the equation are omitted for simplicity).
For the inhomogeneous DMFT calculation, we have to compute the local Green's functions $G_n=\frac{1}{N_k}\sum_k (G_k)_{n,n}$ for the different layers, 
and hence we only need the diagonal elements $(G_k)_{n,n}$ of the momentum dependent Green's function.
Because Eq.~(\ref{realspace-dyson}) is essentially the Dyson equation for a linear hopping chain, the diagonal 
elements $(G_k)_{n,n}$ can be obtained recursively, instead of treating the full $N\times N$ matrix problem:
\begin{align}
\label{Gknn}
&(G_k)_{n,n}^{-1}=i\partial_t+\mu-\epsilon_{k,n}
-\Sigma_n-t_p^2 G^{[n]}_{k,n-1}-t_p^2 G^{[n]}_{k,n+1},
\end{align}
where $(G_k^{[n]})$ denotes the Green's function for a chain with site $n$ removed. 
The latter satisfy the Dyson equations
\begin{align}
(G^{[n\pm1]}_{k,n})^{-1}
&=i\partial_t+\mu-\epsilon_{k,n}
-\Sigma_n-t_p^2 G^{[n]}_{k,n\mp 1},
\label{Delta_R}
\end{align}
for $n=1, \ldots, N$.  Boundary conditions must be defined for $G^{[1]}_{k,0}$ (left lead) and $G^{[N]}_{k,N+1}$ (right lead).
(For a free surface, the hopping $t_p$ is set to $0$ between the surface layer and the vacuum.)
Once the $G^{[n]}_{k,n-1}$ and $G^{[n]}_{k,n+1}$ for a given layer $n$ have been updated, one computes $(G_k)_{n,n}$ using 
Eq.~(\ref{Gknn}), and determines the hybridization function $\Delta_n=\Delta_n[G_n]$ by solving the impurity 
Dyson equation
\begin{equation}
G_{n}=\frac{1}{N_k}\sum_{k} (G_k)_{n,n}\equiv 
[i\partial_t+\mu-\Sigma_n-\Delta_n]^{-1}.
\label{def_lambda}
\end{equation}
The hybridization function is the input for the impurity solver, which in turn yields an updated $G_n$ and $\Sigma_n$. 
A detailed description of the nonequilibrium implementation of inhomogeneous DMFT can be found in \cite{Eckstein2013layer}.

%\input{applications}
%%%%%%%%%%%%%%%%%%%%%%%%%%%%%%%%%%%%%%%%%%%%%%%%%%%%%%%%%%%%%%%%%%%%%%%%%%%%%%%%%%
  \section{Applications}
\label{applications}

  In this section, we review nonequilibrium DMFT results for different 
types of nonequilibrium situations. 
On the one hand, for the theoretical investigation of pump-probe spectroscopy
  for solid-state systems, the effect of time-dependent electric fields on the electrons must be determined 
on the femtosecond time scale
  [Sec.~\ref{subsec:electricfields}].
  On the other hand, a parameter in the Hamiltonian might be changed as
  a function of time, either abruptly (``quench'')
  [Sec.~\ref{subsubsec:quenches}] or gradually (``ramp'')
  [Sec.~\ref{subsubsec:ramps}]. These changes are most readily 
  realized in experiments on cold atomic gases in optical
  lattices, which allow precise control of the interaction and hopping
  parameters and can be very well isolated from the environment
  \cite{Bloch2008a}. 

 \subsection{Electric fields}
 \label{subsec:electricfields}
%  \input{electricfields.tex}
%%%%%%%%%%%%%%%%%%%%%%%%%%%%%%%%%%%%%%%%%
\subsubsection{Overview of field-induced phenomena}

Electron systems in strong electric fields raise a broad range of interesting issues, 
from linear and nonlinear transport to the fundamental question whether it is possible 
to control phase transitions by external fields. 
On the experimental side, the effect of an electric field in solids can be studied either by a nonlinear transport measurement or by studing the nonlinear optical response.
For an insulator, one can roughly distinguish the regimes of nonlinear optics (ac response) and nonlinear transport (dc response) as a function of the field strength $F$ and the laser frequency $\Omega$ by the 
{\em Keldysh line} $F_{\rm K}\sim \Omega/\xi$ ($\xi$ is the correlation length that 
characterizes the length scale of an insulator) (Fig.~\ref{fig:efield_ediagram}): Nonlinear 
transport is characterized by electron-hole production due to field-induced quantum tunneling 
across the gap, while in the regime of nonlinear optics one has generally multi-photon absorptions 
and emissions \cite{Oka2012prb}.
In the following, we  briefly describe various electric field-induced phenomena which are 
of interest in strongly correlated systems, and can possibly be investigated using 
nonequlibrium DMFT.

\begin{figure}[tbh]
\centering 
\includegraphics[width=8.5cm]{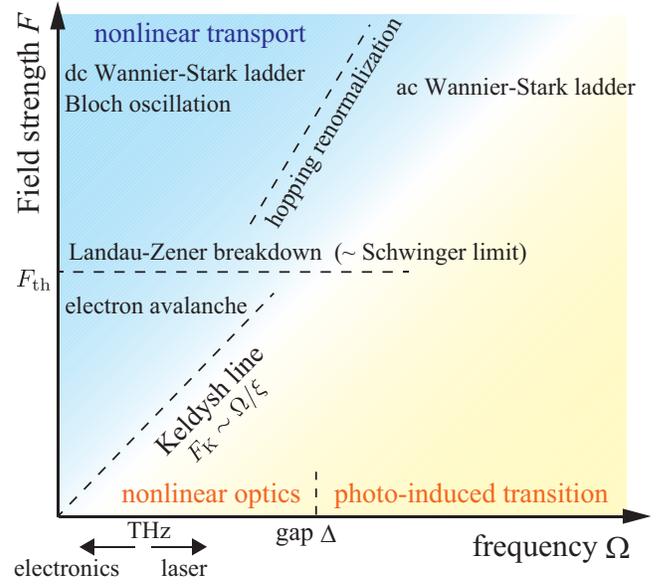}
\caption{Various regimes of electric field induced phenomena
plotted against the field strength $F$ and frequency $\Omega$. 
}
\label{fig:efield_ediagram}
\end{figure}

{\it Non-linear transport}:
Many interesting phenomena have been reported 
on nonlinear transport properties in correlated electron systems, 
such as the colossal electroresistance, which corresponds to a  
large memory effect in the $IV$-characteristics
\cite{Asamitsu1997,Liu2000,Oshima1999,Sawa2004},
the thyristor effect, in which current oscillations emerge
\cite{Sawano2005}, or 
a negative differential resistance, which is observed in many 
correlated insulators \cite{tag,Moriheating,Moriheating2}.

{\it Photo-induced phase transitions}:
Nonequilibrium phase transitions can be 
realized by applying a laser with photon energy exceeding the energy 
gap of a given system 
\cite{YonemitsuNasuPR,NasuPPT,Tokurajpsj06}. 
Since nonequilibrium phase transitions can result in large responses, these 
phenomena are expected to lead to novel devices such as all optical memories. 

{\it Dielectric breakdown of insulators}:
When a dc electric field in an insulator exceeds a critical value, quantum  
tunneling causes  pair-creation of charge carriers.
For a band insulator, this is known as the Landau-Zener breakdown,
which corresponds to the Schwinger effect in non-linear QED.
Another 
possible 
mechanism for the dielectric breakdown is the electron avalanche effect: when the kinetic energy of 
thermal electrons accelerated in an electric field exceeds the pair-creation energy, an exponential 
growth in the carrier density occurs. This was demonstrated in semiconductors irradiated with 
THz lasers \cite{Hirori2011} as well as in a Mott insulator in DC fields \cite{Guiot2013a}.

{\it Non-linear optical responses}:
In the context of correlated electron 
systems, a giant nonlinearity in the optical response
has been reported in 1D Mott insulators \cite{Kishida2000a,Mizuno00}.

{\it Bloch oscillations}:
Bloch oscillations result from coherent periodic motions of 
particles in a lattice system driven by strong electric fields.
In strong electric fields, the electron wave function becomes localized in the direction of the field. 
This phenomenon is called Wannier-Stark localization, and results in a ladder structure in the energy spectrum. 
DMFT allows to study the interplay of interactions and field-induced localization.

{\it Hopping renormalization}:
The hopping parameter is renormalized in strong AC-fields
with a Bessel function, c.f., see Eq.~(\ref{Jeff}). This can lead to a band flipping.

Many of the above-listed strong-field effects have recently been studied with nonequilibrium 
DMFT and in the following subsections, we summarize some key results.

\subsubsection{dc electric fields}
\label{subsec:dcfields}
% \input{DCelectricfields.tex}
%%%%%%%%%%%%%%%%%%%%%%%%%%%%%%%%%%%%%%%%%%
\paragraph{Bloch oscillations}

In the absence of electron scattering, a DC electric field applied to a metallic system will result in an undamped 
oscillating current, a phenomenon known as Bloch oscillations \cite{Bloch1928, Zener1934}. Because the period 
of the Bloch oscillations is typically much longer than the scattering time for experimentally accessible field strengths, 
these oscillations can hardly be observed in metals. On the other hand, they have been studied intensively in 
semiconductor heterostructures \cite{Glueck2002}. Also cold-atom systems \cite{Dahan1996} are ideally suited 
to study this intrinsic nonequilibrium phenomenon, 
and its dependence on the 
underlying lattice structure \cite{Tarruell2012}. 

In a noninteracting tight-binding model, the origin of the oscillations can be understood either as arising from 
a time-dependent
shift of the occupied momentum states in  
${\bm k}$-space with reflections at the Brillouin zone boundary,
or alternatively as a localization of the wave packet in a linear potential gradient. 
An interesting theoretical question is what will happen to this oscillating current if electron-electron scattering is 
taken into account. In particular, one may wonder if and how a dc response 
is established at long times. 
A numerical investigation of Bloch oscillations in the half-filled Falicov-Kimball model (\ref{eq:FK-model}) has been 
undertaken in the pioneering nonequilibrium DMFT papers by Freericks and collaborators 
\cite{Freericks2006, Freericks2008, TurkowskiFreericks2007} (see also \onlinecite{Tran2008}). 
Figure~\ref{fig_bloch_freericks} shows the (rescaled) current induced by a constant electric field $F=1$ (in 
a model with a Gaussian density of states, whose variance sets the unit of energy). This model has a Mott 
transition at $U=\sqrt{2}$, so the curve for $U=0.5$ corresponds to a moderately correlated metal, $U=1.0$ to 
a strongly correlated metal, while $U=1.5$ shows the current induced in a Mott insulating, but nearly critical system. 
One can clearly see the damping of the Bloch oscillations with increasing electron-electron scattering, 
which becomes quite incoherent as one goes across the metal-insulator transition point.
Similar investigations in the metallic phase of the Hubbard model (using second-order weak-coupling perturbation theory and CTQMC as an impurity solver) 
show a sharp crossover between a DC regime, in which the current at long times is given by the linear response 
conductivity, and an AC regime in which Bloch oscillations persist for all times until the system reaches an 
infinite temperature state with zero current \cite{Eckstein2011bloch}.

\begin{figure}[t]
\begin{center}
\includegraphics[angle=0, width=0.9\columnwidth]{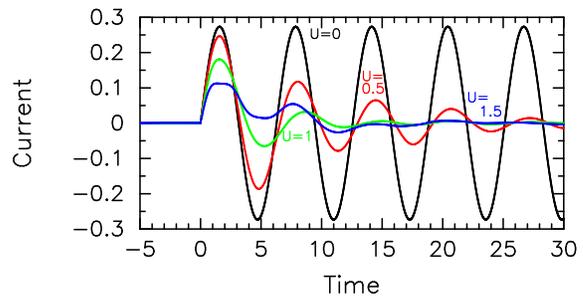}
\caption{
Damped Bloch oscillations in the Falicov-Kimball model with a constant electric 
field $F=1$, and indicated values of the interaction $U$.
(From \onlinecite{Freericks2006}.)}
\label{fig_bloch_freericks}
\end{center}
\end{figure} 

The many-body density of states approaches a steady-state limit characterized by Wannier-Stark 
resonances separated by multiples of the electric field \cite{Joura08,Freericks2008,Tsuji08,FreericksJouraBook,Eckstein2011bloch, Eckstein2012d}.
Figure~\ref{fig_bloch_spectra_freericks} shows results for the Falicov-Kimball model at $U=1$. The Wannier-Stark peaks 
are broadened into bands, whose width is approximately given  by $U$. The central peak is split due to interaction effects, 
which leads to a beating pattern in the time-dependent current \cite{Freericks2008}. The interaction effects are effectively 
enhanced by the electric field, because a steep potential gradient leads to an additional localization of carriers. 
The extreme limit of this localization is the phenomenon of field-induced dimensional reduction, which occurs 
when the electric field along one crystallographic direction is so strong that the potential difference between 
neighboring sites exceeds all other energy scales \cite{Aron11}.

\begin{figure}[t]
\begin{center}
\includegraphics[angle=0, width=0.8\columnwidth]{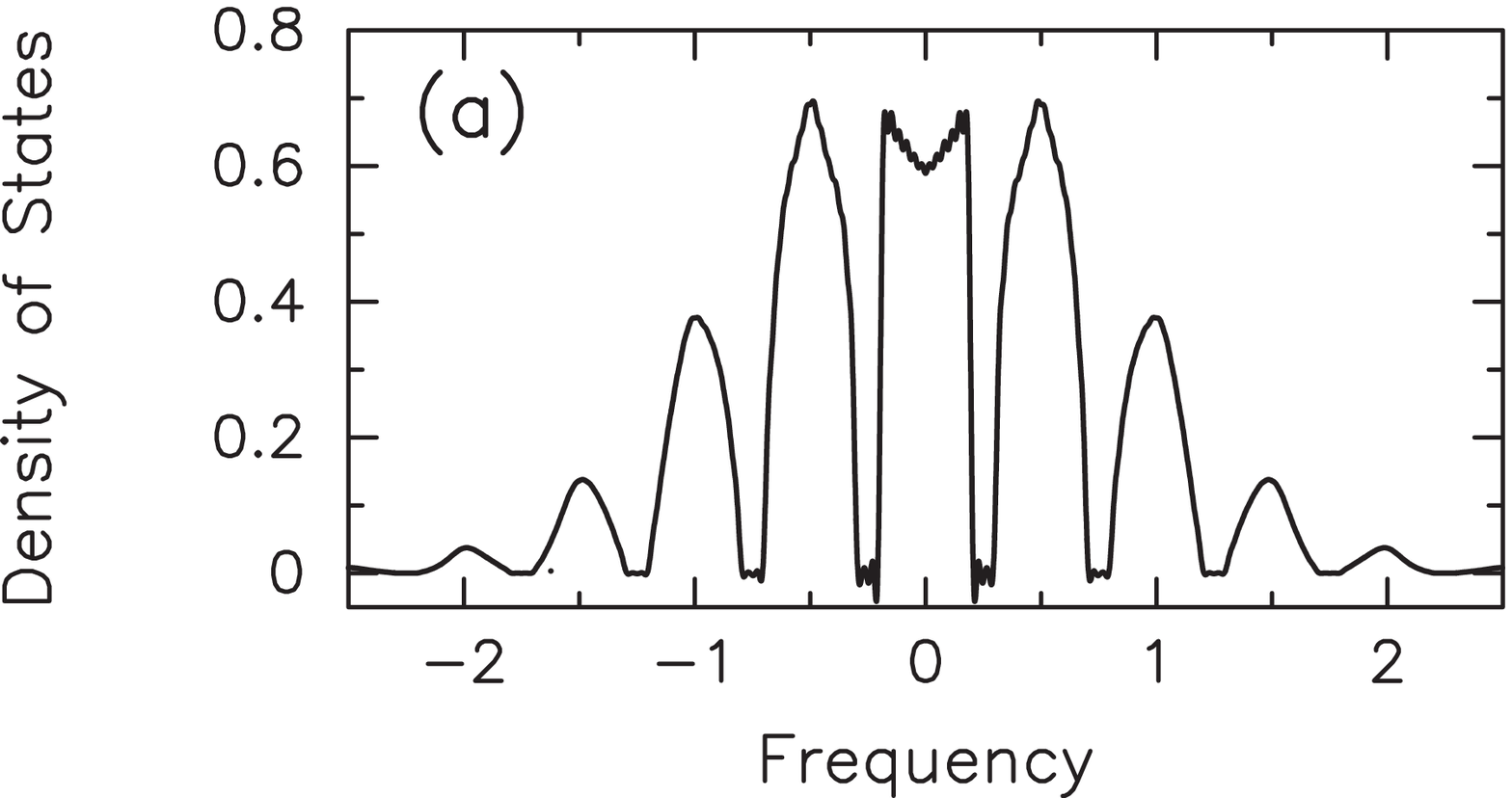}\\
\includegraphics[angle=0, width=0.8\columnwidth]{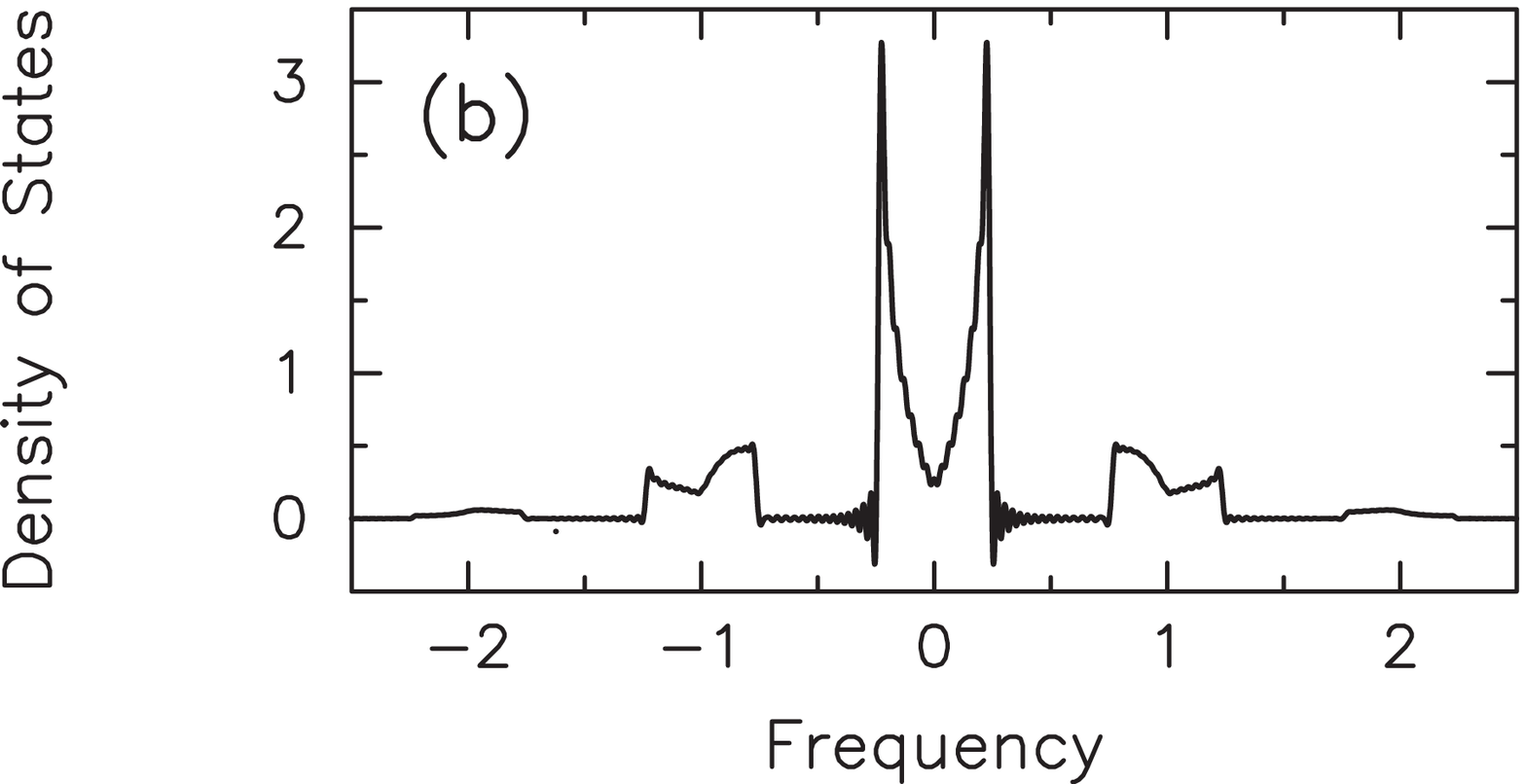}
\caption{
Many-body density of states of the Falicov-Kimball model with $U=0.5$ and a constant electric 
field $F=0.5$ (a) and $F=1$ (b). 
(From \onlinecite{Freericks2008}.)}
\label{fig_bloch_spectra_freericks}
\end{center}
\end{figure} 

%%%%%%%%%%%%%%%%%%%%%%%%%%%
\paragraph{Steady-state current in a dissipative system}

While Bloch oscillations are a typical transient phenomenon, a true stationary state with nonzero current 
in an interacting system can only be reached when the system is coupled to an external heat bath.
(The bath might also be part of the model, as in the situation of a single 
carrier in a many-body background \cite{Mierzejewski11,Golez13}.) 
Otherwise, the Joule heating of the system leads to a time-dependent change in the total energy. 
The dynamics in the DC-driven Hubbard model coupled to a local electron heat-bath of the type described in 
Sec.~\ref{steady-states} was studied by \onlinecite{Amaricci2012}. 
Figure~\ref{fig_current_bath_hubbard} shows the time evolution of the current in a Hubbard model on a square lattice with 
$U=6$ (strongly correlated metal) after a sudden switch-on of a DC field in the diagonal direction. The large initial 
spike in the current is associated with the build-up of a polarization.
Without coupling to the heat-bath, the current eventually decays to zero in an oscillatory manner.
As the coupling $\Lambda$ to the fermionic heat-bath is switched on, 
the current approaches a non-zero stationary value, because
the system relaxes to a nonequilibrium steady state whose momentum distribution is shifted to a position 
at which the electric field driving is balanced with the dissipative effects from interactions and the heat-bath coupling.
The heat-bath thus prevents the electrons from reaching an infinite-temperature distribution,  
so that they contribute to a direct current in the long-time limit. 
For fixed values of $U$ and $\Lambda$, the current is a nonmonotonic function of the field strength: in the weak-field regime
one finds Ohmic behavior ($j\propto F$), whereas in the strong-field regime
the current decreases with increasing $F$, because 
at fixed $\Lambda$ the rate of energy transfer to the bath is limited, while the heat production is proportional 
to the current and to $F$.

\begin{figure}[t]
\begin{center}
\includegraphics[angle=0, width=0.8\columnwidth]{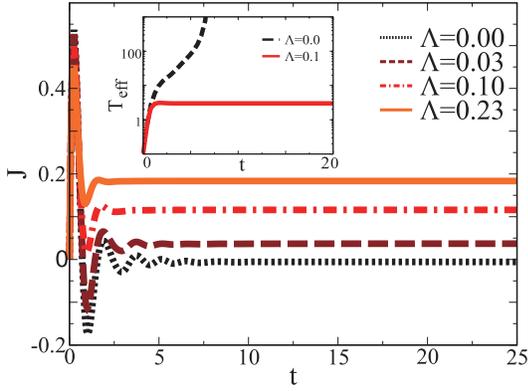}
\caption{
DC field induced current in the Hubbard model with coupling to a thermostat. 
Current for $U=6$ and $F=4.7$ and indicated values of the heat-bath coupling $\Lambda$. 
The field is suddenly switched on at time $t=0$. 
The DMFT impurity problem is solved with 2nd order perturbation theory (IPT).
(inset) time-evolution of the effective temperature 
for $U=6$ and $F=1.9$.
(Adapted from \onlinecite{Amaricci2012}.)}
\label{fig_current_bath_hubbard}
\end{center}
\end{figure}

%%%%%%%%%%%%%%%%%%%%%%%%%%%
\paragraph{Dielectric breakdown}

\begin{figure}[ht]
\centerline{\includegraphics[width=0.63\columnwidth]{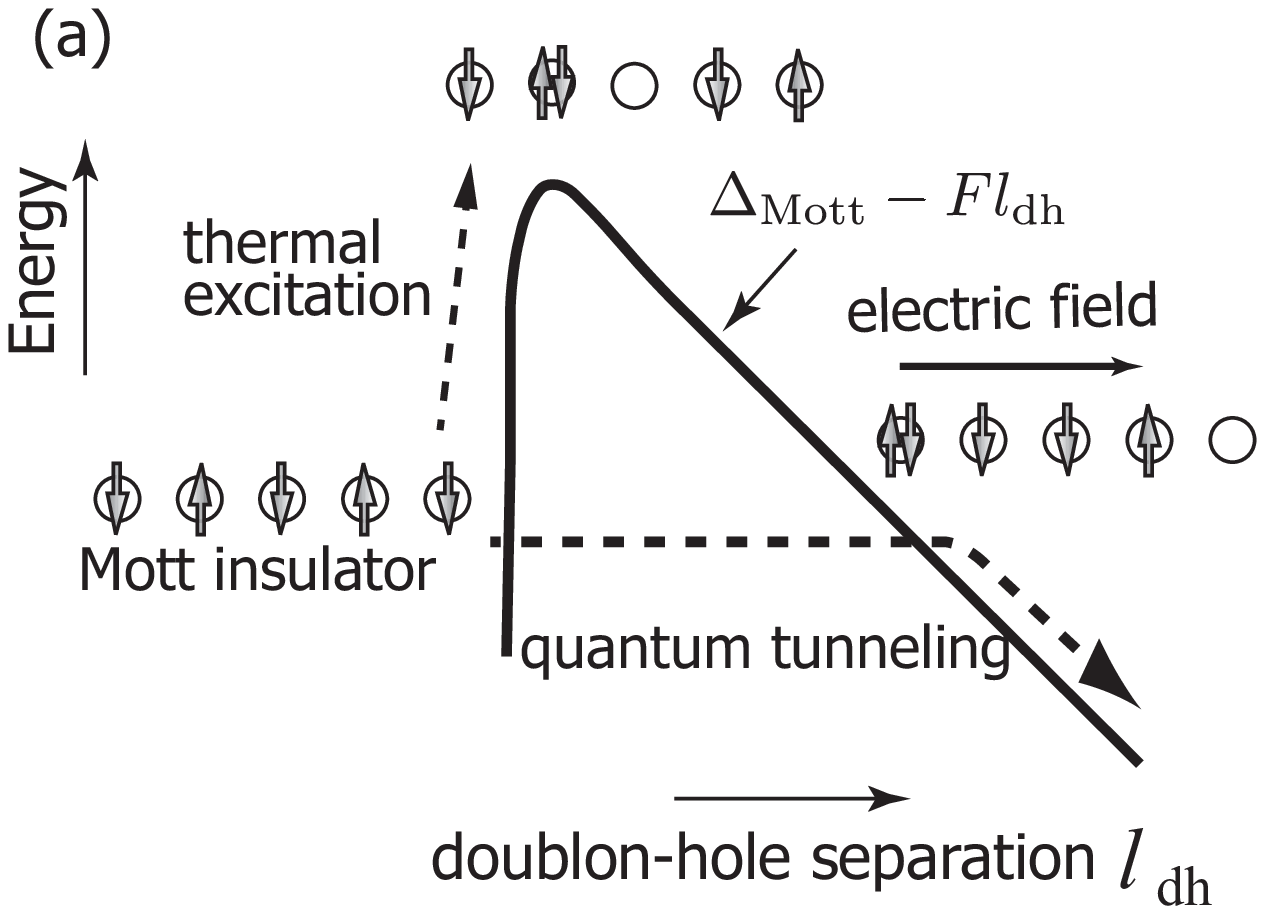}\hspace*{0.05\columnwidth}
\includegraphics[angle=0,bb = 120 0 240 160, width=0.32\columnwidth,clip=true]{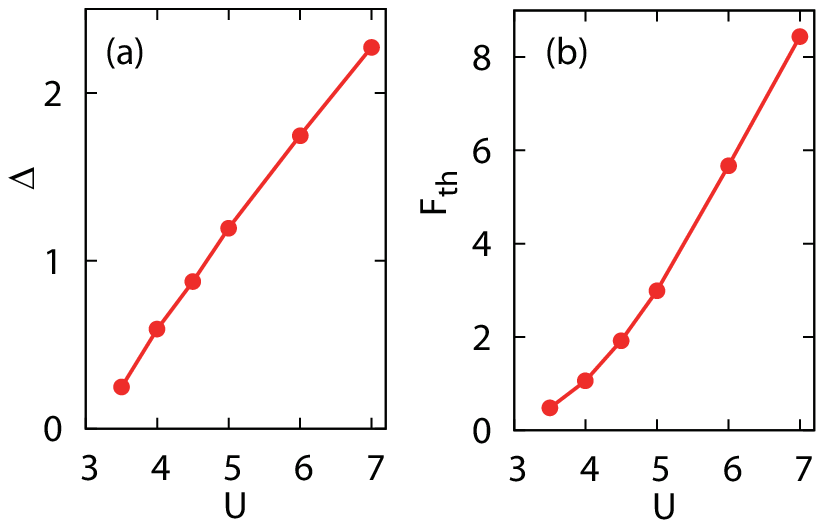}
}
\caption{\label{breakdown_schematic}
Panel (a): Illustration of the dielectric breakdown of a Mott insulator in 
a strong electric field due to many-body quantum tunneling.  
The many-body energy (solid curve) is plotted against 
the separation of doublons and holes, $l_{\rm dh}$.  
For the dielectric breakdown the state has to 
tunnel (dashed line) through an 
energy barrier $\Delta_{\rm Mott}\sim U$ 
related to the creation of charge excitations (doublon-hole pairs).  
Since the electric field
reduces the energy of a pair by $Fl_{\rm dh}$, 
the quantum tunneling among many-body states takes place when  
$\Delta_{\rm Mott}-F l_{\rm dh}\sim 0$, i.e. when $ l_{\rm dh}=\ell_U$.
The tunneling probability depends on the overlap
between the groundstate and excited-state wave functions.  
(Adapted from \onlinecite{okaLMP}.)
Panel (b): The threshold field $F_\text{th}(U)$ from DMFT calculations for the  hypercubic 
lattice with density of states $\rho(E) \propto {\rm exp}(-E^2)$
(OCA results from \onlinecite{Eckstein10}).
}
\end{figure}

The previous examples have focused on correlated metallic systems subject to a DC field. For interactions 
larger than approximately the bandwidth, and at low temperatures, the half-filled Hubbard model is in a Mott 
insulating state. 
A strong electric field $F$ can lead to a metallization of the Mott insulator via a production of doublon-hole 
pairs:
In the scalar potential gauge, the field gives rise to a potential energy difference $F$ 
between two neighboring lattice sites in the direction of the field (in units in which the lattice constant 
$a=1$ and electron charge $e=1$). An electron can thus gain the energy $U=\ell_U F$ from the 
field by tunneling over a distance $\ell_U$,   
which
will produce a doublon and leave a hole behind (Fig.~\ref{breakdown_schematic}a). The tunneling process 
is thus expected to contribute a current 
\begin{equation}
\label{doublon-current}
\jd (t) \sim \ell_U \dot d(t),
\end{equation}
where $\dot d$ is the production rate of doublon-hole (\dh) pairs. 

For $F \ll 1$ the tunneling takes place over many lattice sites, and hence 
at an exponentially small rate.  
The scaling for the electric current is expressed, with a threshold field 
$\fth$, as
\begin{equation}
\label{fth}
j \propto F \exp(
-\fth/F).
\end{equation}
Observables related to the doublon-production rate 
are indeed found to exhibit this threshold behavior (with different powers of $F$ in the pre-factor) 
in studies of the Hubbard model in one dimension \cite{Oka03,Okadmrg05,Heidrich-Meisner2010,Oka2012prb,
Okabethe,Lenarcic2012}, and in infinite dimensions \cite{Eckstein10, Eckstein2012d}.
The steady state properties of
a system coupled to dissipative bath have been studied in Ref.~\cite{Aron12}.
Here we briefly discuss the nonequilibrium DMFT results \cite{Eckstein10, Eckstein2012d}
for the half-filled Hubbard model on an infinite-dimensional hypercubic lattice.

In these calculations, which do not involve a coupling to a thermostat, one observes  a quasi-steady current $j(t)$ 
at long times which is nonzero and strongly field-dependent. The highly nonlinear $j$-$F$ characteristics obtained 
from these quasi-steady values can be well fitted by Eq.~(\ref{fth}), where the threshold field $F_\text{th}$ 
associated with the dielectric breakdown of the Mott insulator extrapolates to zero around the Mott crossover
(see Fig.~\ref{breakdown_schematic}b). (Note that the analysis has been performed above the critical temperature). 

Although the current 
becomes almost stationary at long times, the system is not in a true steady state: energy 
increases at a rate given by $\dot E_\text{tot}(t) = {\bm j}(t) \cdot {\bm F}(t)$, and also the number of doublons 
grows almost linearly with time \cite{Eckstein10, Eckstein2012d}. 
A constant current might be surprising at first sight if doublons are viewed as charge carriers, but the observation has 
a simple interpretation \cite{Eckstein2012d}: One finds that the measured current is given almost entirely  
by the quantum mechanical tunneling current \eqref{fth}. This suggests that in the present of an external 
field both thermally excited carriers and field-induced carriers rapidly reach an infinite temperature state 
(zero kinetic energy), in which their average mobility vanishes. Such a behavior was found for transport 
in various other isolated systems \cite{Mierzejewski2010,Eckstein2011bloch}. For the Mott insulator, the 
picture is confirmed by several findings \cite{Eckstein2012d}, most notably by (i), an analysis of the 
occupation function (which becomes flat in the quasi-steady state), and (ii), the behavior of the 
current in the presence of a thermostat (carriers now maintain a finite temperature, and the current 
thus increases with time proportional to the number of field-excited carriers).
The scenario is also supported by an analytical calculation performed for one-dimensional Hubbard model 
with a nonequilibrium extension of the Bethe ansatz \cite{Okabethe,Oka2012prb}. 

We must mention that the theoretical results obtained from nonequilibrium DMFT and also 1D time-dependent DMRG 
studies of the Hubbard model do not entirely agree with nonlinear transport experiments. In experiments, 
a rather strong temperature dependence is seen in the threshold field \cite{tag}, and a negative differential 
resistance is observed in many correlated  materials \cite{Moriheating,Tokura88}. Also the electron avalanche 
mechanism plays an important role \cite{Guiot2013a}.
The origin of the negative differential resistance is not fully understood yet.
It might be explained by different Joule heating scenarios \cite{Moriheating,Altshuler2009a},
or possibly by a nonequilibrium first-order phase transition, where the negative differential 
resistance is explained through a phase bi-stability \cite{Ajisaka2009}. 
A negative differential resistance is also found in a model in high energy physics, namely the 
supersymmetric QCD in the large-$N$ limit \cite{Nakumura2010,Nakumura2012}. It is hence an 
interesting challenge to develop a microscopic understanding of the nonlinear transport 
properties of correlated systems from a universal viewpoint.

  \subsubsection{Photoexcitations and photodoping}
  \label{subsubsec:photodoping}
%  \input{photodoping.tex}
%%%%%%%%%%%%%%%%%%%%%%%%%%%%%%%%%%%%%%%%%%%%%%%%%%%%%%%
Short laser pulses provide a powerful tool to excite and probe the dynamics of electrons and phonons in correlated materials on the femtosecond 
time scale. For the excitation one mainly uses (i), mid-IR pulses ($\approx 10-100$ THz), which can control the properties 
of complex materials by selectively addressing certain optical phonons \cite{Fausti11,Rini2007a}, (ii), THz pulses, 
which act like static fields on the electron time scale \cite{HiroriAPL2011,Watanabe2011,LiuNat2012},
and (iii), pulses in the eV 
photon-energy range, which can promptly generate electron and hole-like carriers (photo-doping) \cite{Iwai03}. 
Photo-doping in correlated materials can induce, e.g., metal-insulator transitions in Mott and charge-transfer 
insulators, and ultrafast melting of charge and spin order (see~Introduction). 
Previous
studies of photo-induced phase transitions have been performed in 1D using exact diagonalization 
and time-dependent DMRG for the Hubbard model \cite{Takahashi08,OkaTL08}, and for the 
Hubbard-Holstein model \cite{Matsueda} as
well as for a spin-charge coupled system \cite{Matsueda07,Kanamori09,Kanamori11}. Nonequilibrium DMFT can potentially simulate the 
corresponding dynamics in extended higher-dimensional systems, and make predictions for time-resolved 
optical and photo-emission spectroscopy (Sec.~\ref{dmft observables}). So far, DMFT was used to study 
photo-doping and the subsequent relaxation in paramagnetic metals and Mott insulators on short times 
(disregarding lattice dynamics), within the Hubbard model \cite{Eckstein2012c,Eckstein11} and the 
Falicov-Kimball model \cite{Moritz2010,Moritz2012}. Two important questions in this context, which we will briefly 
discuss in the following, concern the validity of the two-temperature model in this regime, and the formation of 
quasi-particles after a photo-induced Mott-insulator to metal transition.

\begin{figure}[t]
\begin{center}
\includegraphics[width=\columnwidth]{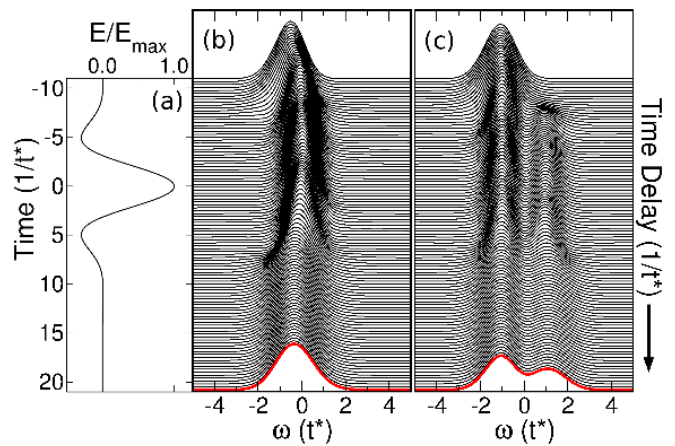}
\caption{ Time-resolved photoemission spectrum of the half-filled, spin-less Falicov-Kimball model, after excitation by a mono-cycle pulse. Panel (a): Pulse shape. Panel (b): metallic system with $U=0.5t^*$. Panel (c): insulating system with $U=2t^*$. (From \onlinecite{Moritz2012}.)}
\label{fig_moritz}
\end{center}
\end{figure}

In simple metals and semiconductors, the photo-excited state can often be understood in terms of the two-temperature 
model of ``hot electrons'' in a colder lattice \cite{Allen1987a}. For this description to work, electron-electron scattering 
must equilibrate the electrons to a quasi-equilibrium state much faster than energy is transferred to the lattice. Thermalization 
in isolated correlated systems, where simple approaches like the Boltzmann equation fail, is a question of fundamental 
interest (see Sec.~\ref{subsubsec:quenches}).
Thermalization after photo-doping is observed in the metallic phase of the Hubbard model \cite{Eckstein11} and to 
excellent accuracy even in the Falicov-Kimball model \cite{Moritz2010,Moritz2012}, where true thermalization of 
single-particle quantities does not occur \cite{Eckstein2008a}. In the Mott phase, on the other hand, the hot-electron 
picture breaks down \cite{Eckstein11,Moritz2010,Moritz2012}. As a result, e.g., photoemission spectra explicitly 
depend on the energy distribution of the excitation pulse, and not only on the total amount of absorbed 
energy \cite{Eckstein11}. 

Figure~\ref{fig_moritz} plots time-resolved spectra of a spin-less Falicov-Kimball model during and after the perturbation by a mono-cycle pulse, as shown in panel (a). In this calculation, it is assumed that the localized particles are uniformly distributed with a density of $0.5$ per site.
The unit of energy is given by the hopping $t^*$ of the conduction electrons, and at half-filling, there is a metal-insulator transition at $U_c=\sqrt{2}t^*$. Panel (b) shows the photoemission response of a metallic system ($U=0.5t^*$). Here, the signal rapidly relaxes back to an almost thermal, but significally broadened spectrum. In the insulating model ($U=2t^*$, panel  (c)), the pulse leads to a significant spectral weight transfer across the gap. While the distribution relaxes within the upper and lower bands, there is no relaxation across the Mott gap and the spectral function is not compatible with a thermal distribution.

\begin{figure}[t]
\begin{center}
\includegraphics[width=\columnwidth]{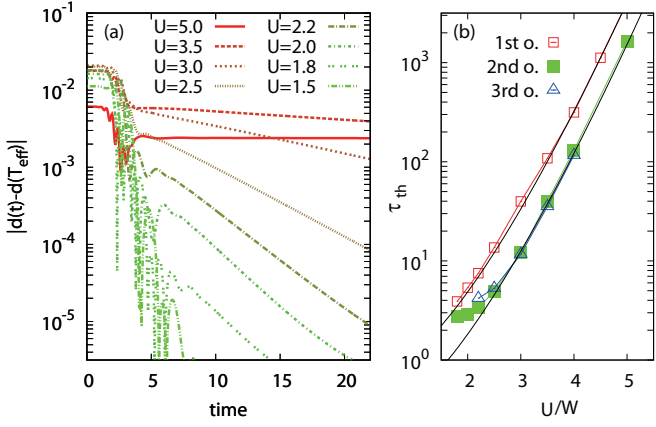}
\caption{(a) Double occupancy in the Hubbard model after photo-doping (hyper-cubic lattice, field along 
the body diagonal of the lattice, see Sec.~\ref{subsubsection:self-consistency}). The system is excited 
with a Gaussian field pulse $F(t) = F_0\cos(\Omega t)e^{-t^2/\Delta t^2}$ (frequency $\Omega=U$). 
The initial temperature is $T=0.2$ (above the critical endpoint of the metal-insulator line), and the field amplitude 
is chosen such that the final effective temperature is $T_\text{eff}=0.5$. (b) Thermalization time, 
obtained from exponential fits to the data in the left panel. Solid lines correspond to the relation
$\tau \propto \exp[ \alpha (U/v)\log(U/v)]$. The strong-coupling expansion 
(see Sec.~\ref{strong-coupling perturbation}) was used to solve the DMFT impurity model, and the results for 
first (NCA), second (OCA) and third order (3rd o.) converge. 
(Adapted from \onlinecite{Eckstein11}.)}
\label{fig docc relax}
\end{center}
\end{figure}

The slow thermalization of a photo-excited Mott-insulator can be related to the long lifetime of 
photo-excited doublons and holes, because changing the interaction 
energy to its quasi-equilibrium value requires the creation or annihilation of doublon-holes pairs. 
This can happen either via the emission of spin excitations (of order $v^2/U$), or by changing 
the kinetic energy (of order $v$) of other charge excitations. In the Hubbard model, for $U\gg v$, 
the process involves many scattering partners and is thus exponentially suppressed \cite{Strohmaier10,
Lenarcic2012preprint}. Figure \ref{fig docc relax}a shows the evolution of the double occupancy $d(t)$ for 
a laser-excited Hubbard model at various $U$ in the metal-insulator crossover range ($U \approx 3$) 
and the Mott insulating phase \cite{Eckstein11}. After an initial transient during and after the pulse, 
$d(t)$ follows an exponential relaxation to a final value $d(T_\text{eff})$, which can be independently 
obtained as the thermal expectation value of $d$ in a system with the same energy as the pump-excited one. 
The time scale extracted from exponential fits to these curves exponentially increases with $U$ 
(Fig.~\ref{fig docc relax}b), in agreement with a Fermi-golden rule type argument \cite{Strohmaier10} 
for high-excitation densities (where the decay to spin-excitations is not important). 

In the Mott regime, where doublons and holes are stable, a long-lived metallic state is created.
Due to its long lifetime up to a few picoseconds, the initially large kinetic energy of photo-exited 
carriers can be dissipated to phonons before recombination occurs. This raises the intriguing 
question whether and on what time scale Fermi-liquid quasi-particles might emerge when the 
kinetic energy is reduced below some ``coherence scale''. A recent DMFT study has investigated
this question by coupling the photo-excited Mott insulator at $U\gg v$ to a dissipative environment 
\cite{Eckstein2012c}. Although a reconstruction of electronic states occurs as the 
kinetic energy is reduced, the formation of a Fermi-liquid like state is not observed on accessible 
time scales.

  \subsubsection{ac electric fields}
  \label{subsubsec:acfields}
%  \input{ac_field.tex}
%%%%%%%%%%%%%%%%%%%%%%%%%%%%%%%%%%%%%%%%%%%%%%%%%%%
\paragraph{General remarks}

In this section, we review some recent nonequilibrium DMFT
studies of correlated systems driven by time-periodic (AC) electric fields.  
In contrast to the pulse excitations discussed in Sec.~\ref{subsubsec:photodoping}, which are designed to induce 
rapid changes of the states, we will consider here time-periodic modulations that allow dynamical, nonequilibrium control of system parameters on a microscopic 
level. 
In real solid-state experiments, one has to use pulsed laser fields in order to attain large field intensities. Even in this case, pulses with many oscillation cycles may often safely be regarded as 
a sinusoidal AC field during irradiation.
In the following we give a brief overview  of theoretical and experimental works on AC-field problems, and then 
move on to nonequilibrium DMFT calculations.
We will distinguish between isolated systems, where the energy accumulates  
in the system, and open systems, where the energy injected by the external field is balanced by the energy flowing out of the system.

\paragraph{Isolated systems}

\begin{figure}[t]
\begin{center}
\includegraphics[width=5cm]{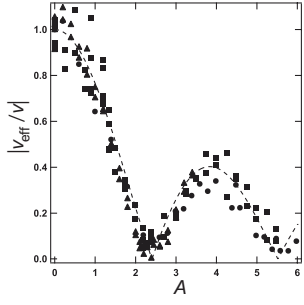}
\caption{Dynamical modification of the hopping amplitude observed for a 
Bose-Einstein condensed cold atoms 
in a periodically shaken optical lattice (from \onlinecite{Lignier2007}).
$A$ is the ratio of the amplitude to the frequency of the driving ac field, and $v_{\rm eff}$ ($v$) is the effective (bare) hopping amplitude of particles.}
\label{dynamical localization}
\end{center}
\end{figure}
An isolated system, in which the total energy and the number of particles are conserved, is ideally realized in cold atom experiments, whereas 
electron systems can
be regarded as 
isolated only on a  
time scale which is short enough that energy dissipation can be ignored.
The effect of AC fields has been theoretically studied 
for a noninteracting tight-binding model \cite{DunlapKenkre1986,Holthaus1992}
in a time periodic electric field $\bm E(t)\equiv \bm E\cos(\Omega t)$.
For simplicity, let us take 
a hypercubic lattice in a field along 
$\bm E=E(1,1,\dots,1)$.
We can then show that the hopping $v_{ij}$ is renormalized 
by the periodic driving 
to an effective hopping \cite{DunlapKenkre1986,Holthaus1992},
\begin{align}
v_{ij}^{\rm eff}=\mathcal{J}_0(A)\,v_{ij},
\label{Jeff}
\end{align}
where $\mathcal{J}_0$ is the zeroth-order Bessel function, and $A=E/\Omega$.
Since $\mathcal{J}_0$ is an oscillating function, 
the effective hopping 
vanishes at zero points, $\mathcal{J}_0(A)=0$,
resulting in immobile particles (dynamical localization).
This effect is analogous to the coherent destruction of tunneling in periodically driven two-level 
systems \cite{Grossmann1991,Grossmann1992}.

The relation (\ref{Jeff}) can be easily understood within the Floquet theory (Sec.~\ref{floquet intro}). 
For this we can take the tight-binding Hamiltonian in the temporal gauge (i.e., in terms of the vector potential, $\bm A(t)=-\bm E\sin(\Omega t)/\Omega$) as 
\begin{align}
H_0(t)=\sum_{\bm k}\epsilon_{\bm k-\bm A(t)} c_{\bm k}^\dagger c_{\bm k},
\label{H0 with ac field A}
\end{align}
where
$\epsilon_{\bm k}=-2 v \sum_{\alpha} \cos k_\alpha$ ($\alpha=x,y,\dots$) 
is the band dispersion. 
The Floquet quasienergy of a single-band tight-binding model is given by 
\cite{Holthaus1992,GrifoniHanggi1998,Tsuji08}
\begin{align}
\langle\!\langle\epsilon_{\bm k}\rangle\!\rangle-n\Omega
\quad (n=0, \pm 1, \pm 2, \dots),
\label{Floquet quasienergy}
\end{align}
where $\langle\!\langle\epsilon_{\bm k}\rangle\!\rangle
=\mathscr{T}^{-1}\int_0^{\mathscr{T}}dt \epsilon_{\bm k-\bm A(t)}$ is the dispersion 
averaged over one period of the ac field (see also Sec.~\ref{floquet-green}).
For the model with nearest-neighbor hopping [Eq.~(\ref{H0 with ac field A})], we end up with 
$\langle\!\langle\epsilon_{\bm k}\rangle\!\rangle=-2v\sum_\alpha\mathcal{J}_0(A)  \cos k_\alpha$, 
hence Eq.~(\ref{Jeff}). The energy (\ref{Floquet quasienergy}) defines
``Floquet quasiparticles'', i.e., the energy dispersion is not only renormalized
due to the coupling to the AC field, but a series of 
``Floquet ($n$-photon dressed) sidebands" appears ($n=0, \pm 1, \cdots$) 
with a spacing $\Omega$. 

The dynamics of systems driven by AC fields has been recently studied in several experiments.
\onlinecite{Lignier2007} used Bose-Einstein condensed (BEC) cold atoms 
trapped in an optical lattice, where the 
AC modulation is induced by shaking the lattice potential in real space.
They observed the dynamical suppression of the absolute value of the 
hopping parameter (Fig.~\ref{dynamical localization}), 
in excellent agreement with the predicted Bessel form (\ref{Jeff}).
The coherent control of single-particle tunneling was also reported
in a driven double-well system \cite{Kierig2008}.
It has been suggested that for interacting systems one can effectively control the 
dimensionless interaction strength $U/W$ ($U$: 
on-site interaction, $W$: bandwidth) using ac fields, and induce a superfluid-Mott insulator phase 
transition in the Bose-Hubbard model \cite{Eckardt2005,Creffield2006,Zenesini2009}. 
One can even reverse the sign of $v_{ij}$  when $\mathcal{J}_0(A)<0$, which 
was used to realize frustrated classical spin systems on a triangular lattice \cite{Struck2011}.

\begin{figure}[t]
\begin{center}
\includegraphics[width=8.5cm]{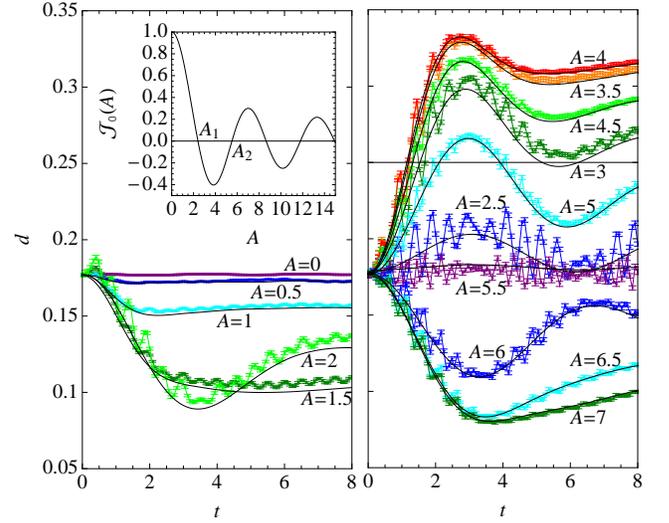}
\caption{(color online). Nonequilibrium DMFT result 
for the time evolution of the double occupancy for the ac-field-driven Hubbard model
(symbols with error bars) with $U=1$, $\Omega=2\pi$, and various values of $A
\equiv E/\Omega$ 
(from \onlinecite{Tsuji11}). Solid curves are the corresponding results
for the interaction quench $U\to U_{\rm eff}=U/\mathcal{J}_0(A)$ with 
time rescaled as $t/|\mathcal{J}_0(A)|$.
The inset shows the Bessel function $\mathcal{J}_0(A)$.}
\label{double occupancy in ac field}
\end{center}
\end{figure}

For fermionic systems, the ac-field problem has been theoretically studied by means of 
the nonequilibrium DMFT for the half-filled Hubbard model [Eqs.~(\ref{dmft_formalism::h-hubbard}) and 
(\ref{dmft_formalism::hint-hubbard})] \cite{Tsuji11}.
Let us consider a hypercubic lattice, with the same 
hopping and field as in Eq.~(\ref{H0 with ac field A}). The system is initially
in thermal equilibrium, and the AC field is suddenly switched on at $t=0$. Figure 
(\ref{double occupancy in ac field}) shows the result for the double occupancy
$d(t)=\langle \hat{n}_\uparrow(t) \hat{n}_\downarrow(t)\rangle$ for various 
values of $A\equiv E/\Omega$ with a 
fixed $\Omega$. Initially $d$ is smaller than the noninteracting value
$\langle \hat{n}_\uparrow \rangle \langle \hat{n}_\downarrow \rangle=0.25$,
due to the repulsive interaction. Switching on an AC field with small amplitude leads to a 
decrease of $d$, accompanied by rapid oscillations with frequency $2\Omega$
due to the nonlinear effect of the ac field. The suppression of $d$ can be interpreted 
as coming from an increased $U/W$ due to the hopping renormalization (\ref{Jeff}).

Remarkably, the double occupancy in Fig.~\ref{double occupancy in ac field} 
{\it exceeds} the noninteracting value of 0.25 in the region where $\mathcal{J}_0(A)<0$ 
(hence $v_{ij}^{\rm eff}<0$). 
This indicates that the many-body interaction indeed turns into an attraction 
[$U_{\rm eff}<0$]. 
This AC-field-induced attractive interaction may lead  
to an $s$-wave superconducting state with high $T_c$.
Note that in equilibrium, the inverted sign of $v_{ij}$ does not change the physics due to the particle-hole symmetry.  
In the present nonequilibrium situation, however, the sign change
in $v_{ij}$ between the initial state and the time-evolving state 
cannot be absorbed by the particle-hole transformation. Physically, it can be interpreted as a dynamical 
band flipping \cite{Tsuji11}. As long as the field is 
ramped up quickly, the occupation in momentum space
does not change significantly, resulting in a population inversion, or so-called ``negative absolute temperature'' ($T_{\rm eff}<0$)
\cite{Ramsey1956,Klein1956} in the flipped band. 
If the system thermalizes to the negative-temperature state, 
the density matrix takes a form of
$\rho(t) \sim e^{-[\mathcal{J}_0(A) H_0+H_{\rm int}]/T_{\rm eff}} 
= e^{-[H_0+H_{\rm int}/\mathcal{J}_0(A)]/[T_{\rm eff}/\mathcal{J}_0(\mathcal{A})]}$.  Hence the AC-quench 
amounts to an interaction quench, 
\begin{align}
U \to U_{\rm eff}=U/\mathcal{J}_0(A).
\label{Ueff}
\end{align}
In other words, a positive (repulsive) $U$ at a negative $T$ 
translates to a negative (attractive) $U$ at a positive $T$.
We can confirm this by comparing the ac-quench results
with those of an interaction-quench calculation (solid curves in 
Fig.~\ref{double occupancy in ac field}), where the interaction parameter is quenched as above.  
For a consistent comparison, time is also rescaled as $t/|\mathcal{J}_0(A)|$
in each interaction-quench simulation.
We can see that the results for the interaction quench and the AC-field driving
agree with each other very well (except for the $2\Omega$ oscillations).

Recently, the negative absolute temperature state has been realized in an experiment
on ultracold bosonic atoms \cite{Braun2013}, following theoretical proposals
\cite{Mosk2005,Rapp2010}.
In real materials, however, it might be difficult to experimentally realize a negative-temperature state using continuous ac fields, 
since a continuous illumination with an intense laser may result in violent heating. 
A different proposal for realizing the negative-temperature state is to use a half-cycle or mono-cycle pulse \cite{TsujiOkaAokiWerner2012}.
If one applies a pulse field $E(t)$ to a noninteracting system, the momentum distribution is shifted 
according to $k\to k+\varphi$, 
where $\varphi=\int_{-\infty}^{\infty} dt\, E(t)$ is the dynamical phase. 
If the momentum shift is about $\pi$, one would have a similar inverted population.
However, in a usual experimental situation Maxwell's equation dictates that $\varphi$ should vanish.  
The nonequilibrium DMFT calculation for the interacting system shows that
with an asymmetrically shaped mono-cycle pulse that satisfies $\varphi=\int_{-\infty}^{\infty} dt\, E(t)=0$, 
the many-body interaction can exert different effects for the 
first half-cycle and the second half-cycle, so that one may end up with 
a nonzero shift with an inverted population \cite{TsujiOkaAokiWerner2012}.
Recently, a similar  phenomenon has been demonstrated in a cold-atom system, where the complex phase of 
the hopping $v_{ij}$ was controlled by a train of sinusoidal pulses separated by a certain waiting time \cite{Struck2012}.

\paragraph{Open systems}

In real materials, the system of interest is usually coupled to an environment with energy and/or particle
dissipation, whose effect should be taken into account in a realistic calculation.
When one continuously drives an open system with
an AC field, the injection of energy from the external field is balanced by the dissipation of energy
into the heat bath, and a nonequilibrium steady state will emerge. This can be thought of as an approximate
description of excited states of materials realized during irradiation 
with a continuous-wave laser or a pulsed laser with many cycles.
Theoretically, the nonequilibrium steady state of strongly correlated fermionic systems driven by AC fields 
has been studied for the Falicov-Kimball model \cite{Tsuji08,Tsuji09} and the Hubbard model 
\cite{LubatschKroha2009} with the nonequilibrium DMFT technique.
The Falicov-Kimball model [Eq.~(\ref{eq:FK-model})] was studied at half-filling with a 
body-diagonal field $\bm A(t)=-\bm E\sin(\Omega t)/\Omega$, 
The heat bath was modeled by free fermions (Sec.~\ref{buttiker}).
One can solve this problem exactly, since the Hamiltonian is quadratic 
in itinerant fermionic operators (Sec. II.C.2).

Figure \ref{fk spectral func}a shows the momentum-resolved single-particle spectral function
averaged over one period of the AC field, $A(\bm k,\omega)$ (``Floquet spectrum''), 
constructed from the gauge invariant Green's function.
A salient feature is that, even in the correlated system, Floquet sidebands appear, on top of the original Mott-Hubbard bands, 
with spacing $\Omega$ [Fig.~\ref{fk spectral func}b].
In particular, for the Mott insulating state with $\Omega<U$, Floquet sidebands penetrate into the original Mott-Hubbard gap, 
generating a ``photoinduced midgap band'' \cite{Tsuji08}.  

\begin{figure}[t]
\begin{center}
\includegraphics[width=8.5cm]{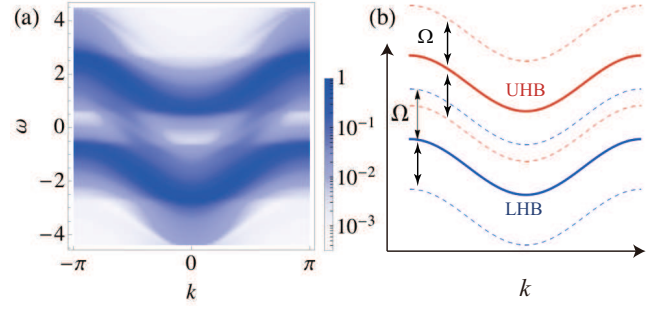}
\caption{(color online). (a) The ``Floquet spectrum'' $A(\bm k,\omega)$
for the AC-field-driven Falicov-Kimball model with $U=3$, $E=0.8$ and $\Omega=1.8$ (from \onlinecite{Tsuji08}).
(b) A schematic band structure in the ac field, where the upper (lower) Hubbard 
band are represented in red (blue), while their Floquet sidebands with a spacing $\Omega$ are shown by dashed lines.
}
\label{fk spectral func}
\end{center}
\end{figure}

The optical conductivity $\sigma(\nu)$ has also been calculated for the AC-field-driven Falicov-Kimball model \cite{Tsuji09}
(Fig.~\ref{fk opt cond}).
As we increase the amplitude $E$, the charge-transfer peak of the Mott insulator at $\nu\sim U$
collapses due to the bleaching effect. 
For $\Omega\lesssim U$ [Fig.~\ref{fk opt cond}b], a broad positive peak appears 
around $\nu\sim 0$, implying that the system is driven into a bad metal state. 
For $\Omega>U$ [Fig.~\ref{fk opt cond}c], a negative optical conductivity appears,
which suggests that the system gains energy from the photon. This comes from 
a partial population inversion within the upper (lower) band of the Mott insulator.
For $\Omega<U$ [Fig.~\ref{fk opt cond}a], a midgap absorption is observed at
$\nu\sim U-\Omega$, which is attributed to an excitation from the Floquet sidebands
to the original band (or vice versa).
Another feature in $\sigma(\nu)$ is that 
there appears kinks [Fig.~\ref{fk opt cond}a] and dips [Fig.~\ref{fk opt cond}b,c]
at $\nu\sim\Omega$. 
By comparing the results with and without vertex-correction for the optical conductivity
(Fig.~\ref{fk opt cond}), 
we can conclude that the vertex correction contributes to $\sigma(\nu)$
significantly around $\nu\sim\Omega$, creating resonance-like spectral structures.
Since there is no such correction in the equilibrium DMFT \cite{Khurana1990},
these features can be considered as a genuine nonequilibrium quantum many-body effect.
\begin{figure}[t]
\begin{center}
\includegraphics[width=7cm]{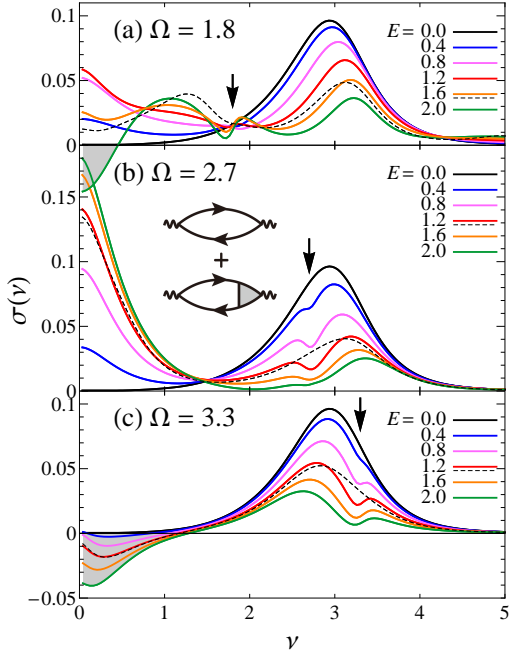}
\caption{(color online). Optical conductivity for the ac-field-driven Falicov-Kimball model
coupled to a fermionic heat bath
with $U=3$, $\Gamma=0.05$, $T=0.05$ and (a) $\Omega=1.8$, (b) 2.7, and (c) 3.3 (from \onlinecite{Tsuji09}).
The dashed curves illustrate (for specific values of $E$) the results without vertex correction.
The arrows indicate the frequency $\Omega$ of the AC field. 
Inset: diagrams of the bubble and vertex correction.}
\label{fk opt cond}
\end{center}
\end{figure}

  \subsection{Time-dependent parameter changes}
  \label{subsec:parameterchanges}
 
  In the following two subsections we discuss two types of parameter
  changes in isolated many-fermion systems: abrupt quenches and
  gradual ramps. For quenches the main interest is in the
  subsequent relaxation of the system, in particular the question how
  the system \emph{relaxes} and whether it \emph{thermalizes}, as
  discussed below.  The energy is typically conserved after the quench
  because the evolution of the isolated system is governed by a time-independent Hamiltonian. For ramps the main
  goal is to understand how the excitation of the system depends on the 
  ramp protocol, e.g., how much energy is transferred to the system
  and how to minimize it.

  \subsubsection{Quenches, relaxation, and thermalization}
  \label{subsubsec:quenches}
%  \input{quenches}
%%%%%%%%%%%%%%%%%%%%%%%%%%%%%%%%%%%%%%%%%%%%%%%%%%
\paragraph{Comparison with the thermal state.}  
Suppose the initial state is given by the density matrix $\rho(0)$ and the time evolution is determined by
  the constant Hamiltonian $H$ after the quench.
  The many-body density matrix
  \begin{align}
    \rho(t)&=e^{-iHt}\rho(0)e^{iHt}
  \end{align}
  will not relax to a stationary limit for long times, but each of its components oscillates forever. On the other 
  hand, quantum-mechanical expectation values of a large system can relax to steady values, because 
  many oscillating components usually contribute to them. This raises the question of whether the
  time-dependent expectation values relax to the thermal expectation
  values as obtained from a standard microcanonical, canonical, or
  grand-canonical ensemble.  If this happens,
  for momentum-integrated as well as momentum-dependent quantities, 
  the system is said to \emph{thermalize}. At first glance it is surprising that this 
  should be possible, as the thermal state depends only on the (constant) mean 
  energy and particle number of the system, whereas the time-evolved state
  may depend on
  details of the initial state. The
  so-called eigenstate thermalization hypothesis (see \citet{Polkovnikov2011RMP} for a review) 
  explains this in terms of the observation that for generic interacting systems, expectation
  values in energy eigenstates usually depend only on the eigenenergy, not on the details 
  of the eigenstate. For integrable systems, on the other hand, a large number of conserved 
  quantities lead to a dependence of expectation values on the individual eigenstates and not 
  just their energy. 
  (Possible criteria for defining integrability in quantum systems have been discussed in \onlinecite{Caux2011a}.)
  As a consequence, integrable systems usually do not thermalize. However, 
  they can often be described by generalized Gibbs ensembles (GGEs) which take the
  conserved constants of motion into account\cite{Jaynes1957b,Jaynes1957c,Rigol2008a}; 
  for reviews see \citet{Polkovnikov2011RMP,Dziarmaga2010a}.
  On the other hand, even small integrability-breaking terms in the
  Hamiltonian can lead to thermalization after sufficiently long times~\cite{Rigol2009t,Rigol2009pra}. %\cite{Rigol2009t,Rigol2010f}.

\paragraph{Interaction quench in the Falicov-Kimball model and the role of conserved quantities}
  
  For an interaction quench 
  in the Falicov-Kimball (FK) model [see Eq.~(\ref{eq:FK-model}) and Sec.~\ref{sec:FKequations}] 
  from $U_-$ to $U_+$ and 
  hopping 
  $v_{ij}$ corresponding to 
  a semielliptical density of states~\eqref{eq:semielliptic} with
  $v_*=1$, the set of equations~(\ref{eq:eqm-d})-(\ref{eq:eqm-b})
  can be 
  solved analytically \cite{Eckstein2008a,Eckstein2008b,Eckstein2008c}.  
  In this setup, a metal-insulator transition occurs at the critical interaction $U_c$ $=$ $2$ for 
  half-filling, $n_c$ $=$ $n_f$ $=$ $\frac{1}{2}$. 
  Time-dependent observables such as the double occupancy $d(t)$ $=$ $-iw_1 \Gweins^<(t,t)$ and the
  momentum distribution $n(\epsilon,t)$ [Eq.~\eqref{neps}] relax
  from their values in the initial state to steady-state values.
  The energy per site $E$ $=$ $ \langle H\rangle/L + \mu n_c$
  is increased by $\Delta E$ $=$ $(U_+-U_-)d(0^-)$ at the quench.
  In general, however, the equilibrium state corresponding to this
  new energy value is not reached by time-evolving the 
  initial state, as discussed below.
  
  \begin{figure}[t]
    \includegraphics[width=0.9\hsize]{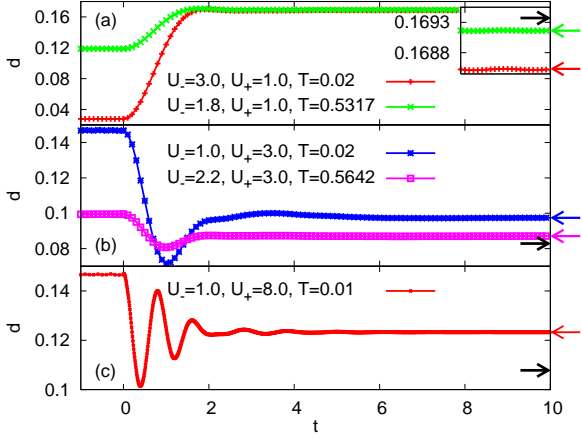}\vspace*{-2mm}
    \caption{Double occupation $d(t)$ 
      for quenches to (a)~$U_+$ $=$ $1$, (b)~$U_+$ $=$ $3$, and (c)~$U_+$
      $=$ $8$, starting from an initial metallic ($U_-<2$) or 
      insulating state ($U_->2$). The half-bandwidth is $2$.
      For (a) and (b) the energy is the same after both quenches, 
      but the long-time limit (left-pointing arrows) is different
      for both and also different
      from the expected thermal value (thick right-pointing arrows).
      The inset in (a) shows a magnification. From~\citet{Eckstein2008a}.
      \label{fig:double}}
  \end{figure}
 Figure~\ref{fig:double} shows the double occupation $d(t)$ for different quenches, both
  within and between the two phases.  The relaxation to a
  new stationary value $d(\infty)$ occurs on the time scale of inverse hopping,
  with damped collapse-and-revival oscillations (Fig.~\ref{fig:double}c)
  after quenches to large interactions.  As
  can be seen from the arrows in Fig.~\ref{fig:double}, the stationary
  value $d(\infty)$ differs from the double occupation in the thermal
  state with the same density and energy. 
  Moreover, the two initial states in Fig.~\ref{fig:double}(a,b) 
  have the same  energy after the quench, and thus the same  thermal state is expected, but the
  steady state maintains a more detailed memory of the initial
  conditions. Thermalization is also lacking for the  momentum occupation
  $n(\epsilon,t)$.

The exact solution of the Falicov-Kimball model after an 
interaction quench allows to make analytic statements about the limit of 
infinite times. One can show for 
$t$ $\to$ $\infty$ that the 
time-dependent occupation function 
$G^<(\omega,t)$ $=$ $\int ds\,e^{i\omega s}G^<(t+s/2,t-s/2)$ 
approaches a steady-state form,
    \begin{align}\label{eq:glpinf}%
      g^<_{\infty} (\omega)
      &=
      2\pi i h(\omega) A_+(\omega)
      \,,
    \end{align}%
  where $A_+(\omega)$ $=$ $-i\,\text{Im}[G^R_+(\omega)]/\pi$ denotes 
  the (temperature-independent) equilibrium spectrum for $U_+$.
Equation \eqref{eq:glpinf} has the same form as the fluctuation-dissipation relation in equilibrium
[cf.~Eq.~\eqref{occupied}], but the Fermi distribution $f(\omega)$ is replaced by a real and positive 
function $h(\omega)$, which can be expressed analytically in terms of the initial-state distribution 
$f(\omega)$ and the equilibrium propagators at $U_+$ and $U_-$ \cite{Eckstein2008a}.
The function $h(\omega)$ determines
steady-state quantities such as the energy, $E(t$ $>$ $0)$ $=$ $\int\,d\omega$ $h(\omega)$
$(\omega+\mu)$ $A_+(\omega)$, the double occupation $d(\infty) $ $=$ $w_1\int\!d\omega$ $h(\omega)$ 
$\text{Im}[\gweins^A_+(\omega)]/\pi$, and the momentum occupation $n(\epsilon,\infty)$ $=$ $\int\!d\omega$
$h(\omega)$ $\text{Im}[(\omega-i0-\epsilon-\Sigma^A_+(\omega))^{-1}]/\pi$.
In general one finds $h(\omega)\neq f(\omega)$, which is
evidence for the absence of thermalization in the Falicov-Kimball model.

This lack of thermalization implies that either the $f$
or the $c$ particles (or both) do not reach their thermal state. The
equilibrium $f$ occupation numbers correspond to annealed disorder,
but in the paramagnetic phase in DMFT the occupation numbers ${\bm
  n_f}\equiv\{f^\dagger_if_i\}$ are independently distributed on all
sites for all temperatures. Therefore the observed nonthermal steady
state must be attributed to the lack of thermalization of the $c$
particles, and indeed, the $c$ Hamiltonian is \emph{quadratic} for given
${\bm n_f}$.
After diagonalization one obtains a set of single-particle states
$|\alpha [{\bm n_f}] \rangle$, and the occupation numbers
$n_{\alpha[{\bm n_f}]}$ are entirely determined by their equilibrium
values before the quench. Nonthermal steady states are therefore
expected. For infinitesimal interaction quenches $\delta U$, a GGE
built from the conserved $n_{\alpha [{\bm n_f}]}$, averaged over ${\bm
  n_f}$ with the statistical weight taken from the initial state,
provides a correct prediction of the final steady state with Green's
function $g^<_\infty(\omega)+ \delta g^<_\infty(\omega)$.  The general
scenario of nonthermal steady states and their description by GGEs,
originally developed for one-dimensional integrable models, thus also
applies to the DMFT solution of the FK model, i.e., for
infinite-dimensional lattices.

  \paragraph{Interaction quench in the Hubbard model, prethermalization and thermalization}

  The Hubbard model [Eqs.~(\ref{dmft_formalism::h-hubbard}) and (\ref{dmft_formalism::hint-hubbard})] 
  at half-filling, with a 
  time-dependent interaction 
  term,  was studied using nonequilibrium DMFT 
  by \citet{Eckstein09,Eckstein2010} 
  for the paramagnetic 
  phase and a semielliptic density of states~\eqref{eq:semielliptic}
  with $v_*=1$.
  The system is prepared in the ground state 
  of the noninteracting Hamiltonian, i.e., $U(t$$<$$0)$ $=$ $0$.  At $t$ $=$ $0$ the Coulomb repulsion is
  switched to a finite value, $U(t$$\ge$$0)$ $=$ $U$. 
The impurity problem is solved with CTQMC,
and observables of the lattice model are computed as described in Sec.~\ref{dmft observables}.
For the noninteracting initial state the weak-coupling method has the advantage that the imaginary 
branch of the contour does not enter the CTQMC calculation, so that even zero-temperature initial 
states can be studied.

  Figure~\ref{fig:quench-hubb-nk} shows the momentum distribution
  \begin{figure*}[bt]
    \begin{center}
      \includegraphics[width=\textwidth]{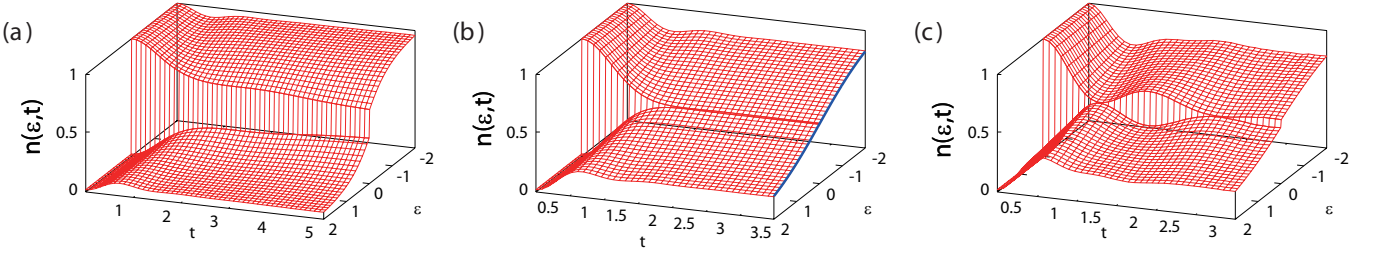}%
    \end{center}
    \caption{Momentum distribution $n(\epsilon,t)$ after an
      interaction quench in the Hubbard model from the noninteracting
      ground state ($U$ $=$ $0$) to interaction (a) $U=2$, (b)
      $U=3.3$, (c) $U=5$; the half-bandwidth is $2$. The solid line at $t=3.5$
      in (b) is the equilibrium expectation value for the momentum
      distribution at the same total energy (temperature $T$ $=$
      $0.84$). Adapted from \citet{Eckstein2010}.}
    \label{fig:quench-hubb-nk}
  \end{figure*}
  $n(\epsilon_\kk,t)=\langle c_{\kk,\sigma}^\dagger(t)c_{\kk,\sigma}(t)\rangle$ 
  for different final values of $U$ as a function of the band energy $\epsilon\equiv\epsilon_\kk$.
  It evolves from a step function for the initial Fermi sea towards a continuous
  function of $\epsilon$. The discontinuity $\Delta n$ at $\epsilon$
  $=$ $0$ does not disappear at once, rather it remains sharp 
    while its size decreases with time. 
  This 
  decrease is directly 
  related to the
  decay of electron and hole excitations which are created at time $t$
  $=$ $0$ at the Fermi surface. This follows from the equation
  \begin{align}
    \label{jump}
    \Delta n(t) = n(0^-,t)-n(0^+,t) =  |G_{\epsilon=0,\sigma}^R(t,0)|^2
    \,,
  \end{align}
  (valid for a noninteracting initial state at half-filling),
  where $G_{\epsilon_{\BK},\sigma}^R(t,0)$ 
  is the retarded component of the momentum-resolved Green's function.
 The double occupation $d$ and the discontinuity $\Delta n$ are shown as 
 a function of the time after the quench in  Fig.~\ref{fig:quench-hubb-double}.
 
  \begin{figure}[b]
    \includegraphics[width=0.95\hsize]{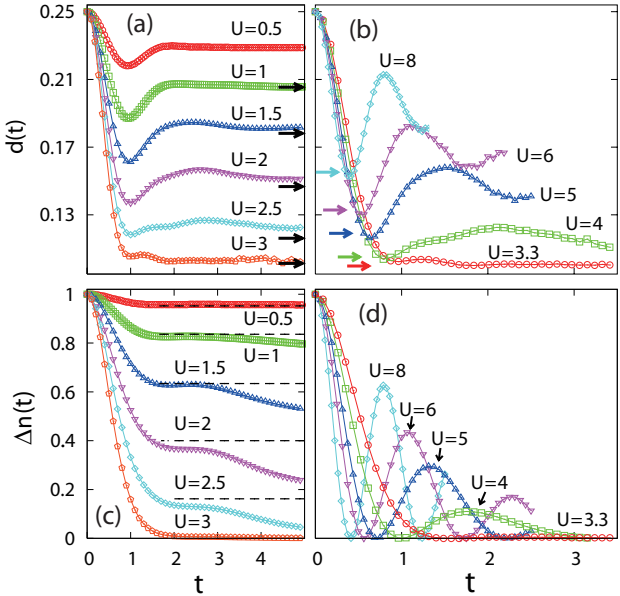}
    \caption{Fermi surface discontinuity $\Delta n$ and double
      occupation $d(t)$ after quenches to $U\le3$ (left panels) and
      $U\ge 3.5$ (right panels).  Horizontal dotted lines in panel (c)
      are the prethermalization plateaus predicted by
      \citet{Moeckel2008a}. Horizontal arrows indicate thermal values
      of the double occupation. Adapted from \citet{Eckstein09}.}
    \label{fig:quench-hubb-double}
  \end{figure}
Three different regimes are apparent in the relaxation 
after the interaction quench: small and large values of $U$, separated 
by a sharp crossover or transition near the intermediate 
scale $U$ $\approx$ $3.2$ $=$ $U_c^{\text{dyn}}$.
Near $U$ $=$ $U_c^{\text{dyn}}$, the momentum distribution quickly  relaxes to the thermal distribution
   for all energies $\epsilon$ 
  (solid line in Fig.~\ref{fig:quench-hubb-nk}b, obtained from a
  grand-canonical DMFT equilibrium calculation for the temperature that
  gives the same total energy $E$),
  and thermalization on the same time scale of $\approx2$
  is observed for the double occupancy (thick arrows in \ref{fig:quench-hubb-double}a,b),
  as well as for dynamical observables like the retarded Green's function $G^R(t+s,t)$
  (as a function of time difference $s$) and the two-time optical conductivity
  $\sigma(t,t+s)$ \cite{Eckstein2010}.
After the relaxation it is thus appropriate to regard the system as thermalized, establishing the
infinite-dimensional Hubbard model as one of the few isolated quantum many-body systems 
for which thermalization can be demonstrated (see \citet{Polkovnikov2011RMP} for other examples).

For quenches to interactions $U$ above and below $U_c^{\text{dyn}}$
the system does not relax directly to a thermal state. Rather, metastable
states are observed on intermediate time scales.  For quenches to
weak coupling, $U$ $\le$ $3$, the double occupation $d(t)$ relaxes
from its initial uncorrelated value $d(0)$ $=$ $1/4$ almost to its
thermal value $d_{\text{th}}$, whereas the Fermi surface
discontinuity $\Delta n(t)$ approaches a quasistationary value and remains
finite for $t$ $\leq$ $5$.  This so-called prethermalization
\cite{Berges2004a} was predicted for a quenched Fermi liquid by
\citet{Moeckel2008a} on the basis of a weak-coupling calculation: while the kinetic
and potential energy thermalize on time scales
$1/U^2$, the Fermi surface continuity only reaches the plateau
$\Delta n_{\text{stat}}$ $=$ $1-2Z$, where $Z$ is the quasiparticle
weight in equilibrium at zero temperature. The momentum
occupations are then redistributed as the thermal state is
approached. 
  In the limit of infinite dimensions for a half-filled symmetric band
  the weak-coupling result of \citet{Moeckel2008a} for the transient
  towards the prethermalization plateau 
   describe the transient behavior and the prethermalization plateau
  well for  $U$ $\lesssim$ $1.5$  \cite{Eckstein09}, even though 
  at the  larger $U$ values the time scales $1/U^2$ and $1/U^4$
  are no longer well separated.
  Prethermalization plateaus 
after an interaction quench are also correctly predicted by a GGE
built from approximate constants of motion \cite{Kollar2011a}. 
Recently a quantum kinetic equation was used to describe the
subsequent crossover from the prethermalization plateau to the thermal
state~\cite{Stark2013}.

  For quenches to strong coupling ($U$ $\ge$ $3.3$ in
  Fig.~\ref{fig:quench-hubb-double}b,d), the relaxation exhibits damped collapse
  and revival oscillations of approximate periodicity $2\pi/U$, due to
  the exact periodicity of the propagator $e^{-iHt}$ without hopping 
  \cite{Greiner2002b}. For large values of $U$ both $d(t)$
  and  $n(\epsilon,t)$ oscillate around nonthermal values. Using
  strong-coupling perturbation 
  theory,~\citet{Eckstein09} 
  showed that the mean
  value of $d(t)$ for these oscillations is $d_{\text{stat}}$ $=$
  $d(0)$ $-$ $\Delta d$
  with $\Delta d$ $=$
  $(1/2U)\expval{H_{\text{kin}}/L}_{t=0}$ (for the quench from $U=0$,
  in infinite dimensions at
  half-filling). The thermal value is obtained
  from a high-temperature expansion as $d_{\text{th}}$ $=$ $d(0)$
  + $(1/U)\expval{H_{\text{kin}}/L}_0$. Hence during the initial stage
  of the relaxation the double occupation relaxes only halfway towards
  $d_{\text{th}}$. Although longer times cannot be accessed with the
  weak-coupling CTQMC method, a relaxation to the thermal state is
  expected after the oscillations have decayed, as in the case of a pump-excited
  Mott insulator \cite{Eckstein11}. In general this crossover will set
  in only on times scales that are exponentially large in the
  interaction $U$ \cite{Sensarma2010a}; see also Sec.~\ref{subsubsec:photodoping}.

  The rapid thermalization at $U\approx U_c^{\text{dyn}}$ occurs at the 
  border between the delayed thermalization either due to
  weak-coupling prethermalization plateaus or strong-coupling
  oscillations around nonthermal values. Indeed, no finite width was
  detected for the width of this crossover region, so that a dynamical
  phase transition might occur at $U_c^{\text{dyn}}$. This sharp crossover was
  unexpected because the corresponding equilibrium temperature $T_\text{eff}$ after
  the quench is much higher than the critical endpoint of the Mott
  metal-insulator transition in equilibrium ($T_c \approx
  0.055$~\cite{Georges96}, but $T_\text{eff}$ $=$ $0.84$ for $U$ $=$ $3.3$).
  Interestingly, a good approximation for the
  critical interaction, $U_c^{\text{dyn}}\approx3.4$,  is obtained from a time-dependent
  variational theory using the Gutzwiller approximation
  \cite{Schiro10,Schiro11}. A similar strong dependence on the quenched interaction was observed
  in Heisenberg chains \cite{Barmettler2008a}. Several possible origins for
  nonequilibrium phase transitions have been proposed
  \cite{Sciolla2010a,Gambassi2011a,Heyl2012a,Karrasch2013,Hamerla2013}.

%\input{afm}
%%%%%%%%%%%%%%%%%%%%%%%%%%%%%%%%%%%%%%%%%%%%%%%%%%%
 \paragraph{Interaction quench in the presence of long-range order}

Correlated lattice systems exhibit various types of long-range order including antiferromagnetism, superconductivity, and charge order,
in the presence of which the relaxation behavior after an interaction quench changes qualitatively. A symmetry-broken state
on a bipartite lattice can be treated within DMFT by solving impurity problems for each sublattice \cite{Georges96}. 
For the antiferromagnetic phase and the semielliptic 
density of states (\ref{eq:semielliptic}), the hybridization function $\Delta_{A,\sigma}$ ($\Delta_{B,\sigma}$) for the $A$ ($B$) sublattice is given
by the self-consistency condition $\Delta_{A,\sigma}=v_*^2G_{B,\sigma}$ ($\Delta_{B,\sigma}=v_*^2G_{A,\sigma}$), where $G$ is the local
lattice Green's function. Together with the relation $\Delta_{A,\sigma}=\Delta_{B,\bar\sigma}$ (for pure N\'eel-type symmetry breaking), this leads
to a single impurity calculation with the self-consistency $\Delta_\sigma=v_*^2G_{\bar\sigma}$.

\begin{figure}[t]
\begin{center}
\includegraphics[width=\columnwidth]{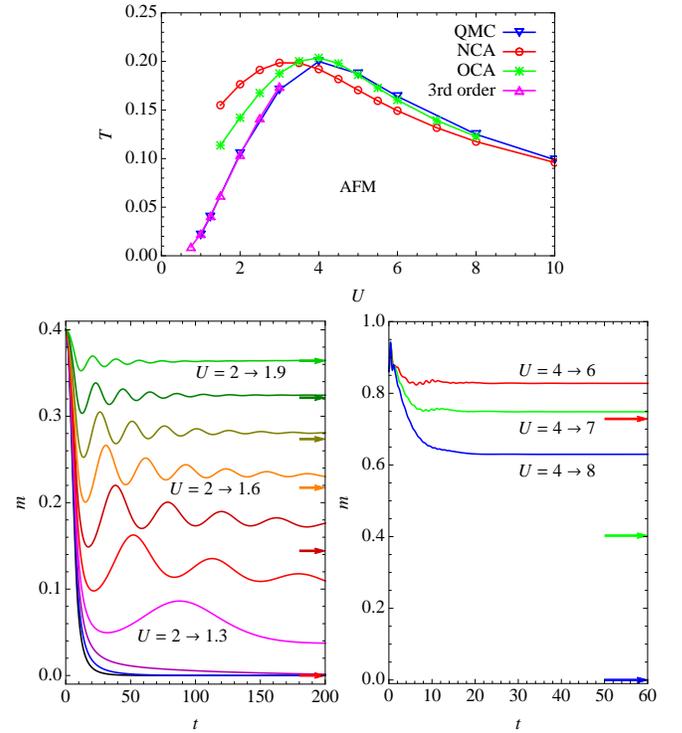}
\caption{
Top panel: Antiferromagnetic phase diagram for the half-filled Hubbard model (semi-elliptic DOS, bandwidth 4). 
The QMC data are taken from \cite{Koga2011}, the NCA and OCA phase-boundaries from \cite{Werner2012afm},
and the third-order weak-coupling perturbation results from \citet{Tsuji2012}. The time evolution of the order parameter (staggered magnetization) 
is shown for quenches $U=2\to 1.9, 1.8,\dots, 1.0$ [bottom left, from \citet{Tsuji2012}], and for quenches $U=4\to 6, 7, 8$ [bottom right, from \cite{Werner2012afm}]. The arrows indicate the corresponding thermal values of the order parameter reached in the long-time limit.
}
\label{afm_phasediagram}
\end{center}
\end{figure}

The DMFT phase diagram for the half-filled, repulsive Hubbard model exhibits an 
antiferromagnetically ordered insulating phase at low temperature (denoted by AFM in the top panel of Fig.~\ref{afm_phasediagram}). 
For attractive $U$, one finds an analogous phase diagram with AFM order replaced by $s$-wave 
superconductivity \cite{Micnas1990,Keller2001} (at half-filling the superconducting state is degenerate with a charge ordered phase
due to a symmetry between the repulsive and attractive models \cite{Shiba1972}, but 
in the doped system, superconductivity is more stable). The nature of the AFM insulating (or $s$-wave superconducting) 
state changes qualitatively as $|U|$ crosses the value corresponding roughly to the maximum in the critical temperature. 
This is known as the ``BCS-BEC" crossover in the literature on cold atomic gases.
Besides the exact CTQMC result \cite{Koga2011}, Fig.~\ref{afm_phasediagram} shows the phase boundaries obtained using 
the third-order weak-coupling perturbation theory (\ref{weak-coupling perturbation}), as well as the first- and second-order strong-coupling perturbation theory (NCA and OCA) (\ref{strong-coupling perturbation}) as an impurity solver. These solvers have been used in \cite{Tsuji2012} and \cite{Werner2012afm} to compute the dynamics of the order parameter after interaction quenches from the AFM insulating phase into the paramagnetic metallic phase. 

In the bottom left panel of Fig.~\ref{afm_phasediagram}, we plot the time evolution of the magnetization $m$ after quenches from $U_i=2$ to $U_f=1.9, 1.8, \dots, 1.0$ \cite{Tsuji2012}. These  nonequilibrium DMFT calculations are based on third-order weak-coupling perturbation theory (\ref{weak-coupling perturbation}). The arrows indicate the values of the magnetization expected for the thermalized state in the long-time limit. After the quench, the order parameter shows coherent oscillations (``amplitude mode''), followed by a slow decay. A remarkable observation is that
even though the system, after quenches to $U_f<1.5$, is highly excited and will eventually relax to a high-temperature paramagnetic state, the magnetization oscillates for a long time around a non-zero value. This nonthermal magnetized state persists up to a ``nonthermal critical point'',
where the frequency of the amplitude mode vanishes, and the dephasing time constant diverges \cite{Tsuji2012}.
The behavior of the system in the trapped state is similar to the Hartree solution,
which is mathematically equivalent to an integrable BCS equation \cite{Barankov2006,Yuzbashyan2006}, so that 
the system may be considered as evolving in the vicinity of the nonthermal ``Hartree'' fixed point. 
The slow relaxation of the nonthermal order is followed by a faster thermalization process.

Trapping phenomena of a different origin are found in quenches to large $U$.
The bottom right panel of Fig.~\ref{afm_phasediagram} shows the magnetization for quenches from $U_i=4$ to $U_f=6$, $7$ and $8$,
obtained from the nonequilibrium DMFT with the NCA impurity solver \cite{Werner2012afm}. 
Again the magnetization does not immediately decay to zero after a quench to $U=8$, but remains trapped at a remarkably large value.
The state after such a quench is similar to a photo-doped state, and the trapping of the magnetization is linked to the exponentially long life time
($\sim Ae^{\alpha(U/2)\log(U/2)}$) of artificially created doublons in a Mott insulator with large gap (see also Sec.~\ref{subsubsec:photodoping}). If the density of frozen-in (or ``photo-doped") carriers is larger than some critical value, the trapping disappears, and the magnetization relaxes to zero exponentially, with a relaxation time which depends like a power law on the distance from the trapped phase \cite{Werner2012afm}.

  \subsubsection{Ramps and nonadiabaticity}
  \label{subsubsec:ramps}
%  \input{ramps}
%%%%%%%%%%%%%%%%%%%%%%%%%%%%%%%%%%%%%%%%%%%%%%%%%%%%%
  \paragraph{Excitation energy after a continuous parameter change}

  The intention in the study of continuous parameter changes (``ramps'') is often to excite an
  isolated quantum system as little as possible. The adiabatic theorem
  of quantum mechanics~\cite{Born1928,Kato1950,Avron1999} 
  states that in the limit of infinitely slow changes the system
  (described by a pure state $\rho(t)$ $=$ $\ket{\psi(t)}\bra{\psi(t)}$) remains in the ground
  state of the Hamiltonian $H(t)$. However, for ramps that take place in a
  finite time, $0\le t \le \tau$, the excitation energy, defined as 
  \begin{equation}
    \label{exe}
    \Delta E(\tau) = E(\tau) - E_0(\tau),
  \end{equation}
  is positive in general, and can only be zero if the final
  time-evolved state $\ket{\psi(\tau)}$ is a ground state of the
  final Hamiltonian $H(\tau)$. Here $E(t)$ $=$
  $\bra{\psi(t)}H(t)\ket{\psi(t)}$ is the energy at time $t$ and
  $E_0(t)$ is the ground-state energy of $H(t)$. Note that the
  energy $E(t)$ is thus always well-defined, in contrast to
  the entropy, which is defined only for equilibrium states.  In
  equilibrium thermodynamics, the entropy remains constant during
  adiabatic processes (defined as quasistatic processes
  without heat exchange with the environment), but increases for
  processes that take place in a finite time and are hence no 
  longer reversible.  The simplest nontrivial quantum system
  that illustrates the crossover from adiabatic to nonadiabatic
  behavior is the exactly solvable Landau-Zener model
  \cite{Landau1932a,Zener1932a}: a two-level system $H_{\text{LZ}}(t)$ $=$ $vt
  \sigma_z + \gamma \sigma_x$ that is driven through an avoided level
  crossing with finite speed $v>0$ ($\sigma_i$ denote the Pauli matrices). If
  the system is in the ground state $\ket{\phi_0(-\infty)}$ $=$
  $\binom{1}{0}$ at time $t$ $=$ $-\infty$, the probability $p$ to
  find the system in the excited state $\ket{\phi_1(\infty)}$ $=$
  $\binom{1}{0}$ at time $t$ $\to$ $\infty$ vanishes exponentially
  when the speed $v$ is small compared to the scale $\gamma^2/\hbar$
  set by the gap $\gamma$ at the avoided crossing, $p$ $\sim$
  $\exp(-\pi\gamma^2/v\hbar)$. This prediction was recently confirmed
  in cold-atom experiments with accelerated optical lattices~\cite{Zenesini2009a}. 
  An analogous
  mechanism
  also explains the amount of energy injected into a system upon crossing a
  quantum-critical point or parameter changes in a gapless
  phase~\cite{Dziarmaga2010a}, as obtained from adiabatic perturbation
  theory~\cite{Polkovnikov2010springer}. In these cases the excitation
  energy typically behaves as $\Delta E(\tau)\sim\tau^{-\eta}$ for
  large ramp times, $\tau$ $\to$ $\infty$, with the positive
  exponent $\eta$ depending on the details of the system and the ramp
  protocol.
  %%%
  The behavior
  can be understood
  in terms of the Kibble-Zurek mechanism~\cite{Dziarmaga2010a}: The
  dependence of $\Delta E$ on $\tau$ is due to excitations that are
  `frozen in' when the rate of change of the Hamiltonian exceeds the fastest possible relaxation rate of the system
  (estimated by the energy gap divided by $\hbar$). 

  As a generic case, suppose a system is in the ground state of a Hamiltonian $H_0$ at
  time $t$ $=$ $0$ and a ramp protocol is given by the Hamiltonian $H(t)$ $=$
  $H_0+\delta\kappa r(t/\tau)W$. Here the ramp shape $r(t)$  starts at
  $r(0)$ $=$ 0 and ends  at $r(1)=1$, i.e., after the ramp time $\tau$
  the operator $\delta\kappa W$ has been added to the Hamiltonian $H_0$,
  where $\delta\kappa$ determines the strength of the perturbation.
  If no phase boundary is crossed by this ramp and the magnitude of
  the ramp is small, the excitation energy can be estimated from
  lowest-order adiabatic  
  perturbation theory (APT) as~\cite{Eckstein2010ramps},
  \begin{align}
    \label{fgr-exe}
    \Delta E(\tau)
    &=
    \delta \kappa^2
    \int\limits_0^\infty \frac{d\omega}{\omega}\,
    R(\omega) F(\omega\tau)
    + \mathcal{O}(\delta \kappa^3),
    \\
    \label{fgr-r}
    R(\omega)
    &=
    \frac{1}{L}\sum_{n \neq 0}
    \big|\expval{\phi_n |W| \phi_0}\big|^2
    \,\delta(\omega-\epsilon_{n0}),
    \\
    \label{adiff}
    F(x)
    &=
    \left| \int_0^1  \!ds\,\, r'(s) e^{i x s} \right|^2,
  \end{align}
  in terms of the initial spectrum, $H_0\ket{\phi_n}$ $=$
  $\epsilon_{n0}\ket{\phi_n}$.  
  The correlation function
  $R(\omega)$ measures the spectral density of excitations created by
  $W$ and depends neither on the ramp time $\tau$ nor on the ramp
  shape $r(x)$, while the ramp spectrum $F(x)$ depends only on the
  ramp shape. In particular for fast ramps, $\tau$ $\lesssim$
  $\tau_\text{quench}$ $\approx$ $1/\Omega$, where $\Omega$ is the bandwidth
  of $R(\omega)$, the ramp shape does not matter, as one can then
  replace $F(\omega\tau)$ by $F(0)=1$ in Eq.~\eqref{fgr-exe}.  On the
  other hand, for large $\tau$ and continuous ramps, $F(\omega\tau)$
  develops a peak at $\omega$ $=$ $0$ and its finite width is
  responsible for the positive excitation energy $\Delta
  E(\tau)$. In general, smoother ramps lead to a faster decay of
  $F(x)$. For a continuous ramp without finite steps, $F(x)$ falls off
  at least as $1/x^2$. The simple linear ramp, $r_1(x)=x$, corresponds to $F_1(x)
  =2(1-\cos(x))/x^2$, but this can  usually be improved by making one
  or more of its derivatives continuous \cite{Eckstein2010ramps}, or
  by allowing $r(x)$ to oscillate~\cite{Eurich11}.
  
  \paragraph{Linear ramps in the Falicov-Kimball model.}
  
  The solution for quenches in the Falicov-Kimball (FK) model
  [Eq.~\eqref{eq:FK-model}] can also be adapted for continuous changes
  $v(t)$ or $U(t)$ \cite{Eckstein2010ramps}.  For a semielliptic
  density of states a Mott metal-insulator transition occurs at $U_c$
  $=2v_\ast$ in equilibrium DMFT, and for a time-dependent bandwidth
  [Eq.~\eqref{eq:semielliptic} with $v_*=v(t)$] the self-consistency
  equation is given by Eq.~\eqref{eq:selfcons-timedep}.

  For linear ramps of the hopping parameter from $v_i$ to $v_f$ the
  resulting excitation energy depends on the thermodynamic phase,
  namely \cite{Eckstein2010ramps}
  \begin{align}
    \label{exe-fkm}
    \Delta E(\tau)
    \stackrel{\tau\to\infty}{\sim}
    \begin{cases}
      \tau^{-{\textstyle\frac12}}
      & \text{across the transition},
      \\[0mm]
      \tau^{-1}
      & \text{in the metallic phase,}
      \\[0mm]
      \tau^{-2}
      & \text{in the insulating phase.}
    \end{cases}
  \end{align}

  Here the exponents for ramps inside one of the phases can be
  explained in terms of the perturbative result~\eqref{fgr-exe}. For
  an insulating initial state, the function $R(\omega)$ can be
  approximated by a sharp peak near $U$ due to the charge gap,
  $R(\omega)\propto\delta(\omega-U)$, and hence $\delta E(\tau)$
  $\propto$ $F_1(\omega\tau)$ $\propto$ $\tau^{-2}$. 
  This indicates that for such a ramp the excitation as a function of $\tau$ 
  strongly depends on the ramp shape, which is confirmed by the numerical 
  analysis. Also the asymptotic power-law behavior 
  for large $\tau$ is then determined by the ramp shape rather than by 
  intrinsic properties of the system.

  For ramps inside a gapless phase, on the other hand, not only the
  ramp shape $r(x)$, but also the excitation spectrum $R(\omega)$
  matters. Following \citet{Eckstein2010ramps}, we suppose a ramp is
  performed inside the metallic phase of the FK or, for comparison,
  the Hubbard model, i.e., starting from $U$ $=$ $0$ we turn on a
  (small) interaction $U$ linearly during the ramp time $\tau$.  In
  DMFT it follows from perturbation theory in $U$ that for this ramp
  $R(\omega)$ $\propto$ $\omega$ for the FK model, while $R(\omega)$
  $\propto$ $\omega^3$ for the Hubbard model, i.e., more excitations
  are created in the former non-Fermi liquid than the latter Fermi
  liquid. It now depends on the ramp shape whether this behavior
  $R(\omega)\sim\omega^\nu$ can be observed. Suppose the ramp shape is
  such that $F(x)\sim1/x^\alpha$ (e.g., $\alpha$ $=$ 2 for the linear
  ramp). Then only for sufficiently smooth ramps with $\alpha$ $>$
  $\nu$ the excitation energy is indeed given by $\Delta
  E\sim\tau^{-\nu}$, otherwise the `intrinsic' exponent $\nu$ is
  hidden and $\Delta E\sim\tau$ is determined by $\alpha$. Hence for
  the FK model, any continuous interaction ramp leads to $\Delta
  E\sim\tau^{-1}$. On the other hand, for the Hubbard model the
  excitations caused by a linear ramp, $\Delta E\sim\tau^{-2}$ (also
  obtained by \citet{Moeckel2010a}), already mask the intrinsic
  behavior of $R(\omega)$. This `universal' dependence of the
  excitation energy on the ramp time due to the ramp shape was
  also discussed for a veriety of systems by
  \citet{Haque2013a}.

\paragraph{Oscillating ramps in the Hubbard model}

Interaction ramps from $0$ to $U$ with arbitrary ramp shapes for the
Hubbard model in DMFT were studied by \citet{Eurich11}, with the aim 
of minimizing the excitation energy by suitably shaping the ramp for  
a given ramp time $\tau$. By optimizing the expression~\eqref{fgr-exe} 
it was found that ramps with oscillations, e.g., $r(x)=x+a\sin(2\pi nx)$ 
can lead to a lower excitation energy $\Delta E$ than linear ramps, where 
$n$ is an integer and $a$ is on the order of unity. For such ramps the function $F(x)$ first decays
from $F(0)$ $=$ $1$ for small $x$, then grows again to large values
for larger $x$. Nevertheless the excitation energy remains small
because $R(\omega)$ has a finite bandwidth and, depending on $\tau$,
may collect only a part where $F(\omega\tau)$ is
small, as supported by  DMFT weak-coupling CTQMC data, which agree with
the APT estimates [Eq.~\eqref{fgr-exe}].
Physically this means that if one cannot take sufficient time to slowly change the
Hamiltonian, it may in some cases be better to change it so quickly that the system
cannot follow at all.

In practice, not only a small excitation energy may be desirable, but also a speedy 
thermalization after the ramp is finished. \citet{Eurich11} found that a similar critical value 
of $U_c$ 
for rapid thermalization exists as in the case of a sudden quench (Sec.~\ref{subsubsec:quenches}),
for which rapid thermalization occurs after the ramp. For fixed ramp time $\tau$ $=$ 1.25 they 
obtained $U_c^\text{linear}\approx 3.75$ for linear ramps, while $U_c^\text{oscillating}\approx 4.25$ 
for oscillating ramps with $r(x)=x+0.87\sin(4\pi x)$. This may be compared to variational results 
by \citet{Sandri2012a}, who found that $U_c$ increases towards the equilibrium critical value for 
the Mott transition as $\tau$ tends to infinity. The study of ramps may thus help to understand the 
relation between the critical interaction values in equilibrium and nonequilibrium.

%\input{summary}
%%%%%%%%%%%%%%%%%%%%%%%%%%%%%%%%%%%%%%%%%%%%%%%%%%%%%%%%%%%%%
\section{Concluding remarks and prospects}

We have given an overview of the nonequilibrium DMFT formalism for the study 
of nonequilibrium phenomena in correlated fermionic lattice systems. 
While the extension of the DMFT formalism to nonequilibrium situations 
involves, formally, only the replacement of the imaginary-time interval with 
a suitable contour, we have shown that   
the solution of the DMFT equations on this contour is 
different and requires significantly elaborate techniques.   
While in thermal equilibrium the Green's functions and self-energies are time-translation invariant, these quantities
depend on two time variables in nonequilibrium, and the DMFT equations 
either acquire an additional Floquet matrix-structure (for periodically driven systems), or they become integral-differential 
equations on the contour (for temporal evolutions 
after a disturbance). The integral-differential equations 
are best solved by implementing a step-by-step propagation on the real-time axis, starting from some equilibrium DMFT solution. 
%As another important difference 
%from the equilibrium physics, where distribution functions and spectral functions are related by the 
%fluctuation-dissipation relation, the relation in nonequilibrium is only fixed by the initial condition, or the balance between the external 
%driving and the dissipation. 
Furthermore, while distribution functions and spectral functions in
equilibrium are related by the fluctuation-dissipation relation, their
relation in nonequilibrium is only fixed by the initial condition or by
the balance between the external driving and the dissipation.
In nonequilibrium DMFT, one thus has to solve a set of coupled equations for the lesser, 
retarded and greater Green's functions, known as the Kadanoff-Baym equations.

A challenge for the implementation of the contour-equations is that typical applications require a high accuracy. 
For example, one often wants to characterize the relaxation and thermalization of some excited state that involves 
tiny differences between the time-dependent observable and an independently computed 
expectation value in thermal equilibrium. 
In order to achieve a high accuracy, while keeping a relatively large time step, 
it is important to use a high-order scheme for the solution of the integral-differential equations. 

While insights into the nonequilibrium DMFT formalism can be gained from the Falicov-Kimball model, which has the virtue of being exactly solvable within 
DMFT, generic models of correlated electron systems such as the Hubbard model and its multi-orbital extensions require a numerical solution of the effective quantum impurity model.  
Various techniques familiar from the study of 
impurity problems in equilibrium have recently been adapted to nonequilibrium 
situations, and tested as impurity solvers for nonequilibrium DMFT. 
One of these techniques is the continuous-time Monte Carlo approach, which 
is the method of choice for most equilibrium applications. 
This method has the advantage that it is numerically exact and can cover the weak, intermediate and strong 
correlation regimes. However, the implementation on the Keldysh contour leads to a severe sign problem, 
which effectively restricts the applicability of this technique to nonequilibrium situations where 
the interesting dynamics is very fast ($\sim$ femto-seconds in the case of electron systems). 

We have also discussed and illustrated the use of low-order weak-coupling and strong-coupling perturbation theories as impurity solvers for DMFT. 
These methods are computationally less demanding than the Monte Carlo approach, and give reliable results in 
the weak and strong-coupling regimes. In the weak-coupling perturbative method, it is 
often better to use a diagrammatic expansion based on bare (rather than renormalized) propagators. 
While the approximation is then not conserving, it can still accurately reproduce the dynamics in the weak-coupling regime. 
The strong-coupling perturbative method, on the other hand, involves renormalized propagators (and is hence conserving), but still shows good convergence 
with the order of the approximation. The perturbative solvers have been important in the study of phenomena that require simulation times $\sim 100$ fs, as in relaxation processes in a photo-doped Mott insulator, or the calculation of the 
evolution of order parameters in symmetry-broken states. While the lowest-order implementations 
of the perturbative solvers are similar, in terms of computational effort, to the solution of the contour equations, 
the effort increases polynomially with the order of the approximation.  

The impurity solvers described in this review have 
lead to interesting insights into the nonequilibrium dynamics of correlated 
systems as typically described by the Hubbard model in quite a wide range of applications,  
and have thus been essential for establishing the formalism as a viable tool for the description 
of electronic excitation and relaxation phenomena in strongly correlated 
systems. 
We have reviewed the major topics which have been successfully addressed with nonequilibrium 
DMFT over the past several years. These applications can be grouped into (i) simulations of phenomena 
occurring in correlated lattice systems driven by strong electric fields, and (ii) the study of relaxation phenomena after time-dependent 
parameter changes. The former are relevant to pump-probe experiments on correlated electron systems, while the latter is more directly related to 
cold-atom systems, where parameters such as the depth of the lattice potential or the interaction strength can actually be varied. 
Both types of applications demonstrate that nonequilibrium DMFT calculations 
not only reproduce phenomena seen in experiments but also 
theoretically expected ones, 
such as collapse-and-revival oscillations after a quench into the strong-coupling regime \cite{Eckstein09}, or Bloch oscillations in metallic systems subject to a strong DC field \cite{Freericks2006}.  
More importantly, these calculations gave new insights into phenomena which emerge 
specifically in  nonequilibrium correlated electron systems. Prime examples are the 
numerical demonstration of a dynamical phase transition in the relaxation dynamics after an interaction quench \cite{Eckstein09}, or the finding 
that the effective Coulomb interaction can be tuned and even become attractive by the application of periodic electric fields \cite{Tsuji11}.   

The technical challenge in coming years will be to develop 
powerful and flexible impurity solvers for nonequilibrium DMFT.  
Besides the obvious extension to multi-orbital systems, an important direction for future 
will be the study of electron-phonon coupled systems \cite{Werner2013phonon}. 
A remarkable experimental result has recently been obtained 
by coherently exciting phonons in cuprates \cite{Fausti11}, and nonequilibrium DMFT may provide insights into such phenomena. 
Cluster extensions of DMFT \cite{Tsuji2013cluster} will enable the study of the dynamics of $d$-wave 
superconductors. 
The extension of the DMFT formalism to spatially 
inhomogeneous systems will enable us to 
simulate nonequilibrium phenomena around surfaces or interfaces \cite{Eckstein2013layer}.   
Let us also mention that equilibrium DMFT has recently been applied with 
remarkable success to bosonic systems \cite{Anders2010, Anders2011}, so that it will be desirable to extend 
it to a nonequilibrium bosonic DMFT (with bosonic impurity solvers).   

For realistic materials calculations, the combination of electronic-structure input with DMFT is becoming an established and powerful  
method \cite{Kotliar2006, Held2007}. Thus a desirable direction 
is to combine the nonequilibrium DMFT formalism  with 
the first-principles electronic structure, which will enable us 
to quantitatively analyse e.g. time-resolved photoemission spectra. 
This step raises intriguing and fundamental questions, 
e.g., how the down folding of the electronic structure 
into an effective lattice model should be done 
in nonequilibrium, and how one can treat electromagnetic fields 
far beyond linear response.

Finally, we emphasize that the study of nonequilibrium 
quantum systems has a long and very {\it interdisciplinary} history.  
Many important concepts were indeed developed in parallel in condensed 
matter physics and field theory, and some of them were put forward  
at the emerging stage of quantum mechanics.  
(i) For instance, the application of intense lasers may drive a nonequilibrium phase transition of the vacuum 
(in the field-theoretic language) \cite{PIF}, 
and this is intimately related to the photo-induced phase transitions 
discussed here, although there are orders of magnitude differences in the relevant energy scales.  
(ii) The dielectric breakdown of the Mott insulator is related to the Schwinger mechanism for the breakdown 
of the QED vacuum in high-energy physics \cite{HeisenbergEuler,Schwinger51}. The latter refers to a quantum 
tunneling across the mass gap of the electron 
(the energy required to create an electron-positron pair), which 
is $\Delta=2m_{e}c^2\sim10^6 \mbox{eV}$ with 
$m_{e}=5\times 10^5\mbox{eV}$ the electron mass and 
$c$ 
the speed of light.  The threshold field strength $F_{\rm th}$, 
at which $\xi eF_{\rm th} \sim \Delta$, with $\xi$ the size of an electron-positron pair $\sim$ Compton wavelength $\hbar/m_e c$, is given in QED by
$
F^{\rm QED}_{\rm th}=m_{e}^2c^3/e\hbar \sim 10^{8}\;\mbox{V/\AA}.
$
This is gigantic, although the possibility of realizing it with free-electron lasers is being discussed, while 
in condensed-matter physics, the energy gap, $\Delta\sim 1\;\mbox{eV}$, is orders of magnitude smaller. 
In strongly-correlated systems, where the gap is a many-body (Mott) gap $\Delta_{\rm Mott}$, a Mott insulator in an intense electric field is predicted to become metallic with the threshold 
field \cite{Okabethe,Oka2012prb}
$E^{\rm Mott}_{\rm th} \sim \Delta_{\rm Mott}/\xi\sim 0.1 \mbox{V/\AA}
$,
where $\xi \sim 10\AA$ in this case is the size of a doublon-hole pair 
for a typical Mott insulator with doublon-hole recombination corresponding to pair 
annihilation in field theory.  The maximum intensity of ultrashort laser 
pulses currently available is well above this condensed-matter version of the Schwinger limit.  
(iii) Floquet picture described in the present review also has 
a field-theoretic counterpart as Furry picture.  
(iv) The interdisciplinary concepts extend to the relaxation processes.  
In fact, the concept of prethermalization \cite{Berges2004a} was originally proposed in the study of the quark-gluon 
plasma production in hadrons out of equilibrium, as typically realized experimentally in the relativistic heavy-ion 
colliders (RHIC). 
%This is no wonder, since the hadron physics is a quantum many-body problem.  
Another example is the Kibble-Zurek mechanism \cite{Kibble1976,Zurek1985} originally 
proposed for phase transitions in the early universe, which is now being studied in connection with 
quench dynamics near quantum critical points \cite{Dziarmaga1999}, 
in cold atom systems \cite{Horiguchi2009,Weiler2008,Sadler2006,Saito2007}
and even in real materials \cite{Griffin2012}.

\acknowledgments
We wish to thank R.~Arita, P.~Barmettler, P.~Beaud, J.~Berges, S.~Biermann, A.~Cavalleri, E.~Demler, A.~Dirks, N.~Eurich, J.~Freericks, L.~Fu, A.~Georges,
D.~Greif, E.~Gull, H.~Hafermann, C.~Hauri, G.~Ingold, S.~Ishihara, C.~Jung, S.~Johnson, S.~Koshihara, A.~Lichtenstein, S.~Kaiser, S.~Kehrein, H.~Kishida, T.~Kitagawa, 
M.~Knap, A.~Komnik, N.~Konno, H.~Matsueda, A.~Millis, S.~Miyashita, M.~Moeckel, L.~M\"uhlbacher, Y.~Murakami, S.~Nakamura, K.~Nasu, D.~Nicoletti, T.~Ogawa, H.~Okamoto, O.~Parcollet, V.~Pietil\"{a}, R.~Shimano, K.~Tanaka, L.~Tarruell, 
T.~Tohyama, S.~Tsuneyuki, K.~Ueda, D.~Vollhardt, K.~Yonemitsu 
for stimulating discussions and collaborations on topics related to this review. 
H.A., T.O. and N.T. were supported in part by Grants-in-Aid for Scientific
Research on Innovative Areas from JSPS, Grant Nos. 23740260, 23104709 and 23110707.
M.K. was supported in part by Transregio 80 of the Deutsche Forschungsgemeinschaft.
M.E., N.T. and P.W. acknowledge support by the Swiss National Science Foundation (Grant No. PP0022-118866) 
and FP7/ERC starting grant No. 278023.

%\input{appendices}
%%%%%%%%%%%%%%%%%%%%%%%%%%%%%%%%%%%%%%%%%%%%%%%%%%%%%
\appendix

\section{Numerical solution of Volterra integral-differential equation}
\label{Appendix A}

In this supplementary material, we briefly discuss 
the numerical implementation of the Volterra integral-differential equation
\begin{align}
\frac{d}{dt} y(t) = q(t) + p(t)y(t) + \int_0^t \!d\bar t\,  k(t,\bar t) y(\bar t).
\label{appendix:Volterra}
\end{align}
This type of equation frequently appears in nonequilibrium 
DMFT calculations (Sec.~\ref{Sec_DMFT}, Eq.~(\ref{volterra diff})), 
in particular in solving the nonequilibrium Dyson equation (Sec.~\ref{quantum boltzmann}). 
One also encounters a Volterra integral equation of a form
\begin{align}
y(t)=q(t)+ \int_0^t \!d\bar t\,  k(t,\bar t) y(\bar t),
\label{appendix:Volterra integral}
\end{align}
which is a special case of Eq.~(\ref{appendix:Volterra}), and can be solved in the same way as Eq.~(\ref{appendix:Volterra}).  

Various numerical algorithms to solve Eq.~(\ref{appendix:Volterra}) are found in the  literature
\cite{NumericalRecipesC, LinzBook, BrunnervanderHouwenBook}. Here we present
the implicit Runge-Kutta method (or the collocation method) \cite{Tsuji2013},
which may not be the most efficient one,
but it allows us to discuss the relevant issues. 
In practice, we discretize the time with equal spacing,
$t_i=i\times \Delta t$ ($i=0, 1, \dots, n$), with $\Delta t=t_{\rm max}/n$.
It is crucial to employ higher-order schemes to accurately simulate the long-time evolution.
The $m$th order scheme has numerical errors of $O(n(\Delta t)^{m+1})=O(t_{\rm max}(\Delta t)^m)$.
Typically we require $m\ge 2$ to control the errors. In the following, we 
explicitly give expressions for the second-order and fourth-order schemes.

Equation (\ref{appendix:Volterra}) can be solved by increasing $t_{\rm max}=t_n$
step by step on the discretized grid from the initial condition $y(t_0)=y(0)$
due to the causality. 
To get $y(t_n)$, we replace the differential 
operator on the left hand side of Eq.~(\ref{appendix:Volterra}) 
by an integral, which is numerically evaluated by an appropriate
numerical integration formula,
\begin{align}
y(t_n)-y(t_0)
&=
\int_{t_0}^{t_n} d\bar t\, y'(\bar t)
\approx
\Delta t\sum_{i=0}^n w_{n,i} y'(t_i)
\label{appendix:t_n}
\end{align}
with $w_{n,i}$ ($i=0,1,2,\dots,n$) the corresponding weights. Since $y(t_{n-1})$ is
already known from the previous calculation, we also use the relation
\begin{align}
y(t_{n-1})-y(t_0)
&=
\int_{t_0}^{t_{n-1}} d\bar t\, y'(\bar t)
\approx
\Delta t\sum_{i=0}^{n-1} w_{n-1,i} y'(t_i)
\label{appendix:t_n-1}
\end{align}
By subtracting Eq.~(\ref{appendix:t_n-1}) from Eq.~(\ref{appendix:t_n}), we get
\begin{align}
y(t_n)-y(t_{n-1})
&=
\Delta t\sum_{i=0}^{n-1} (w_{n,i}-w_{n-1,i}) y'(t_i)+\Delta t w_{n,n}y'(t_n).
\label{appendix:y(t_n)-y(t_n-1)}
\end{align}
Here  
$y'(t_n)$ is evaluated from Eq.~(\ref{appendix:Volterra}) as
\begin{align}
y'(t_{n})
&=
q(t_{n})+p(t_{n})y(t_{n})+\int_0^{t_{n}} d\bar{t}\, k(t_n,\bar{t}) y(\bar{t})
\nonumber
\\
&\approx
q(t_{n})+p(t_{n})y(t_{n})+\Delta t\sum_{i=0}^{n} w_{n,i} \,k(t_{n},t_i)y(t_i),
\label{appendix:y'(t_n)}
\end{align}
Equations (\ref{appendix:y(t_n)-y(t_n-1)}) and (\ref{appendix:y'(t_n)})
consist of a set of linear equations for $y(t_{n})$, so that one can explicitly solve them,
\begin{align}
y(t_n)
&=
\left[1-\Delta t w_{n,n}p(t_n)-(\Delta t w_{n,n})^2k(t_n,t_n)\right]^{-1}
\nonumber
\\
&\quad\times
\left\{y(t_{n-1})+\Delta t\sum_{i=0}^{n-1} (w_{n,i}-w_{n-1,i}) y'(t_i)\right.
\nonumber
\\
&\quad
\left.+\Delta t w_{n,n}\left[q(t_n)+\Delta t\sum_{i=0}^{n-1} w_{n,i} \,k(t_{n},t_i)y(t_i)\right]\right\}.
\label{appendix:y(t_n)}
\end{align}
As we will see below, $w_{n,i}-w_{n-1,i}$ vanishes for most $i$'s
so that one has to store $y'(t_i)$ for only a few $i$ ($i=n-1$ in the second-order
scheme, and $i=n-3, n-2, n-1$ in the fourth-order scheme).
For the use in the next steps ($t_{\rm max}=t_{n+1}, \dots$), we calculate
$y'(t_n)$ from Eq.~(\ref{appendix:y'(t_n)}) with $y(t_n)$ substituted with the result of (\ref{appendix:y(t_n)}). To avoid repeated calculations of the sum in Eqs.~(\ref{appendix:y'(t_n)}) and (\ref{appendix:y(t_n)}),
it is efficient to store them in memory.

In the same way, the Volterra integral equation (\ref{appendix:Volterra integral}) is solved as
\begin{align}
y(t_n)&=
\left[1-\Delta t\, w_{n,n}k(t_n,t_n)\right]^{-1}
\left[q(t_n)+\Delta t\sum_{i=0}^{n-1}w_{n,i}k(t_n,t_i)y(t_i)\right].
\label{appendix:y(t_n) int}
\end{align}

In the $m$th order scheme, we employ the numerical integration formula with
numerical errors of $O(n(\Delta t)^{m+1})=O(t_{\rm max}(\Delta t)^m)$.
In the second-order scheme, one can use the trapezoid rule with weights
\begin{align}
w_{n,i}
&=
\begin{cases}
1/2 & i=0, n, \\
1 & 1\le i \le n-1.
\end{cases}
\end{align}
In the fourth-order scheme, one can use Simpson's rule for $n=2$,
\begin{align}
w_{2,i}
&=
\begin{cases}
1/3 & i=0, 2, \\
4/3 & i=1,
\end{cases}
\end{align}
Simpson's $3/8$ rule for $n=3$,
\begin{align}
w_{3,i}
&=
\begin{cases}
3/8 & i=0, 3, \\
9/8 & i=1, 2,
\end{cases}
\end{align}
the composite Simpson's rule for $n=4$,
\begin{align}
w_{4,i}
&=
\begin{cases}
1/3 & i=0, 4, \\
4/3 & i=1, 3, \\
2/3 & i=2,
\end{cases}
\end{align}
and the fourth-order Gregory's rule for $n\ge 5$,
\begin{align}
w_{n,i}
&=
\begin{cases}
3/8 & i=0, n, \\
7/6 & i=1, n-1, \\
23/24 & i=2, n-2, \\
1 & 3 \le i \le n-3.
\end{cases}
\end{align}

The remaining task is to get the starting value $y(t_1)$. Since the higher-order integral formulae need
at least three points, the above approach cannot be directly applied for $n=1$.
One way to get around this is to take very fine grids on $t_0\le t\le t_1$, 
and use a lower-order integral formula (trapezoid rule). Another way is to take
the middle point $t_{1/2}=\Delta t/2$ \cite{LinzBook}, and apply Simpson's rule to the integral from $t_0$ to $t_1$,
\begin{align}
y(t_1)-y(t_0)
&\approx
\frac{\Delta t}{6}\left[y'(t_0)+4y'(t_{1/2})+y'(t_1)\right].
\label{appendix:y(t_1)-y(t_0)}
\end{align}
The value at the middle point is obtained from the quadratic interpolation,
\begin{align}
y'(t_{1/2})
&\approx
\frac{3}{8}y'(t_0)+\frac{3}{4}y'(t_1)-\frac{1}{8}y'(t_2),
\end{align}
which has an error of $O((\Delta t)^3)$ for the smooth function $y(t)$.
Since $y'(t_{1/2})$ is multiplied with $\Delta t$ in Eq.~(\ref{appendix:y(t_1)-y(t_0)}), the overall error is
of $O((\Delta t)^4)$, which is compatible with that for the fourth-order scheme.
$y'(t_0)$ is known from the initial condition.
$y'(t_2)$ is derived from Eq.~(\ref{appendix:y'(t_n)}).
$y'(t_1)$ is calculated by Simpson's rule with the middle point,
\begin{align}
y'(t_1)
&\approx
q(t_1)+p(t_1)y(t_1)
\nonumber
\\
&\quad
+\frac{\Delta t}{6}[k(t_1,t_0)y(t_0)+4k(t_1,t_{1/2})y(t_{1/2})+k(t_1,t_1)y(t_1)].
\end{align}
One can repeat the quadratic interpolation to get the middle-point values,
\begin{align}
y(t_{1/2})
&\approx
\frac{3}{8}y(t_0)+\frac{3}{4}y(t_1)-\frac{1}{8}y(t_2),
\\
k(t_1,t_{1/2})
&\approx
\frac{3}{8}k(t_1,t_0)+\frac{3}{4}k(t_1,t_1)-\frac{1}{8}k(t_1,t_2).
\label{appendix:kernel}
\end{align}
Thus, the equations for $y(t_1)$ depend on $y(t_2)$. On the other hand, $y(t_2)$
can be determined from $y(t_0)$ and $y(t_1)$ as described above.
In total, we have a combined set of linear equations that determines $y(t_1)$ and $y(t_2)$
simultaneously.

The middle-point approach has a subtle problem when it is applied to a retarded kernel that has a causality, $k(t,t')=0$ ($t<t'$).
Since $k(t,t')$ is smooth only for $t\ge t'$, the quadratic interpolation (\ref{appendix:kernel}) using $k(t_1,t_2)$ ($t_1<t_2$) is inapplicable in the present form. 
This problem can be avoided by taking the mirror image for $k(t,t')$ in $t<t'$ to realize a function that is smooth in the entire $(t,t')$
and is equal to $k(t,t')$ for $t>t'$ \cite{Tsuji2013}.

\section{Sample programs}

For a pedagogical purpose, we provide sample program codes for the nonequilibrium DMFT, in both C++ and Fortran, as
Supplemental Material. To enhance the readability of the codes,
we focus on a particular set-up: the program solves an interaction-quench problem (Sec.~\ref{subsubsec:quenches})
for the single-band Hubbard model (\ref{dmft_formalism::h-hubbard}) with the semicircular density of states
at half-filling. It assumes a paramagnetic phase with no long-range orders.
The impurity solver is the second-order weak-coupling perturbation theory (iterated perturbation theory)
(Sec.~\ref{weak-coupling perturbation}), where the self-energy is given by
\begin{align}
\Sigma(t,t')=U(t)U(t')\mathcal G_0(t,t')\mathcal G_0(t',t)\mathcal G_0(t,t')
\end{align}
with $\mathcal G_0(t,t')$ the Weiss Green's function.
To solve the Dyson equation (Sec.~\ref{quantum boltzmann}), 
we use the second-order scheme (Appendix \ref{Appendix A}) for the Volterra integral-differential equation.

To install the codes, download the tar file from Supplementary Material and untar them in a certain working directory:
\\

{\tt \$ tar zxvf noneq-dmft.tar.gz} 
\\

\noindent
It generates the subdirectories `{\tt cxx}' and `{\tt fortran}', which contain the C++ and Fortran codes, respectively.
To compile them, one needs the FFTW library for fast Fourier transformation, which can be downloaded.\footnote{See {\tt http://www.fftw.org}.} 
One should specify the path for the FFTW library in the make file, which is 
by default set to {\tt /usr/local}.
To build the code, execute make in the directory in which it is installed:
\\

{\tt \$ make}
\\

\noindent
If the build is successful, it generates an executable file {\tt a.out}. It requires input parameters,
which are listed in the file {\tt parm.sh}. In the sample programs, the parameters `{\tt dos}' (density of states) 
and `{\tt solver}' (impurity solver) are restricted to be `{\tt semicircular}' and `{\tt IPT}' (iterative perturbation theory), respectively,
while the other parameters can be freely changed. After choosing the parameters, run the program by typing:
\\

{\tt \$ ./parm.sh}
\\

\noindent
During the execution, it outputs a measure of the DMFT convergence ({\tt |G0\_new-G0\_old|})
as well as the time up to which the system has been evolved. After the simulation has finished,
it automatically creates the following output files:
\\

{\tt density}

{\tt double-occupancy}

{\tt interaction-energy}

{\tt kinetic-energy}

{\tt total-energy}
\\

\begin{figure}[tb]
\begin{center}
\includegraphics[width=8cm]{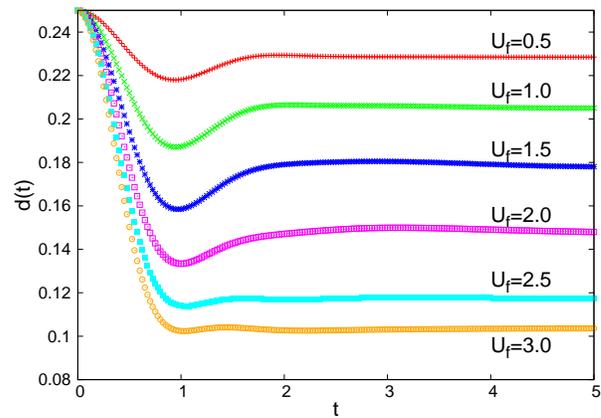}
\caption{The results of the sample program for the time evolution of the double occupancy $d(t)$
after interaction quenches from $U_i=0$ to $U_f$ with $\beta=16$.}
\label{appendix:double occupancy}
\end{center}
\end{figure}

\noindent
In Fig.~\ref{appendix:double occupancy}, we show some results obtained with the sample program.
These results can be used to check the correctness of the program output.

The sample programs have been designed and implemented by N. Tsuji, one of the authors.
The codes can be used and modified for non-commercial purposes, but their use must be acknowledged in publications with a citation to this review article.

\bibliographystyle{apsrmp}
\bibliography{rmpbib}

\end{document}